%% file: thesis.tex
\documentclass[12pt,a4paper,openright,oneside]{book} 

\usepackage{fancyheadings}

\usepackage{thesis}
\usepackage{setspace}
\usepackage{epsfig}
\usepackage{graphics}
\usepackage{amsmath}
\usepackage{bm} 
\usepackage{bbm} 
\usepackage{psfrag}
\usepackage{cite}

\usepackage{caption}
\usepackage{makeidx}

\newcommand{\ket}[1]{\lvert#1\rangle}
\newcommand{\bra}[1]{\langle#1\rvert}

\newcommand{\vek}[1]{{\mathbf{#1}}}
\newcommand{\vekk}[1]{{\bm{#1}}}

\newcommand{\braket}[2]{\langle#1|#2\rangle}
\newcommand{\im}{{\rm i}}

\newcommand{\abs}[1]{|#1|}
\newcommand{\iden}{\mathbbm{1}}

\newcommand{\half}{\frac{1}{2}}

\DeclareMathOperator*{\Tr}{Tr}

\makeindex
\usepackage{eurosans} 

\usepackage{afterpage}

\reversemarginpar

\begin{document}

\include{titlepage}
\include{newfrontmatter}

\pagestyle{fancyplain} 
\renewcommand{\chaptername}{Chapter} 
\part{Introductions}\label{part1}
\include{c-1}         
\include{c-2}          
\part{Robust Entanglement Generation in  Atoms and Cavities}\label{part2}
\include{c-3}
\include{c-4}
\include{c-5}

\part{Entanglement Distillation  with Linear Optics}\label{part3}
\include{c-6}
\include{c-7}
\include{c-8}
\input{c-9}
\include{conc}

\include{backmatter}
\include{acknow}
\end{document}

%% file: titlepage.tex
\pagestyle{empty}

\begin{center}
  {\sc{University of London}}\\[1.0cm]
  \centerline{\includegraphics[height=3.5cm]{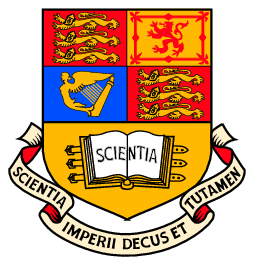}
}
  \vspace{1.0cm}
  Imperial College of
  Science, Technology and Medicine\\ The
  Blackett Laboratory\\ Quantum Optics \& Laser Science Group\\[1.5cm]
  
  {\Huge Generation and Manipulation}\\[0.5cm]
  {\Huge of Entanglement in}\\[0.5cm]
  {\Huge Quantum Optical Systems}\\[2cm]

  {\Large{Daniel Edward Browne}}\\[1.5cm]

  Thesis submitted in partial fulfilment of the \\
  requirements for the degree of\\
  Doctor of Philosophy\\
  of the University of London\\
  and the Diploma of Membership of Imperial College.\\[.8cm]
  \vfill
December 2004
\end{center}


%% file: newfrontmatter.tex
\clearpage \pagestyle{empty}


\begin{center}
  {\Huge Abstract}\\[1cm]
\end{center}

In this thesis, we introduce and evaluate several new procedures for the generation and manipulation of entangled states in quantum optical systems. Each  is evaluated in terms of realistic models of the imperfect apparatus of any real laboratory implementation.

First we consider atoms interacting with optical cavities.
We describe a  method of generating maximally entangled states in  two separate microwave cavity modes by the passage of an atom between them and its subsequent measurement. We demonstrate the natural robustness of this approach against a variety of experimental imperfections. 

We then introduce a further scheme for the generation of maximally entangled states between pairs atoms or ions trapped in  separate cavities. The ions are weakly driven by an external field and a leaking photon  from one cavity is  detected, after passing through a beam splitter to erase which-path information. In the ideal case, this leads to the deterministic generation of high fidelity states, and when more realistic conditions are assumed, still allows high fidelity states to be generated in a heralded non-deterministic manner.

Finally, we turn our attention to the manipulation of entangled states of light pulses. Entanglement distillation has many applications in quantum information science. 
Gaussian states are the natural states of light pulses produced in laser experiments, and  standard optical operations preserve the Gaussian form.
 Entanglement distillation on Gaussian states is impossible with these operations alone. We illustrate that the additional element of photon detection enables one to circumvent this restriction, but produces highly non-Gaussian states. We then introduce a simple  protocol, which we call ``Gaussification'', which allows more highly entangled approximately Gaussian states to be distilled from a non-Gaussian supply. Together, these form an entanglement distillation scheme for Gaussian states, whose convergence characteristics and realistic operation we analyse.

\tableofcontents


%% file: c-1.tex
\chapter{Quantum Mechanics and Entanglement}
\label{ch:1}

Entanglement is the feature of quantum mechanics which sets it apart most starkly from the classical theories that preceded it. It has long been seen as an interesting curiosity, but  the establishment of \emph{quantum information science} has given it a new role. 
The goal of {quantum information science} is to exploit the features of quantum mechanics which differ from classical physics, in order to accomplish tasks which conventional reasoning, based implicitly on classical physics,  would deem   difficult or even impossible.

In this thesis, we will introduce several new proposals for the generation and manipulation of entanglement. Each proposal is practically motivated and based in the physics of real quantum optical systems. We will assess the feasibility of each with detailed simulation of the most important imperfections which one would expect to encounter in the laboratory.

In this initial chapter, we will introduce the non-classical properties of entanglement and discuss its quantification and some of its applications in quantum information science. First, though, we will briefly review the most important quantum mechanical concepts,
 focusing on the features of the theory which have most importance for quantum information science.
We will  assume that the reader has a basic knowledge of linear algebra. Good introductions to linear algebra as applicable to quantum mechanics can be found, for example, in \cite{cohenbook,nielsenchuang}. Our discussion will be restricted to an overview of the most important concepts. More detailed introductions can be found in the references just cited, and many other quantum mechanics textbooks.

\section{Quantum mechanics}
\subsection{Quantum systems and quantum states}

A \emph{quantum system} is a number of physical degrees of freedom in a physical object or set of objects which is to be described quantum mechanically. For example, this can include the position or  momentum of a particle,  the internal energies of several particles or the energy of an electro-magnetic field. 

The \emph{quantum state} of a system is a mathematical object which represents the knowledge we have about the system,
and from which all measurable physical quantities relating to the system can be calculated.




A special class of quantum states are called the \emph{pure states}. Pure states, denoted in Dirac's notation by a ``ket'', written $\ket{\psi}$,  represent a ray in a  Hilbert space $\mathcal{H}$, i.e. a complex vector space with an inner product. The dimension of $\mathcal{H}$ is a property of the degrees of freedom being described. For example, the state of a spin-half particle lives in a two-dimensional Hilbert space, while the Hilbert space for its position has infinite dimensions. Due to their simplicity and its importance for quantum information science, we will often use a two-dimensional systems for illustration. Such a systems is called a quantum bit or \emph{qubit}, and its  basis vectors are labelled $\ket{0}$ and $\ket{1}$.

When we express a pure state in terms of a basis in $\mathcal{H}$ it takes the form
\begin{equation}
\ket{\psi}=\sum_i \alpha_i \ket{i}\ ,
\end{equation}
where $\ket{i}$ are normalised basis states. Thus pure states are sometimes called \emph{state vectors}. The dual space  $\mathcal{H}^*$ also plays an important role. Its elements are called ``bras'' and written $\bra{\psi}$. The dual of a vector in $\mathcal{H}$ represents a linear functional on $\mathcal{H}$ and thus the inner product $\langle\psi|\phi\rangle$ is a complex number while $\ket{\psi}\bra{\phi}$ is an operator on $\mathcal{H}$.


The pure states are only a very special class of state, whose properties will be described below. More general states cannot be described by a simple state vector, but require a \emph{density matrix} \cite{voneumanndens,landaudens}. A density matrix is a positive Hermitian operator  on $\mathcal{H}$ with finite trace, and a vector in the product space $\mathcal{H}\times\mathcal{H}^*$ It is usually denoted $\rho$. A basis in $\mathcal{H}\times\mathcal{H}^*$ is formed from the product of the set of basis vectors of $\mathcal{H}$ with the basis of dual space $\mathcal{H}^*$. Thus a general state may be expressed
\begin{equation}
\rho=\sum_{i,j}\rho_{i,j}\ket{i}\bra{j}\ .
\end{equation}
 The state must be normalisable, i.e. capable of being rescaled so that it has unit trace. This is especially important for infinite dimensional states as it means that the trace of $\rho$ must be finite, or in other words $\rho$ must be \emph{trace-class}. When working with infinite dimensional states it is sometimes useful to formulate ``states'' which are not trace-class. Such states are called improper and cannot exist in nature, but can be useful mathematical tools. It is usually convenient to work with \emph{normalised} states. Normalisation is the multiplication of the state with a real factor such that the trace is unity. In this thesis we shall assume all states are normalised unless otherwise specified.

\subsection{Combining and disregarding sub-systems}

Given two quantum systems with state spaces  $\mathcal{H}_1$ and $\mathcal{H}_2$, their joint state is described by a state on the tensor product space  $\mathcal{H}_1\otimes\mathcal{H}_2$. The description of the state then proceeds as above within this larger space. For example, the joint state of two qubits, the first in state $\ket{0}$ and the second in $\ket{1}$ is written $\ket{0}\otimes\ket{1}$. 
This tensor product structure has the consequence, however, that there exist states in the joint space which did not exist when the systems were described separately. These are the  \emph{entangled states} which will be discussed at length below.

Given a quantum system made up of several sub-systems, it is natural to ask how the system should be described when one or more sub-system is out of reach of the experimenters. Can a state of the remaining system be described in a consistent way? The answer is yes, as long as these sub-systems interact at no later time with the sub-systems left behind. The procedure to construct this \emph{reduced state} of the remaining sub-systems is simple. One performs a \emph{partial trace} over the sub-space corresponding to the sub-system which is out of reach,

\begin{equation}
\rho=\sum_{i,j,k,l}\rho_{i,j,k,l}\ket{i}\ket{j}\bra{k}\bra{l}\rightarrow \sum_j\sum_{i,l}\rho_{i,j,k,j}\ket{i}\bra{k}\ .
\end{equation}
In fact,  the partial trace is the only way of constructing the reduced state which satisfies the following physically reasonable demand. Once the lost sub-system is out of reach, no operations on it may affect the measurement statistics of the remaining sub-systems\footnote{It is this condition in fact which means that even though entangled states have apparently ``non-local'' characteristics, no operation is possible which allows instantaneous signalling between distant parties, and thus there is no conflict between quantum mechanics and special relativity.}.
As we shall see below, the partial trace is an important tool for the quantification of entanglement in pure states.

\subsection{Dynamics}\label{sec:dynamics}

The evolution of a quantum system is determined by its Hamiltonian $\hat{H}$, the energy observable for the system, according to the von-Neumann equation
\begin{equation}
\dot{\rho}=-\frac{i}{\hbar}[\hat{H},\rho]\ ,
\end{equation}
the density matrix form of the famous Schr\"{o}dinger equation. Formally integrated, this gives,
\begin{equation}
\rho(t)= U(t) \rho(t=0) U(t)^\dag\ ,
\end{equation}
where $U(t)=\mathcal{T}\exp[-i\int_0^t\hat{H(t')}dt'/\hbar]$  and $\mathcal{T}$ is the time-ordering operator, which is needed since Hamiltonians at different times may not always commute. We see that $U(t)$ is a unitary operator since $U(t)U^\dag(t)=\iden$.  Thus the most general evolution for a closed quantum system is unitary evolution.

This  does not apply for \emph{open} quantum systems, such as an atom interacting with surrounding electro-magnetic field modes, where only  a sub-system of the unitarily evolving system is described. In this case, the evolution of the state is described by a completely positive map. A  positive map is a map which preserves the positive nature of a state and thus transforms it into another  physical state. Complete positivity demands that $\iden\otimes\rho$ is also positive. An example of an operation which is positive and not completely positive is the transpose operation, which takes a state's matrix elements $\bra{i}\rho\ket{j}$ to their diagonally opposite counterparts $\bra{j}\rho\ket{i}$. Consider the (non-normalised) two-qubit state $\rho=\ket{0,0}\bra{0,0}+ \ket{0,0}\bra{1,1}+ \ket{1,1}\bra{0,0}+ \ket{1,1}\bra{1,1}$. If we perform the transpose operation on the second qubit \emph{alone} we obtain the \emph{partially transposed} density matrix $\rho^{PT}=\ket{0,0}\bra{0,0}+ \ket{0,1}\bra{1,0}+ \ket{1,0}\bra{0,1}+ \ket{1,1}\bra{1,1}$. This density matrix is no longer positive since it has a negative eigenvalue $-1$, and therefore no longer describes a physical states. Thus positivity alone is not sufficient for an operation to be physical. It must be completely positive.

Any completely positive map may be expressed in \emph{Kraus form} \cite{krausoperators},

\begin{equation}\label{eq:kraussop}
\rho\rightarrow\sum_i\hat{K}_i \rho\hat{K}_i^\dag
\end{equation}
where $\sum_i \hat{K}_i^\dag \hat{K}_i=\iden$. Physically, the Kraus operators $\hat{K}_i$ of a completely positive map on an open system may be obtained if the unitary evolution on the system and its environment is known, by performing a partial trace over the state of the environment\footnote{This is only possible if the state and environment are initially in a product state. For an example of the difficulties of characterising the dynamics of an open system initially entangled with its environment see \cite{buzekcposmap,hayashi}.}. Alternatively, given a set of Kraus operators one may always construct  such a unitary operation over the system and an environment, although this construction is not unique.

\subsection{Measurement of quantum systems}\label{sec:measurement}

The traditional way in which measurements on quantum systems are described is in terms of \emph{observables}. Observables are Hermitian operators which correspond to physically measurable quantities such as energy, momentum, spin, etc. Any Hermitian operator has a complete set of real eigenvalues corresponding to orthogonal eigenspaces. 

A \emph{projective measurement} of observable $\hat{O}$ on a system in state $\rho$ is described in the following way. We assume that some measuring device exists which returns, as a measurement result, one of the eigenvalues of $\hat{O}$. Once a result $\lambda$ has been registered, the state of the system is transformed by a projector onto the eigenspace belonging to eigenvalue $\lambda$, $P_\lambda$, i.e.

\begin{equation}
\rho \rightarrow P_\lambda \rho  P_\lambda\ .
\end{equation}

The probability of that measurement outcome occurring is $\textrm{Tr}[P_\lambda \rho ]/\textrm{Tr}[\rho]$.
In quantum information science it has proven beneficial to introduce more generalised concepts of measurement as a counterpart to the generalised evolution described by equation~(\ref{eq:kraussop}). However, these will not play a role in this thesis and  will not be discussed here. Excellent introductions to generalised measurement exist in \cite{peresbook} and \cite{nielsenchuang}.



\subsection{Von Neumann entropy}

The introduction of measurement outcome probability leads us naturally to a definition of entropy for a state. In classical statistical mechanics, the Boltzmann entropy of a classical discrete probability distribution $p_i$ is defined $S_\textrm{B}=-\sum_i p_i \log_2 p_i$. In an analogous way, an entropy for normalised quantum states was introduced by von Neumann and now bears his name,
\begin{equation}
S_\textrm{vN}=-\textrm{Tr}[\rho\log_2\rho]=-\sum_i(\lambda_i\log_2\lambda_i)\ ,
\end{equation}
where $\lambda_i$ are the eigenvalues of $\rho$. If the observable which shares the same eigenspaces as $\rho$ is measured, $\lambda_i$ are the probability of outcome $i$ and the von Neumann entropy reduces to the Boltzmann entropy of this probability distribution. The von Neumann entropy is an extremely important quantity in quantum information science. As well as its importance for quantifying entanglement (see below) it replaces Shannon's entropy for classical information as the fundamental unit for the quantification of information  in quantum information theory (see \cite{nielsenchuang}, chapters 11 and 12 for a review).

The von Neumann entropy sheds more light on the special status of pure states, as these are the only states of zero entropy. Thus  only for pure states does a projective measurement exist whose outcome can be predicted with certainty. On the other hand, density operators that are proportional to the identity have the maximal entropy for a system of their dimension.
Since mixed states can always be written in diagonal form as the convex sum\footnote{By ``convex sum'' we mean here a sum of normalised operators with positive coefficients which sum to unity.} of a set of pure states,

\begin{equation}
\rho=\sum_i \lambda_i\ket{\psi_i}\bra{\psi_i} \ ,
\end{equation}
a mixed state $\rho$ is sometimes interpreted as being a probabilistic mixture of this set of states, i.e. with probability $\lambda_i$ the system is in the pure state $\ket{\psi_i}$. However, it is important to recognise that while such a state is indeed indistinguishable from a state prepared in this way, this decomposition is not unique. In fact, there are infinitely many possible decompositions for any density operator \cite{schrodens}. For example,
\begin{equation}
\frac{1}{2}\left(\ket{0}\bra{0}+\ket{1}\bra{1}\right)=\frac{1}{4}\left((\ket{0}+\ket{1})(\bra{0}+\bra{1})+
(\ket{0}-\ket{1})(\bra{0}-\bra{1})\right)
\end{equation}

Thus one should be wary of assigning special status to any particular pure state decomposition of $\rho$.

\subsection{Complementary observables and the uncertainty \mbox{principle}}
While individual measurement outcomes are probabilistic, mean or \emph{expectation values} for the operators are straightforwardly calculated in the density matrix formalism using the following expression,
\begin{equation}
\langle \hat{O}\rangle=\textrm{Tr}[\hat{O}\rho]\ .
\end{equation}
Uncertainty in measurement outcomes can therefore be quantified in terms of the variance of $\hat{O}$ written $[\Delta(\hat{O})]^2$, 
\begin{equation}
[\Delta(\hat{O})]^2=\langle\hat{O}^2\rangle-\langle\hat{O}\rangle^2\ 
\end{equation}
and its positive root $\Delta(\hat{O})$, often referred to as the ``uncertainty'' of $\hat{O}$.

The value of the commutator between observables,  $[\hat{O}_1,\hat{O}_2]=\hat{O}_1\hat{O}_2-\hat{O}_2\hat{O}_1$ plays an important role in the relationship between the uncertainties of different observables for a given state. If two operators commute, an eigenstate of one is an eigenstate of the other, so the system may be prepared in states of zero uncertainty for both observables. If a pair of observables do not commute, the converse is true, and a system may never be in a state of zero uncertainty for both.
Thus there is a relationship between the uncertainties of observables which depends upon their commutator. This is the general form of the Heisenberg uncertainty relation,
\begin{equation}
\Delta(\hat{O}_1)\Delta(\hat{O}_2)\ge\frac{|\langle[\hat{O}_1,\hat{O}_2]\rangle]|}{2}\ .
\end{equation}

\section{Entanglement}\label{sec:entanglement}

As we have mentioned above, entanglement is one of the features of quantum mechanics most at odds with a classical world view. In this section  we will introduce  definitions of entanglement, and discuss how it is incompatible with a world-view based in classical physics. We will then discuss how entanglement may be quantified, before describing some of the important applications of entangled states which have been proposed in quantum information science.
Since this thesis will focus on \emph{bi-partite} entanglement (between pairs of sub-systems) we shall restrict our discussion to this simplest case, and not mention more general multi-partite entanglement, which has a more complicated structure that is still far from well understood.

A pure state on  $\mathcal{H}_1\otimes\mathcal{H}_2$ is entangled when it cannot be written as a product state -- a tensor product of a pure state on  $\mathcal{H}_1$ times a pure state on $\mathcal{H}_2$. For example, the two-qubit state $\ket{0}_1\otimes\ket{0}_2$ is a product state. Here we have chosen to label the states on both qubits $0$ and $1$ and distinguish between them via the suffixes 1 and 2. In this thesis we shall often omit the symbol $\otimes$ and write $\ket{0}_1\ket{0}_2\equiv\ket{0}_1\otimes\ket{0}_2$. Furthermore  for compactness we shall often omit the suffixes and differentiate the different systems by the order in which they are written, using the concise forms $\ket{0,0}\equiv\ket{0}\ket{0}\equiv\ket{0}_1\ket{0}_2$.

A mixed state is entangled when it cannot be written as a \emph{separable state}, i.e. as a convex sum of product states,
\begin{equation}
\rho_\textrm{sep}=\sum_{n,m}p_{i}\ket{\psi_i}\bra{\psi_i}\otimes\ket{\phi_i}\bra{\phi_i}\ .
\end{equation}

The definitions are motivated by the concept of Local Operations and Classical Communication, LOCC\cite{wernerlocc}. We assume that the sub-systems we are considering are in spatially separated laboratories where general quantum operations including  measurements may be performed, which are connected to each other via a ``classical'' communications channel such as a telephone connection. The separable states are those states which, starting from sub-systems in a product state, can be generated by LOCC, the entangled states are those that cannot. Thus entanglement can be seen as a kind of non-classical correlation between the sub-systems.

It is natural to ask whether, given a particular density matrix, it is easy to ascertain whether the state it describes is separable or not.  In fact, algorithms  \cite{dohertyseparability} to check this for an arbitrary matrix can take a long time to compute, belonging to the class of algorithms known by computer scientists as NP-HARD\cite{sepnphard}. Fortunately, a simple criterion exists which heralds the entanglement present in some states, and indeed, for some simple cases,  is a necessary and sufficient condition for entanglement.
This is known as the positivity of partial transpose or PPT criterion, and was introduced  by Peres \cite{peresppt} and the Horodeckis \cite{horodeckippt}. It is based upon the fact that transposition is a positive but not completely positive operation. Thus, if we consider a bipartite quantum system in a separable state $\rho_\textrm{sep}$, and apply the transposition operation to only one of the systems (thus partial transpose), the partially transposed state has the following form;
\begin{equation}
\rho_\textrm{sep}^{PT}=\sum_{n,m}p_{i}(\ket{\psi_i}\bra{\psi_i})^T\otimes\ket{\phi_i}\bra{\phi_i}\ .
\end{equation}
Since transposition is positive,  $(\ket{\psi_i}\bra{\psi_i})^T$ is still a positive operator and corresponds to a physical state, and therefore $\rho_\textrm{sep}^{PT}$ is positive as well. However, this argument does not hold when the state is entangled and entangled states may possess a negative partial transpose. While possessing a negative partial transpose is a sufficient condition for entanglement, it is not in general a necessary condition, and certain states exist which do have positive partial transpose\cite{boundent}.

Nevertheless, for certain systems of low-dimensionality the breaking of the PPT condition is a necessary and sufficient condition for entanglement \cite{horodeckippt}. These include pairs of qubits and systems of one qubit and one three-level system as well as the bi-partite Gaussian states, an important class of infinite dimensional states which will be introduced in chapter~\ref{ch:6}.

\subsection{EPR argument and the Bell inequality}
Now that we have formally defined entanglement, it is time to investigate some of the properties which make it incompatible with the  expectations of classical physics. We will reproduce the argument of Einstein, Podolsky and Rosen (EPR), who, in their seminal 1935 paper \cite{EPR} argued that quantum mechanics is incompatible with some basic intuitions about the physical world.

EPR argued that physical quantities which can be predicted with certainty must be ``elements of reality'' and for a theory to be ``complete'' these values must therefore be incorporated into the theoretical description of the system.  If quantum mechanics is complete in this sense, it follows that non-commuting observables can never be simultaneously ``elements of reality'' as if one property can be predicted with certainty, the other cannot.
EPR used the following argument to show that entanglement in quantum mechanics leads to an apparent contradiction with this assertion and that quantum mechanics is an incomplete and thus unsatisfactory theory.

We shall use the version of their argument introduced by Bohm\footnote{For their example, EPR used infinitely entangled two-mode Gaussian states which are, strictly speaking, unphysical non-normalisable states. However, this does not affect the validity of their argument.} \cite{bohmepr} and consider the following entangled two-qubit state,
\begin{equation}
\ket{\psi_\textrm{EPR}}=\sqrt{\frac{1}{2}}\left(\ket{0}\ket{1}-\ket{1}\ket{0} \right)\ , 
\end{equation}
and assume that the two qubits are located in two spatially separated laboratories; one belonging to Alice and one to Bob\footnote{The names Alice and Bob arose in discussions of classical communication theory and have been adopted whole-heartedly by the quantum information science community, especially in the discussion of spatially separated entanglement. Needless to say, they will appear throughout this thesis, and can be taken as shorthand meaning two spatially separated experimenters.}.

We will introduce a very important family of single qubit observables, the Pauli operators.
\begin{equation}
\sigma_X=\ket{0}\bra{1}+\ket{1}\bra{0}\qquad\sigma_Y=-i\ket{0}\bra{1}+i\ket{1}\bra{0}\qquad\sigma_Z=\ket{0}\bra{0}-\ket{1}\bra{1}
\end{equation}

Together with the identity operator, the Pauli operators form a basis  for two-qubit Hermitian operators, i.e., any single qubit operator can be written as a linear combination of $\sigma_X$,  $\sigma_Y$,  $\sigma_Z$, and $\iden$. The Pauli operators fulfill commutation relations of the form $[\sigma_X,\sigma_Z]=-2i\sigma_Y$ and cyclic permutations thereof.

Imagine that Alice measures the observable $\sigma_Z$ on her qubit, and obtains the measurement outcome +1 or -1. Using the measurement postulate outlined in section~\ref{sec:measurement}, we see that after the measurement Bob's qubit will be in state $\ket{0}$ or  $\ket{1}$ depending on Alice's measurement outcome. Thus, now Alice (but not Bob!) knows with certainty the outcome of a $\sigma_Z$ measurement on Bob's qubit. In EPR's language this value must be an ``element of reality''.

Now imagine that Alice measures $\sigma_X$ instead. This leads to Bob's state being projected into either state $(1/\sqrt{2})(\ket{0}+\ket{1})$ or  $(1/\sqrt{2})(\ket{0}-\ket{1})$, the eigenstates of $\sigma_X$,  depending on the measurement outcome. Now the value of $\sigma_X$ on Bob's qubit is an ``element of reality''. However, as we have seen, if we consider quantum mechanics to give a complete description of reality in the EPR sense, the value of non-commuting observables can never simultaneously be predicted with certainty. Thus there is a contradiction, and EPR conclude that quantum mechanics is not a complete description of reality.

\subsection{Bell inequalities}

These ideas upon which EPR based their argument were made more precise by John Bell. John Bell showed that the  implicit assumption behind EPR's argument is that physics should posess \emph{local realism}, that is;
\begin{enumerate}
\item ``Realism'' -- Every measurable quantity should have a definite value regardless whether the measurement is made or not.
\item ``Locality'' -- These quantities must be features of ``local systems'' in the relativistic sense, thus no property can be shared by two spatially separated systems.
\end{enumerate}
Bell showed that a theory constructed under these two assumptions must fulfill certain inequalities,  and that quantum mechanics violates these inequalities\cite{bellineq}. We will  construct such a \emph{Bell inequality} here. This example is one of a class of such inequalities proposed by Clauser, Horne, Shimony and Holt \cite{CHSH}.

To derive a Bell inequality one must assume that physics obeys local realism. The class of theories for which this is the case are called \emph{local hidden variable} models. Imagine that Alice and Bob each possess a particle upon which they can make two different measurements  corresponding to two different physical properties which the particles possess. Let us label these properties $A_1$ and $A_2$ on Alice's particle and  $B_1$ and $B_2$ for Bob and assume that each property  can have only two values $+1$ or $-1$. 

Now consider the value of the composite quantity,
\begin{equation}
A_1 B_1 + A_2 B_1 +A_2 B_2 - A_1 B_2=(A_1+A_2)B_1+(A_2-A_1)B_2\ .
\end{equation}
Since $A_1$ and $A_2$ take the values $\pm1$ this expression must equal either $\pm 2B_1$ or  $\pm 2B_2$. Either way the quantity is equal to either +2 or -2.

Now let us imagine that the particles belong to some ensemble prepared such that the probability that $\{A_1,A_2,B_1,B_2\}=\{a_1,a_2,b_1,b_2\}$ is $p(a_1,a_2,b_1,b_2)$ where $a_i$ and $b_i$ take the values +1 and -1. Now we define $E(M)$ as the expectation value for the measurement of $M$, thus
\begin{equation}\label{eq:bell}\begin{split}
&E(A_1 B_1) + E(A_2 B_1) +E(A_2 B_2) - E(A_1 B_2)\\&\qquad=E(A_1 B_1 + A_2 B_1 +A_2 B_2 - A_1 B_2)\\
&\qquad\qquad=\sum_{a_1,a_2,b_1,b_2}p_{a_1,a_2,b_1,b_2}(A_1 B_1 + A_2 B_1 +A_2 B_2 - A_1 B_2)\le 2\ .
\end{split}
\end{equation}
This is an example of a Bell inequality, an inequality that must be satisfied by the measurement results of a theory if it fulfils both the realism and locality criteria above. To see that in quantum mechanics Bell inequalities can be violated 
consider  measurements on the entangled state $\ket{\psi_\textrm{EPR}}$ introduced above. Let Alice's measurements  $A_1$ and $A_2$ be  observables $\sigma_Z$ and  $\sigma_X$, and let Bob's be  the linear combinations $B_1=\sqrt{1/2}(-\sigma_Z-\sigma_X)$ and  $B_2=\sqrt{1/2}(\sigma_Z-\sigma_X)$  (remember that the Pauli operators form a basis for single-qubit observables), which one can easily confirm have eigenvalues $\pm1$.
\begin{equation}\begin{split}
&E(A_1 B_1) + E(A_2 B_1) +E(A_2 B_2) - E(A_1 B_2)\\&=
\sqrt{\frac{1}{2}}\bigl(-\langle(\sigma_Z)_A(\sigma_Z)_B\rangle-\langle(\sigma_Z)_A(\sigma_X)_B\rangle\bigr)+
\bigl(-\langle(\sigma_X)_A(\sigma_Z)_B\rangle-\langle(\sigma_X)_A(\sigma_X)_B\rangle\bigr)\\&\mbox{}+
\bigl(\langle(\sigma_X)_A(\sigma_Z)_B\rangle-\langle(\sigma_X)_A(\sigma_X)_B\rangle\bigr)-
\bigl(\langle(\sigma_Z)_A(\sigma_Z)_B\rangle-\langle(\sigma_Z)_A(\sigma_X)_B\rangle\bigr)
\end{split}
\end{equation}

Since $\langle(\sigma_X)_A(\sigma_X)_B\rangle=\langle(\sigma_Z)_A(\sigma_Z)_B\rangle=-1$ and the other expectation values are zero, this gives
\begin{equation}
E(A_1 B_1) + E(A_2 B_1) +E(A_2 B_2) - E(A_1 B_2)=2\sqrt{2}>2\ ,
\end{equation}
so the Bell inequality~(\ref{eq:bell}) is violated.

This shows that quantum mechanics is not consistent with both assumptions 1) and 2), and if quantum mechanics is believed to be correct, then these basic assumptions about physical reality need to be revised. Thus either 
quantum mechanics must be considered intrinsically non-local or we must accept a theory whose measured quantities do  not correspond to elements or reality.

The Bell inequalities are not merely a philosophical device. They provide a direct test for this strange and surprising aspect of quantum mechanics since entangled states such as $\ket{\psi_\textrm{EPR}}$ can be created in the laboratories and these expectation values measured directly. Notable early experiments to attempt this were performed by Clauser and Freedman \cite{clauserexp} and  Aspect and co-workers \cite{aspect,aspectinbook}. Although there were many loopholes\footnote{Among other  loopholes in Aspect and co-workers' experiments, an important one was that, unavoidably in the design of the experiment,  many of the  measurement results were lost. Since it is, in principle, possible that the missed results, would have changed these average quantities such that no Bell inequality was violated, this is a serious loophole. In fact,  a local hidden variable model consistent with Aspect and co-workers' measurements has actually been constructed \cite{santoslhv}. For a discussion of other loopholes and some healthy scepticism of quantum mechanics see \cite{loopholeref}.}
 in the experiments, which meant that in principle
it did not categorically rule out local realism,
these and later results (see \cite{zeilingerrmp} for a historical review) have agreed emphatically with the predictions of quantum mechanics. In fact, implementing a loophole-free test of Bell inequality violation has proven very technically challenging; however  no experiments have contradicted quantum mechanics, and experiments which claim to close all loopholes are being attempted at the moment \cite{fryepr}.

We must emphasise that entanglement is fundamental to the Bell inequality violation since if one considers a quantum mechanics where only separable states are allowed, it is clear that the  assumptions of local realism  can indeed be satisfied and no violation will occur. Thus, 
not only are the EPR debate and the Bell inequalities  indications that quantum mechanics can lead to results at odds with \emph{reasoning} based on intuition from classical physics, but they emphasise the special r\^ole of entanglement in this non-classical effect.


\subsection{Quantifying entanglement}
It is clear that in some sense state $\ket{\psi_\textrm{EPR}}$ has a high degree of entanglement since measurement results on the two particles are completely correlated and yet individually completely random, and on the other hand,   separable states have no entanglement. Entanglement quantification attempts to provide a number to express the degree of entanglement of the state. 

Quantifying bi-partite entanglement in pure states is very simple. If a pure bipartite state is not entangled, when we take a partial trace over one sub-system, the state of the remaining sub-system will be pure. However, if the state is entangled the reduced state is necessarily mixed. It thus makes sense to associate the degree of entanglement of the whole bi-partite state with how mixed the reduced state is. The measure of mixedness we choose is the  von Neumann entropy of the reduced state, since, as we shall see below, this gives the entanglement measure a direct operational interpretation. The \emph{pure state entanglement measure} $E_{\textrm{vN}}$ can thus be written,
\begin{equation}
E_{\textrm{vN}}=-\textrm{Tr}[\rho^A \log_2 \rho^A]\ ,
\end{equation}
where $\rho^A$ is the reduced state of one of the sub-systems. Since the entropy of a pure state is zero,  the entropies of both sub-systems of a bi-partite pure state must be equal.

The pure state entanglement has the following important properties, which make it suitable as a measure of entanglement \cite{vedralrippin};
\begin{enumerate}
\item It is zero for separable (product) states.
\item It is invariant under local unitary transformations.
\item It is not increased on average by local operations with classical communication (LOCC) on the two sub-systems.
\end{enumerate}

A quantity which satisfies these three properties is called an \emph{entanglement monotone} \cite{pleniovedral,pleniovedralreview}.
Further properties of the pure state entanglement are \emph{additivity}, i.e. the entanglement of the two copies $\ket{\psi}\otimes\ket{\psi}$ is double the entanglement of a single copy $\ket{\psi}$, and \emph{asymptotic continuity}, which means roughly that small changes in the state lead to small changes in the entropy.

Since the reduced density matrix of the sub-systems of  $\ket{\psi_\textrm{EPR}}$ is proportional to the identity, the  entropy of its sub-systems is maximum, and  the state  can be called \emph{maximally entangled}. The maximally entangled two qubit states have many important properties and  play a  central role in quantum information science. For example, since entanglement can not increase under LOCC, these are the only states from which any  other state can be obtained asymptotically  by LOCC alone\footnote{The general problem of which pure states may be converted under LOCC to which other pure states was solved by Nielsen \cite{nielsentheorem}.}. One can form a basis from four maximally entangled states, and these states are called the \emph{Bell states}, named in honour of John Bell,
\begin{equation}\label{eq:bellbasis}\begin{split}
\ket{\Phi^\pm}&=\sqrt{\frac{1}{2}}\bigl(\ket{0}\ket{0}\pm \ket{1}\ket{1}\ ,\\
\ket{\Psi^\pm}&=\sqrt{\frac{1}{2}}\bigl(\ket{0}\ket{1}\pm \ket{1}\ket{0}\ .
\end{split}
\end{equation}
The Bell state basis has many important applications in quantum information science and will feature prominently in the examples below.

Although $E_\textrm{vN}$ is a satisfactory entanglement measure for pure states, one must consider whether other measures may exist, which may even impose a different ordering of the states. In fact, one can show \cite{popescurohrlich,hororudolph} that the entanglement measure for pure states is unique, in the sense that all other additive and continuous  entanglement monotones are necessarily monotonic functions of $E_\textrm{vN}$.

For mixed states, however, the picture becomes much more complicated. Several entanglement monotones for mixed state have been proposed (see references within \cite{pleniovedralreview} and more recent proposals such as \cite{christandl}). These measures do not always give the same ordering of states \cite{virmaniplenio}, making the study of mixed state entanglement a complicated affair.

Here we will focus on several important mixed state entanglement monotones,  the \emph{entanglement of distillation}, the \emph{entanglement cost} and the \emph{logarithmic negativity}.
The first two of these compare the state to a maximally entangled Bell state in different ways. The first asks, given $n$ copies of the state $\rho$, what is the maximum number $m$ of Bell states which can be made (or distilled\footnote{Entanglement distillation\cite{bennettdistillation} will be discussed in more detail in Section~\ref{sec:distillsec}}) on average with LOCC? The entanglement of distillation is the ratio $m/n$ in the asymptotic limit, where $n$ tends to infinity. The entanglement cost\cite{haydencost} goes the other way, asking  how many Bell states  would one need (asymptotically) to prepare $\rho$ with LOCC.
Both of these measures are equal for pure states and coincide with the pure state entanglement $E_\textrm{vN}$, which gives $E_\textrm{vN}$ a double operational meaning \cite{bennettent}. The two measures are not, in general, equal for mixed states.

 For an arbitrary state, particularly of high dimension, these quantities can be very difficult to calculate. For this reason, the  \emph{logarithmic negativity} has been introduced \cite{zukowneg,eisplenneg,vidallogneg}.
This quantity is inspired  by the Peres-Horodecki PPT separability criterion, described above, and provides a way of quantifying to how great an extent this criterion is violated.

The logarithmic negativity is defined
\begin{equation}
E_\mathcal{N}=\log_2||\rho^{T_A}||_1\ ,
\end{equation}
where $||A||_1$ is the trace-norm of matrix $A$, defined as $\textrm{Tr}[\sqrt{A^\dag A}]$. The trace-norm is equal to the sum of the absolute value of the eigenvalues, and thus also singular values of $A$.
Since singular value decompositions of large matrices may be accomplished efficiently $\mathcal{N}$ and $E_\mathcal{N}$ can readily calculated even for high dimensional systems. 
%
%
%
%
The logarithmic negativity fulfils the three criteria  for an entanglement monotone \cite{vidallogneg,shashthesis}
%
%
%
. Furthermore, it is an additive quantity and is an upper bound to the entanglement of distillation  \cite{vidallogneg}. It also has an interpretation as the entanglement cost of certain states under operations which preserve the positive partial transpose of a state   \cite{audenaertlogneg}. Its tractability makes it very useful in the investigation of entanglement in mixed states, and we shall use it throughout part~\ref{part3} of this thesis.

\section{Entanglement in quantum information \mbox{science}}

As we have seen, measurements on entangled states of spatially separated sub-systems allow physicists to test fundamental notions about the nature of the  physical world. However,  quantum information science gives us a new way to think about entanglement, namely as a resource to carry out tasks in data processing and communication  which would otherwise be very hard or even impossible. In this section we will briefly outline some of the most important of these proposed applications.

Here we will focus on qubit entanglement, but most of the applications can be generalised to higher dimensions. In particular, we will leave the discussion of applications of entangled infinite-dimensional states until section~\ref{sec:gaussinqip}. In the following applications, maximally entangled states play a special role. They are a direct resource for quantum teleportation, quantum key distribution

\subsection{Quantum key distribution}
The only system of public cryptography which has been proven unbreakable\cite{shannononetime} is the ``one-time-pad'' \cite{vernam}. Alice and Bob
share a secret random string called a \emph{key}. Alice encrypts her message, encoded as a bit-string, by bitwise addition modulo two of the key and bit string. She sends this string of data to Bob, and can even broadcast it publicly. Without the key, no information about Alice's message can be recovered, but with the key Bob can reconstruct it easily, by the same bitwise addition procedure. As long as they only use the key once, the one-time-pad cipher is completely unbreakable\cite{shannononetime}. The problem with the system is that Alice and Bob need a secure way of generating and sharing the key. 

\emph{Quantum key distribution} (QKD) solves this problem. By using quantum systems, Alice and Bob can  generate a key which is completely random, and known to no outsiders. The first quantum key distribution was proposed by Bennett and Brassard \cite{BB84} and required no entanglement.  Ekert proposed an entanglement-based QKD approach  \cite{ekertprotocol}, which we shall describe here. 

Alice and Bob share a supply of known maximally entangled states, let us assume they are the Bell state $\ket{\Phi^+}$. Alice and Bob make measurements on the qubits of each entangled pair in one of two bases, the eigenbases of $\sigma_Z$ $\{\ket{0},\ket{1}\}$ and $\sigma_X$ $\{(1\sqrt{2})(\ket{0}+\ket{1}),(1\sqrt{2})(\ket{0}-\ket{1})\}$. For each pair they each choose the basis at random. They keep a record of each measurement outcome. They then make public the measurement bases that they used for each measurement. After they have discarded all outcomes where their bases did not match, they compare a subset of the remaining data publicly. If the results correlate as expected, they are assured that no eavesdropper has intercepted or tampered with their entangled states. They then use the remaining data as a secret key.
The Ekert protocol  was the first protocol to explicitly use an entangled state as a resource. Soon afterwards a further application employing entangled states  was suggested, \emph{quantum teleportation}.

\subsection{Quantum teleportation}
 Unlike many other applications in quantum information science, which utilise entanglement to improve on what is possible with classical physics or with reasoning based on classical physics, quantum teleportation achieves a surprising purely quantum mechanical task. An unknown  quantum state $\ket{\psi}$ is transferred  from one system to another, via measurements and classical communication alone. Without employing entanglement, if Alice has an unknown quantum state and wishes to transfer it to her friend Bob at a remote site, the only way to do so is to physically transport the quantum system carrying the state. Since measurement of a single copy only reveals minimum information about the state and at the same time destroys it, no scheme based on measurement and classical communication will allow a faithful transfer of the state.

On the other hand, if Alice and Bob share a maximally entangled state, say the Bell state $\ket{\Psi^+}$, this is not the case. They can indeed ``teleport'' the state via measurement and classical communication.  Alice  makes a two qubit measurement in the basis of Bell states in equation~(\ref{eq:bellbasis}). This projects Bob's qubit onto the state $(\sigma_X)^{j}(\sigma_Z)^{k}\ket{\psi}$, where $i,j\in\{0,1\}$ depend upon the measurement outcome. Once Alice has broadcast the measurement outcome to Bob, he can correct for these extra Pauli transformations and  reconstruct $\ket{\psi}$.

While transferring unknown quantum states does not, at first sight, appear to be an application of much utility, in fact teleportation has had immense importance for \emph{quantum computation} which will be described below, both on the conceptual level \cite{gottesmanchuang,childsleungnielsen} and practically, e.g. for transferring data between remote sites within a quantum computer.

\subsection{Other applications of entanglement}
The field of quantum information science has produced many other applications of entanglement. Entanglement between spatially separated sites can be used to increase classical data transfer rates through so-called ``dense-coding'' \cite{densecoding} and separated parties can use entangled states to improve the performance in certain competitive games \cite{jensgames} or distributed computational tasks \cite{caslav}.
Entanglement between systems  in a single laboratory also has many applications. Entangled states can allow more precise measurements in, for example, frequency standards \cite{winelandstand,freqstan} and lithography \cite{koklitho}, than would otherwise be possible. Additionally, entanglement is of fundamental importance in \emph{quantum computation}.

\subsection{Quantum computation}

It has been long known that simulating quantum systems is difficult, as the complexity of their description grows exponentially with the system size.  Feynman was one of the first to suggest that this complexity could be countered if quantum systems were used as an integral part of the simulation devices themselves \cite{feynmanquantcomp}. An early idea of a quantum computer was proposed by Benioff \cite{benioff}, 
and this was developed by Deutsch into the  modern idea of a universal quantum computer \cite{deutschquantcom}. David Deutsch postulated that a computer based on the laws of quantum mechanics would be so powerful that it could simulate  any physical system efficiently.  The true power of quantum computation was then demonstrated when Peter Shor showed that factoring large numbers could be achieved efficiently on a quantum computer \cite{shoralg}. The best known classical algorithm increases exponentially in length with increasing input size. The number of steps in Shor's algorithm increases only polynomially giving an enormous ``speed-up'' over the classical case.

Entanglement plays an important role in quantum computation. Firstly, Jozsa and Linden showed that entangled states must be present in a quantum computer at some stage of an algorithm for it to exhibit exponential speed-up compared to classical computers \cite{jozsalinden}. Secondly, information must be encoded in entangled states if quantum computers are going to be able to operate in the presence of noise \cite{shorfault}. Finally, paradigms for quantum computation have been proposed based solely on measurements made on certain  entangled states, which act as a resource for quantum computation \cite{raussenbriegel,raussenbrowne}. 

\section{Contexts and overview of this work}

In this chapter we have seen the power and importance of entangled states, both as a fundamental testing ground for quantum mechanics and as a resource for quantum information science. Although entanglement occurs almost whenever quantum systems interact, the particular quantum states we have identified above as having the most practical applications, the maximally entangled states, such as the Bell states of equation~(\ref{eq:bellbasis}) are not observed in nature. There are several reasons for this, the most important is that as these states are  pure, they can only exist in systems which have no entanglement with any other system, and thus must be isolated to a degree which does not occur naturally. 

Controlled generation of entangled states has therefore been an important goal of experiments in quantum information science. Close to maximally entangled states have been generated in light pulses \cite{oumandelpdc,kwiatpdc,shapiropdc} and between pairs of atoms and between atoms  and cavities\cite{harochereview2} (an optical or microwave cavity is a system of mirrors which confines and concentrates a field). Current schemes for entanglement generation in \emph{cavity QED}\footnote{Cavity Quantum Electrodynamics (QED) refers to experiments where the light field of a cavity  is probed and manipulated through interactions with individual atoms or ions.} have several disadvantages which we shall describe below in part~\ref{part2}, in particular, requiring a very precise control of the evolution of the sub-systems. Other proposals have been made, in particular \cite{bose99tele}, which require less precise control of the systems, depending instead  on the detection of photons emitted from the cavities. These proposals, however, are not expected to be robust to common experimental errors. In part~\ref{part2} we will introduce schemes for the generation of Bell states in so-called ``cavity QED'' experiments. The schemes we introduce are  based on a powerful but simple idea, using weak interactions between systems followed by a measurement on one of the systems, and this makes them robust against a wide range of experimental imperfections and far more suitable for experimental implementation than earlier schemes.

While the generation of entangled states of  light is now very far advanced, manipulating these  states for quantum information processing has remained a challenge. The elements of linear optics, beam splitters and phase-shifters, which we will describe in detail in chapter~\ref{ch:6} are the most readily available operations for the manipulation of light in the laboratory. However, in order to implement, for example,  quantum computation in light, the linear optical components must be augmented by photon counters in a complicated scheme \cite{klm} which is much too complicated to be implemented with present technology. Light pulses are quantum systems of infinite dimension. They can thus carry so-called continuous variable states, which will be introduced in chapter~\ref{ch:6}. As we shall describe in section~\ref{sec:gaussinqip}, many quantum information science applications have been proposed for the continuous variable domain. However, entanglement distillation for continuous variable states, an extremely important application, had so far been unattainable with realistic operations. In part~\ref{part3},  after a detailed introduction of an important class of continuous variable states of light, the Gaussian states, we introduce an original scheme for the distillation of continuous variable entanglement, a demonstration experiment of which could be implemented in a  current linear optics laboratory.

Throughout this thesis, quantum mechanical electro-magnetic radiation plays a central role. With this in mind, the second chapter  is devoted to a review of the quantum mechanical description of light.


%% file: c-2.tex
\chapter{A Quantum  Description of Light}
\label{ch:2}

In this chapter, we will show how a quantum theory of electro-magnetic radiation may be constructed by ``quantisation'' of the classical Maxwell equations.
Of course, as fundamentally different creatures, quantum theories can never be \emph{derived} from their classical counter-parts. Quantisation is rather essentially an \emph{ad-hoc} process, the aim of which is the  creation of a  quantum mechanical
 theory, which  in certain limits matches the classical theory's predictions. The quantum theory of light described here, originally developed by Born, Heisenberg and Jordan\cite{bornheisenbergjordan} and  Dirac \cite{dirac27}, has been verified by  a large number of  experiments, many of which, especially in recent years, have explicitly tested the theory's non-classical predictions. The treatment presented here is slightly more general than that found in textbooks such as \cite{plkbook,mandelbook,meystrebook,loudonbook} as we shall explain below.

\section{Classical electro-magnetic waves}

Classically, the electro-magnetic field in a vacuum is completely  described by the \emph{in-vacuo} Maxwell equations \cite{maxwell},
\begin{subequations}
\begin{align}
\vek{\nabla} \times \vek{E}&=-\frac{\partial \vek{B}}{\partial t}\label{maxwell1}\\
\vek{\nabla} \times \vek{B}&=\mu_0\epsilon_0\frac{\partial \vek{E}}{\partial t}\label{maxwell2}\\
\vek{\nabla}\cdot \vek{E}&=0\label{maxwell3}\\
\vek{\nabla}\cdot \vek{B}&=0\label{maxwell4}
\end{align}
\end{subequations}
where $\vek{E}$ and $\vek{B}$ are electric field and magnetic induction and $\mu_0$ and $\epsilon_0$ are the permeability and permittivity of free space.
It is convenient to introduce the vector potential $\vek{A}$ which is defined as
\begin{equation}\label{potentialB}
\vek{B}=\nabla\times \vek{A}
\end{equation}
and choose a convenient gauge in which to work,  the Coulomb gauge, where
\begin{equation}
\nabla\cdot \vek{A}=0
\end{equation}
and 
\begin{equation}\label{potentialE}
\vek{E}=-\frac{\partial \vek{A}}{\partial t} \ .
\end{equation}

Since there are no external sources, the scalar potential can be set to zero without loss of generality. Substituting these equations into the second Maxwell equation \eqref{maxwell2} and employing the identity $\vek{\nabla}\times\vek{\nabla}\times\vek{A}\equiv\vek{\nabla}(\vek{\nabla}\cdot\vek{A})-\nabla^2 \vek{A}$, results  in the following wave equation for $\vek{A}$,

\begin{equation}\label{maxwellwave}
\nabla^2\vek{A}=\frac{1}{c^2}\frac{\partial^2 \vek{A}}{\partial t^2}
\end{equation}
where $c=\sqrt{1/(\mu_0\epsilon_0)}$ is the speed of light.
This wave equation can be solved by the method of separation of variables  which, in the spatial case, leads to the Helmholtz equation, 
\begin{equation}\label{helmholtz}
\vek{\nabla}^2 u(\vek{r})=-\frac{\omega_\vek{k}^2}{c^2} u(\vek{r})
\end{equation}
where, for later convenience we choose the separation constant $-\omega_\vek{k}^2/c^2$ and where $\vek{k}$ is a yet unspecified label.
The  solutions to equation~(\ref{helmholtz}),  $u_\vek{k}(\vek{r})$  are called \emph{mode functions} and  depend on the  boundary conditions. 
Usually, when the quantisation of the electro-magnetic field is presented one usually  considers only the simplest case of mode-functions that are one-dimensional and sinusoidal. We take a more general approach in order to show that the same quantisation method is valid for any set of mode-functions which satisfy equation~(\ref{helmholtz}), and  shall thus leave the mode-functions unspecified. 
The label $\vek{k}$ can take continuous or discrete values, but here we consider the discrete case which applies, for example, to the field in a cavity. A discussion of the continuous case may be found in, e.g. \cite[page 512]{mandelbook}.

In classical optics, the  mode functions represent resonant modes, and in the quantum case  provide the basis for the quantisation of the field. 
We choose an orthonormal set of mode functions\footnote{This will always exist as $\vek{\nabla}^2$ is a Hermitian operator on a given region of $\mathbbm{R}^3$ under the condition that  functions and their first derivatives vanish on its boundary, as in this case.} and can thus expand $A$ as the sum

\begin{equation}\label{modeexpansion}
\vek{A}(\vek{r},t)=i \sum_\vek{k}\left( \vek{A}_k(t) u_\vek{k}(\vek{r})  -\vek{A}^*_k(t) u^*_\vek{k}(\vek{r}) \right)
\end{equation}
where we have guaranteed that $\vek{A}(\vek{r},t)$ remains real, while allowing complex mode functions which are often convenient for calculations.
Substituting this mode expanded solution back into equation~(\ref{maxwellwave}) produces the following equations of motion for the coefficients  $\vek{A}_k(t)$;

\begin{equation}\label{sho}
\vek{A}_k(t)=-\omega^2_\vek{k}\frac{\partial^2 \vek{A}_k(t)}{\partial t^2}\ .
\end{equation}

This equation is similar to the equation of motion for a simple harmonic oscillator and as we will shortly see,  the quantisation of the electro-magnetic field is based  upon this correspondence.
Choosing positive frequency solutions of equation~(\ref{sho})\footnote{This does not restrict generality, because the mode functions always exist in complex conjugate pairs  the negative solutions will appear in association with the complex conjugate of $u_\vek{k}$.}, and employing the unit polarisation vector $\vekk{\epsilon}_\vek{k}$, $\vek{A}_k(t)=a_\vek{k} \exp[-i\omega t] \vekk{\epsilon}_\vek{k}$ we write the general solution to equation~(\ref{maxwellwave})
\begin{equation}\label{classicala}
\vek{A}(\vek{r},t)=i\sum_\vek{k}\left( a_\vek{k} e^{-i\omega_{\vek{k}} t}
u_\vek{k}(\vek{r}) \vekk{\epsilon}_\vek{k} -a^*_\vek{k} e^{+i\omega_{\vek{k}} t}
u^*_\vek{k} (\vek{r}) \vekk{\epsilon}^*_\vek{k}
\right)\ .
\end{equation}
Thus, from equations \eqref{potentialB} and \eqref{potentialE}, we obtain solutions for the classical fields $\vek{E}$ 
\begin{equation}\label{classicalE}
\vek{E}(\vek{r},t)=\sum_\vek{k} \omega_\vek{k}\left(a_\vek{k} e^{-i\omega_{\vek{k}} t}
u_\vek{k}(\vek{r}) \vekk{\epsilon}_\vek{k} +a^*_\vek{k} e^{+i\omega_{\vek{k}} t}
u^*_\vek{k} (\vek{r}) \vekk{\epsilon}^*_\vek{k}
\right),
\end{equation}
and $\vek{B}$
\begin{equation}\label{classicalB}
\vek{B}(\vek{r},t)=i \sum_\vek{k}\left( a_\vek{k} e^{-i\omega_{\vek{k}} t}
 \bigl(\vek{\nabla}\times(u_\vek{k}(\vek{r})\vekk{\epsilon}_\vek{k})\bigr) -a^*_\vek{k} e^{+i\omega_{\vek{k}} t}
u^*_\vek{k} (\vek{r}) \bigl(\vek{\nabla}\times(u^*_\vek{k}(\vek{r})\vekk{\epsilon}^*_\vek{k})\bigr)
\right)\ .
\end{equation}

%
%
The classical Hamiltonian of the electro-magnetic field is

\begin{equation}
H = \frac{1}{2}\int dr^3\left[\epsilon_0 {E}^2(\vek{r},t)+\frac{1}{\mu_0} {B}^2(\vek{r},t)\right]
\end{equation}
where the integration is over the whole space as defined by the boundary conditions.
We substitute the mode expansions for $\vek{E}$ and $\vek{B}$ (equations~(\ref{classicalE}) and (\ref{classicalB})) into this Hamiltonian. After some manipulation\footnote{The integral over $B$ is brought into a suitable form in several steps, first employing  integration by parts and the  identity $(\vek{A}\times\nabla)\cdot(\nabla\times\vek{B})=\vek{A}\cdot(\nabla\times\nabla\times\vek{B})$. Then the identity introduced above allows one to use the zero divergence of $u_\vek{k}(\vek{r}) \vekk{\epsilon}_\vek{k}$ and the Helmholtz function to remove all differential operators from within the integral.}, this integral can be  solved by using the orthonormality of the mode functions which generates the following expression,

\begin{equation}\label{classicalHam}
H=2\epsilon_0\sum_\vek{k}\omega_{\vek{k}}^2|a_k|^2\ .
\end{equation}

To show that this is equivalent to the Hamiltonian for a simple harmonic oscillator associated with each mode, let us make the linear transformation of variables,
\begin{equation}
a_\vek{k}\equiv\sqrt{\frac{1}{4\epsilon_0}}\left(x_\vek{k}+i\frac{p_\vek{k}}{\omega}\right)
\end{equation}
\begin{equation}
a^*_\vek{k}\equiv\sqrt{\frac{1}{4\epsilon_0}}\left(x_\vek{k}-i\frac{p_\vek{k}}{\omega}\right)
\end{equation}
%
%
%
in terms of which the classical field Hamiltonian is
\begin{equation}\label{classicalhamiltonianshos}
H=\frac{1}{2}\sum_\vek{k}(p_\vek{k}^2+\omega_\vek{k}^2x_\vek{k}^2)\ .
\end{equation}

Thus each mode of the field can be associated with a simple harmonic oscillator with position $q_\vek{k}$ and momentum $p_\vek{k}$. The quantisation of the  field then proceeds by replacing these classical variables with non-commuting  quantum operators, whose commutators are specified by the Poisson brackets of the classical theory.

\section{The quantum harmonic oscillator}

The Hamiltonian for the quantum harmonic oscillator with unit mass and  spring constant $\omega^2$ is

\begin{equation}
\hat{H} = \frac{1}{2}\left[ \hat{p}^2 + \omega^2\hat{x}^2\right],
\end{equation}
where  $\hat{x}$ and $\hat{p}$ represent position and  momentum respectively and obey the commutation relation $[\hat{x},\hat{p}]=i\hbar$.
The Hamiltonian can be diagonalised by introducing the \emph{ladder operators};
\begin{equation}\label{ladderannihilation}
\hat a=\sqrt{\frac{1}{2 \hbar\omega}}\left(\omega \hat{x}+i\hat{p}\right)
\end{equation}
\begin{equation}\label{laddercreation}
\hat a^\dag=\sqrt{\frac{1}{2 \hbar\omega}}\left(\omega \hat{x}-i\hat{p}\right)
\end{equation}
with commutation relation  $[\hat{a},\hat{a}^\dag]=1$.
Written in terms of the ladder operators, the Hamiltonian takes on a simple form
\begin{equation}
\hat{H}=\hbar\omega \left(\hat{a}^\dag \hat{a}+\frac{1}{2}\right).
\end{equation}
The eigenstates of this Hamiltonian are called \emph{Fock states}. Their properties can be derived as follows. Let $\ket{\phi}$ be an energy eigenstate with  energy $E_\phi$
\begin{equation}
\hat H \ket{\phi}=\hbar \omega \left(\hat a^\dag\hat{a}+\frac{1}{2}\right)\ket{\phi}=E_\phi\ket{\phi}.
\end{equation}

By applying the operator $\hat a$ to both sides of this equation, and using the commutation relation, we find 
\begin{equation}
\hat H \hat a\ket{\phi}=(E_\phi-\hbar\omega)\hat{a}\ket{\phi}\ .
\end{equation}
Thus, the state  $\hat{a}\ket{\phi}$ is itself  an (as yet unnormalised) energy eigenstate with a lower energy  $E_\phi-\hbar\omega$. This is why $\hat{a}$ is often called a \emph{lowering operator} and repeated application of $\hat a$ will lead to lower and lower energy eigenstates.

That this lowering cannot go on indefinitely, i.e. there is a \emph{ground state} of lowest energy,  can be seen by remembering that any state must have a non-negative norm, including the state $\hat a |\phi\rangle$. Therefore $\langle \phi |\hat{a}^\dag \hat{a}|\phi\rangle\ge0$, which bounds the expected energy from below,  $\langle n |\hat{H}|n\rangle\ge\hbar \omega/2$. Thus there exists (at least one) state of lowest energy, which we will label $\ket{0}$. Since action of $\hat{a}$ cannot lead to another physical state, this state must fulfil $\hat a \ket{0}=0$. This allows us to calculate its energy, and we find $\hat H \ket{0}=\hbar\omega/2$.

Just as operator $\hat a$  is a lowering operator, its conjugate $\hat{a}^\dag$ is a \emph{raising operator} which maps one energy eigenstate to another with energy raised by $\hbar\omega$. Thus, applying $\hat{a}^\dag$ $n$ times to the ground state is $(1/2)\hbar\omega$ generates an energy eigenstate of energy $\hbar\omega(n +(1/2))$. We label this  eigenstate $\ket{n}$,

\begin{equation}
\hat H \ket{n}=\hbar\omega\left(n+\frac{1}{2}\right)\ket{n}
\end{equation}
and thus 
\begin{equation}
\hat a^\dag \hat a \ket{n}=n\ket{n}\ .
\end{equation}

We thus name $\hat a^\dag \hat a$ the number operator and label it $\hat n$.
So far we have not considered the normalisation of $\ket{n}$, since the eigenvalue equations are invariant when the state is multiplied by a scalar. We write the state $\ket{n}=\alpha_n\hat{a}^\dag\ket{n-1}$, where $\alpha_m$ is a normalisation constant which will be determined inductively. If we demand that $\ket{n}$ has unit norm,
\begin{equation}
|\alpha_n|^2 \langle n-1| \hat{a}\hat{a}^\dag \ket{n-1}= |\alpha_n|^2\langle n-1| (\hat n+1) \ket{n-1}=|\alpha_n|^2 n=1\ .
\end{equation}

Thus $\alpha_n=1/ \sqrt{n}$ and by induction

\begin{equation}
\ket{n}= \sqrt\frac{1}{n!}(\hat a^\dag)^n\ket{0}\ .
\end{equation}
It also follows that 
\begin{equation}
\hat{a}\ket{n}=\sqrt{n}\ket{n-1}
\end{equation}
and
\begin{equation}
\hat{a}^\dag\ket{n}=\sqrt{n+1}\ket{n+1}.
\end{equation}

To show that the vectors $\ket{n}$ form the basis of an infinite-dimensional Hilbert space, one still needs to show that each eigenstate is non-degenerate. This can be proven under the assumption that only analytic functions of $\hat{a}$ and $\hat{a}^\dag$ act on the space, since it can then be shown that  the only operators on the space which commute with $\hat{n}$ are functions of $\hat{n}$ and therefore the eigenstates of $\hat{n}$ must be non-degenerate (see \cite{messiah} volume I, page 436 and Problem XII.1, page 460).

\section{The quantum light field}

The quantised version of the electro-magnetic field Hamiltonian, equation~\eqref{classicalhamiltonianshos}, is thus
\begin{equation}
\begin{split}
\hat{H}&=\frac{1}{2}\sum_\vek{k}(\hat{p}_\vek{k}^2+\omega_\vek{k}^2\hat{x}_\vek{k}^2)\\
&=\sum_\vek{k}\left(\hbar\omega_\vek{k}\hat{a}^\dag_\vek{k}\hat{a}_\vek{k}+\frac{1}{2}\right)\ .
\end{split}
\end{equation}
Since the contribution of the sum of ground state energies to this Hamiltonian is infinite, we shall adopt here the conventional practise of omitting it and redefining~$\hat{H}$,
\begin{equation}\label{quantumHam}
\hat{H}=\sum_\vek{k}\hbar\omega_\vek{k}\hat{a}^\dag_\vek{k}\hat{a}_\vek{k}\ .
\end{equation}

An infinite dimensional Hilbert space is associated with each mode, and the state of the field as a whole inhabits an infinite tensor product of these spaces.
By comparing the quantum Hamiltonian in equation \eqref{quantumHam} with the classical Hamiltonian in equation \eqref{classicalHam}, we see that the quantum case can be obtained from the classical by replacing coefficients $a_\vek{k}$ with the operators $\sqrt{\hbar/(2\epsilon_0\omega_\vek{k})}\hat{a}_\vek{k}$.
Thus by performing the same substitution in equation~(\ref{classicalE}) we can write a quantum electro-magnetic field operator $\hat{\vek{E}}$, 
\begin{equation}\label{quantumE}
\hat{\vek{E}}=\sum_\vek{k} \varepsilon_\vek{k}\left(\hat{a}_\vek{k} 
u_\vek{k}(\vek{r}) \vekk{\epsilon}_\vek{k} +\hat{a}^\dag_\vek{k} 
u^*_\vek{k} (\vek{r}) \vekk{\epsilon}^*_\vek{k}
\right)
\end{equation}
where $\varepsilon_\vek{k}=\sqrt{(\hbar \omega_\vek{k})/(2\epsilon_0)}$.

Note that, unlike its classical counterpart, this operator is time-independent. This is because we are working  in the Schr\"{o}dinger representation where this time dependence is carried by the state.

\section{Quantum states of a light mode}\label{sec:states}

States of a single light mode of a  quantum harmonic oscillator are  density matrices on an infinite-dimensional Hilbert space

\begin{equation}
\rho=\sum_{p,q}\alpha_{p,q}\ket{p}\bra{q}\ .
\end{equation}

Since full description of the state requires specification of an infinite number of coefficients, it is often convenient to express states in terms of \emph{functions in phase space}. 
Rather than working directly in the phase space defined by the position and momentum variables $\hat{x}$ and $\hat{p}$, we shall use the quantum optics  convention of  defining dimensionless and rescaled operators $\hat{X}=\sqrt{\omega/\hbar}\hat{x}=\sqrt{1/2}(\hat a + \hat{a}^\dag)$ and $\hat{P}=\sqrt{/(\hbar\omega)}\hat{p}=-i\sqrt{1/2}(\hat{a}-\hat{a}^\dag)$. These operators are known in quantum optics as quadrature operators\footnote{Note that the precise definition and normalisation of the quadrature operators is a matter of convention and, unfortunately, there is little uniformity in the literature. For a comprehensive discussion of all the different degrees of freedom one has in these conventions see \cite{krugerthesis}, section~A-2.} and obey the commutation relation $[\hat X, \hat P]=i$.

A variety of phase-space functions are used in the literature, and each may be obtained from the other by the appropriate transform. However, in this thesis we  will restrict our discussion to the Weyl function or \emph{characteristic function} of the state.

First we define the \emph{Weyl operator} $W(X,P)$ as follows, 

\begin{equation}
W(X,P)=e^{i(X\hat{P}-P\hat{X})}
\end{equation}
where $X$ and $P$ are real valued variables which represent a point in the rescaled position and momentum phase space. Apart from a change in the argument, the Weyl operator is the same as the phase space displacement operator introduced by Glauber \cite{glauberdisp}, $\hat{D}_\alpha$, where $\alpha=\sqrt{1/2}(X+i P)$, and we shall employ several of the properties derived by Glauber in part~\ref{part3}.

%
%
The characteristic function of the state $\rho$, $\chi_\rho(X,P)$ is then defined
\begin{equation}
\chi_\rho(X,P)=\Tr(\rho W(X,P))\ .
\end{equation}
This function is uniquely defined by and is in one-to-one correspondence to the state $\rho$ and therefore provides an alternative means of representing the state.
The characteristic function plays an important role in the description of the \emph{Gaussian States}, an important and widely occurring class of state in quantum optical systems, which will be discussed in more detail in part~\ref{part3}  of this thesis.

\section{Coherent states}\label{coherentstates}

An important class of single-mode states are  the  \emph{coherent states}. These states are parametrised by a single complex number $\alpha$ as follows,

\begin{equation}
\ket{\alpha}=\sum_n \frac{\alpha^n}{\sqrt{n!}}\ket{n}\ .
\end{equation}

They have many important properties, but in particular they behave in many ways most closely to classical light. For example, the expectation value of the operator $\hat{a}$, $\langle \alpha |\hat{a}|\alpha\rangle=\alpha$. Since under free evolution, these states evolve in time as $\ket{\psi(t)}=|e^{-i\omega t}\alpha\rangle$, it is easy to verify that the expectation value of the electro-magnetic field for these states fulfils the equation of motion for the classical field mode in equation~\eqref{classicalE}. Additionally, one can transform the states and operators of the mode using the displacement operator $\hat{D}_{-\alpha}$. In this new picture, the state is transformed to the vacuum state and field operators are now the sum of a complex number $\alpha$ and a bosonic operator. Since this operator acts on the vacuum, its effect can often be neglected, especially when $|\alpha|$ is large, and the mode can be represented to good approximation by the ``classical'' complex amplitude alone. In such a scenario, the mode is often described as ``classical''.

\section{The quantum state of laser light}\label{sec:laserlight}
A detailed description of the generation of laser light is beyond the scope of this introduction, and many textbooks, for example \cite{lambbook} or \cite{mandelbook} provide a clear and detailed treatment. Here we will restrict ourselves to commenting on how laser may be described quantum mechanically. In the most general of terms,  lasers work via the following principle. Laser light is generated by stimulated emission of light from material  in a cavity resonator (see the next chapter for a detailed discussion of optical cavities) which is continuously re-excited so that the stimulated emission carries on. The resonator typically has one semi-silvered mirror so that the cavity radiation leaks out to form a beam of light.

Under ideal conditions the state of the cavity mode is a Poissonian mixture of states of Fock states

\begin{equation}
\rho=e^{-\lambda^2}\sum_n \frac{\lambda^{2n}}{n!}\ket{n}\bra{n}
\end{equation}
where $\lambda$ is a real parameter. However, this may be re-written using the following identity,

\begin{equation}
e^{-\lambda^2}\sum_n \frac{\lambda^{2n}}{n!}\ket{n}\bra{n}\equiv\frac{1}{2\pi}\int d\phi \ket{\lambda e^{i\phi}} \bra{\lambda e^{i\phi}}\ .
\end{equation}
where $\ket{\lambda e^{i\phi}}$ is a coherent state. 
What this means is that the state of the cavity is indistinguishable from a coherent state of known  amplitude but completely unknown  \\ phase\footnote{This is not, however, to ascribe either the latter or former interpretation of the state. \emph{Both} interpretations are equally valid.}.

To a first approximation, the  state of the light-beam which emerges from the cavity, may then be written \cite{blow,vanenkfuchs}
\begin{equation}
\rho_\textbf{beam}=\frac{1}{2\pi}\int d\phi ( \ket{\lambda e^{i\phi}} \bra{\lambda e^{i\phi}})^{\otimes N}
\end{equation}
where the light beam is treated as a sequence of $N$ wave-packets,  each with a travelling  mode function $u(x,y)S(z-ct,w)$ where we assume that the beam is travelling along the $z$-axis. The function $u(x,y)$ describes the transverse profile of the beam, while $S(z,w)=1$ when $w/2\le z\le w/2$ describes a sawtooth function of width $w$\cite{vanenkfuchs}. The width $w$ of the wave-packets can take any value but must be much greater than both the (mean) wavelength of the light and the size of the laser cavity.

The correlation in the phase  between different wave-packets here is called \emph{phase coherence}.
In  any real laser,  noise in the lasing process means that the phase coherence does not last throughout the entire beam. Rather, the phase becomes gradually decorrelated such that beyond the \emph{coherence length} separated wave-packets become completely decorrelated.

In part~\ref{part3} of this thesis, we will consider the manipulation of the quantum state of light pulses. Laser pulses of femtosecond duration are generated in the laboratory from a mode-locked broadband laser such as a Titanium Sapphire laser (see \cite{ultrafast} for the latest developments in the generation of ultra-fast pulses). The laser pulses can be treated as individual wave-packets with an independent state space and, ignoring phase decorrelation effects, the state of the pulses may be written;
\begin{equation}
\rho_\textbf{pulses}=\frac{1}{2\pi}\int d\phi ( \ket{\lambda e^{i\phi}} \bra{\lambda e^{i\phi}})^{\otimes N},
\end{equation}
where we now quite naturally divide the beam to wave-packets consisting  of individual pulses.
 Since current pulsed laser sources can have a  coherence length of up to many 10s of metres \cite{christinapriv}
and pulses have a typical separation on the order of a metre, pulse trains of many pulses can still retain excellent phase correlation, as was demonstrated for neighbouring pulses in a recent experiment \cite{pulsecoherenceexp}.

In quantum optical experiments, one typically splits the beam at its source with a beam-splitter into an ``object beam'', that is to be probed and manipulated further, and a \emph{phase reference} beam.

\begin{figure}
\psfrag{a}{$\hat{a}^\dag$}
\psfrag{b}{$\hat{b}^\dag$}
\psfrag{aout}{$-R^*\hat{a}^\dag+T^*\hat{b}$}
\psfrag{bout}{$T\hat{a}^\dag+R\hat{b}$}
\begin{center}\includegraphics[width=8cm]{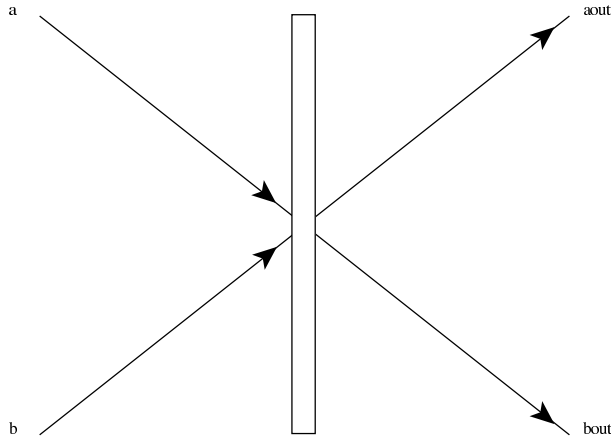}\end{center}
\caption{\label{beamsplitter} A schematic diagram of the action of a beam splitter on a pair of input modes, $\hat{a}$ and $\hat{b}$.}
\end{figure}

A beam splitter is a semi-silvered mirror which mixes two incident light beams together. Under ideal circumstances, when the two modes are perfectly matched in position, timing, frequency and mode envelope, it has a simple action, namely that the mode operators, $\hat{a}^\dag$ and $\hat{b}^\dag$  are linearly transformed to $(\hat{a}^\dag) '=T\hat{a^\dag}+R\hat{b}^\dag$, $(\hat{b}^\dag) '=-R^*\hat{a}^\dag+T\hat{b}^\dag$, where $\{T,R\}\in\mathbbm{C}$ and $|T|^2+|R|^2=1$. The magnitude and phase of the transmittivity $T$ and reflectivity $R$ are properties of the beam splitter. For a more detailed discussion see section~\ref{sec:passive}.
When a coherent state $\ket{\lambda e^{i\phi}}$ is mixed with a vacuum mode on a beam splitter the output modes are a tensor product of coherent states $\ket{T\lambda e^{i\phi}}\otimes\ket{R\lambda e^{i\phi}}$ with a well-defined relative phase.
This means that the state of a split laser pulse may be written,
\begin{equation}\label{rhobeam}
\rho_\textbf{split pulse}=\frac{1}{2\pi}\int d\phi  \ket{T\lambda e^{i\phi}} \bra{T\lambda e^{i\phi}}\otimes \ket{R\lambda e^{i\phi}} \bra{R\lambda e^{i\phi}}\ .
\end{equation}

This state is indistinguishable from  two phase correlated pulses of unknown phase. It is impossible to measure the phase of a coherent state absolutely. It has to be measured relative to some reference beam. When one uses the reference pulse for all phase measurements, measurements will be consistent with the object  pulse having been in a  coherent state.

In fact, the correspondence goes further. Imagine that some quantum operation with Kraus operators $K_i$ is applied to the object pulse
\begin{equation}
\sum_i K_i\rho_\textbf{split pulse}K_i^\dag=\frac{1}{2\pi}\int d\phi [\sum K_i\ket{T\lambda e^{i\phi}} \bra{T\lambda e^{i\phi}}K_i^\dag]\otimes \ket{R\lambda e^{i\phi}} \bra{R\lambda e^{i\phi}}\ .
\end{equation}
When the phase of the object pulse is measured relative to the reference beam, the measurement outcome will be indistinguishable from measurement on the state  $[\sum_i K_i\ket{T\lambda e^{i\phi}} \bra{T\lambda e^{i\phi}}K_i^\dag]$.
Since the Kraus operator formalism includes as a special case projective measurements, this means that as long as all phase measurements on the object beam are performed in relation to the reference beam from the same laser source, the state of the beam behaves \emph{as if it were a coherent state} \cite{molmer}.

For this reason, it is standard practice to treat the light generated from a laser source as a coherent state, as this usually simplifies both the mathematical analysis and the conceptual interpretation of experiments. This is the convention we shall adopt here, and  we shall thus write, for example, $\ket{\lambda e^{i\theta}}$ when we mean 
\begin{equation}
\int d \phi \ket{\lambda e^{i (\phi+\theta)}}\bra{\lambda e^{i (\phi+\theta)}}\otimes \ket{\Lambda e^{i \phi}}_\textbf{ref}\bra{\Lambda e^{i \phi}}\ .
\end{equation}
The presence of the reference beam will always be implicit. We will introduce phase decorrelation as a source of noise when we consider imperfect experimental conditions in chapter~\ref{c-9}.

This is the end of part~\ref{part1}, the general introductory section of this thesis. In part~\ref{part2}, we focus on  entanglement generation with atoms and cavities. In part~\ref{part3}, we shall consider manipulating the entanglement properties of light pulses, in particular introducing a scheme for entanglement distillation.

%% file: c-3.tex
\chapter{Cavity QED and Atom-Field Interaction}
\label{ch:3}

In this part of the thesis, we shall present two novel approaches for generating entanglement in the domain of cavity quantum electrodynamics or cavity QED. Cavity QED is the name given to experiments which probe the quantum nature of the electro-magnetic field trapped in a \emph{cavity}, which usually consists of two or more mirrors. These mirrors can be of an extremely high quality, so quantum states of the resonant cavity modes with lifetimes up to a millisecond \cite{harochermp} can be observed. Additionally, single atoms \cite{kimbletrap} and ions \cite{walthertrap} can be trapped inside the cavities, such that their interactions with the cavity modes can be manipulated and observed.

In this chapter we shall review the basic theory of cavity QED, including  description of the interaction between an atom and a mode of radiation, especially a cavity mode. The methods and results presented  here were collected from a number of sources,  especially \cite{mandelbook}, and also including  \cite{knightrabi}, \cite{cohenphotat} and \cite{loudonbook}.

\section{Cavity modes}\label{sec:cavmodes}
An optical cavity is a system of  mirrors  confining an electro-magnetic field. The mode functions of the cavity are solutions of the Helmholtz equation (\ref{helmholtz}) under boundary conditions defined by the geometry of the cavity mirrors.
Many experiments in cavity QED employ pairs of spherical mirrors, in what is known as a Fabry-Perot configuration, and it is this setting on which this discussion will centre.

\begin{figure}
\begin{center}
\psfrag{rr}{$R$}
\psfrag{ll}{$L$}
\psfrag{zz}{$z$}
\psfrag{z0}{$z=0$}
\includegraphics[width=13cm]{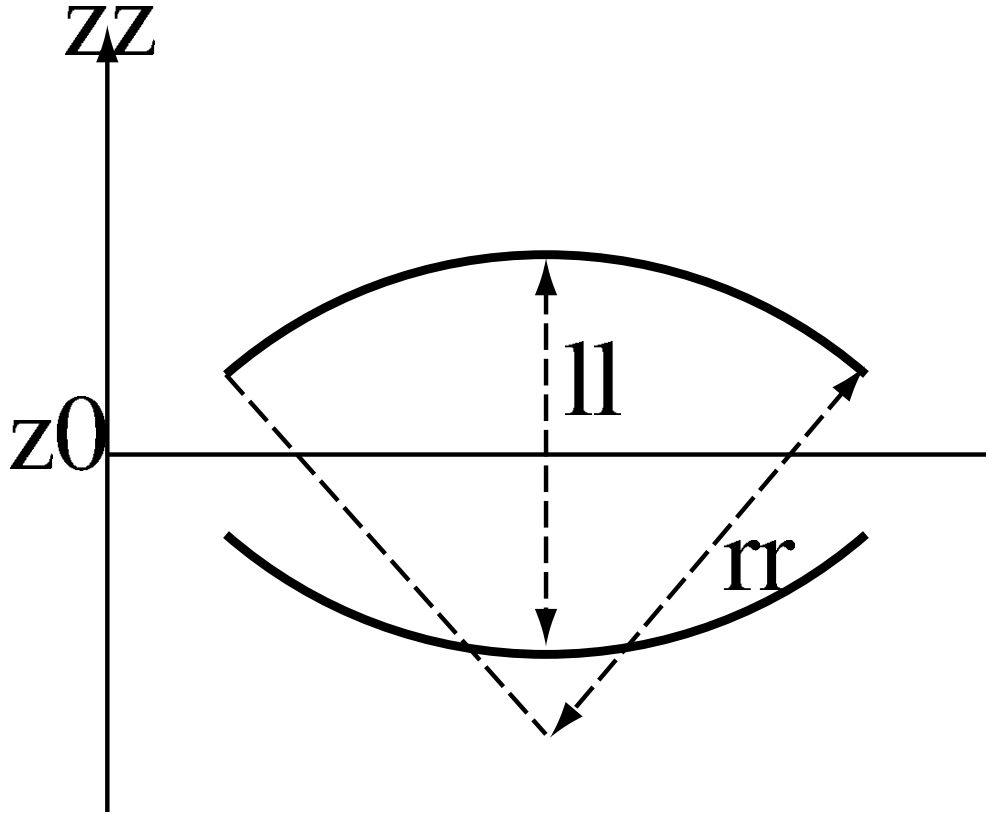}
\end{center}
\caption{\label{cavlayout}The layout of a Fabry-Perot cavity with spherical mirrors of radius of curvature $R$ separated by distance $L$.}
\end{figure}

Approximate  mode-functions for a Fabry-Perot cavity were first calculated numerically by Fox and Li in \cite{foxandli} in 1961, and  approximate analytic solutions soon followed, a  comprehensive review of which was presented  by Kogelnik and Li in 1966\cite{kogelnik66}. For circular spherical mirrors in a Fabry-Perot configuration, the mode-functions are  Laguerre-Gauss functions (products of Laguerre polynomials and Gaussians). The exact form of these mode functions are given in \cite{kogelnik66}. The modes which are most often coupled to atoms in cavity QED experiments are known as TEM$_{n00}$  modes, the transverse electric and magnetic modes which have a Gaussian spatial profile. For two spherical mirrors sharing a central axis and with radius of curvature $R$ whose centres are separated by distance $L$, the  TEM$_{n00}$ mode-function is

\begin{equation}
u_{n00}(z,r)=\frac{w_0}{w(z)}\cos[\Phi(r,z)]e^{-\frac{r^2}{w(z)^2}}
\end{equation}
where cylindrical coordinates are employed. The $z$-axis lies along the line connecting the centres of the two mirrors and the origin is the centre of the cavity, as illustrated in figure~\ref{cavlayout}. 

The function $\Phi(r,z)$ is
\begin{equation}
\Phi=kz+\frac{2 r^2 z}{4z^2+L(2R-L)}-\arctan\left(\frac{2z}{\sqrt{L(2R-L)}}\right)\ ,
\end{equation}
where the wave-number $k$ is
\begin{equation}\label{wavenumber}
k=\frac{(n +1)\pi}{L}+2\arctan\left(\frac{L}{\sqrt{L(2R-L)}}\right)\ .
\end{equation}

The function $w(z)$ is the cavity \emph{beam radius}
\begin{equation}
w(z)^2=w_0^2\left(1+\left(\frac{4 z^2}{L(2R-L)}\right)\right)\ ,
\end{equation}
with
\begin{equation}\label{waist}
w_0^2=\frac{\lambda\sqrt{L(2R-L)}}{2 \pi}\ .
\end{equation}

The frequency of the mode $\nu$ is given by $\nu=c k/(2 \pi)$ and we can see from equation~\eqref{wavenumber} that the frequency difference between TEM$_{n00}$ modes will be $c/(2L)$. For a microwave cavity where $L$ is on the order of mm, this frequency difference is on the order of GHz or THz\footnote{The frequency difference between non-Gaussian modes for a given principle mode number $n$ is much smaller than this, actually  of the order $(c/\pi) \arctan\left(L/(\sqrt{L(2R-L)})\right)$ (see \cite{kogelnik66}) and there is degeneracy in the non-Gaussian modes. However, this frequency difference is still on the order of GHz. That the modes are so widely spaced in frequency will allow us to simplify significantly the description of an interaction with an atom which is resonant with a particular mode - see section~\ref{multimode}.}.
In the literature (see for example \cite{thompsonkimble,hoodcavqed}),  the following approximation  mode function is often employed
\begin{equation}
u_{n00}\approx e^{-\frac{r^2}{w_0^2}}\cos(kz).
\end{equation}
Since this eases calculation considerably and this is a good approximation when the atom remains close to the plane at $z=0$, we shall use this in our calculations in chapter~\ref{ch:4}.
From basic symmetry arguments, it is clear  that the mode polarisation vectors  must lie perpendicular to $z$-axis.

\section{The interaction between an atom and a light mode}

A full treatment of the interaction between an atom and a mode of the electro-magnetic field (see, for example \cite{mandelbook}, chapter 15 or \cite{cohenphotat}) is  rather complex and beyond the scope of this thesis. For the parameter ranges of cavity QED, one can  employ  the \emph{dipole approximation} which simplifies the description of the interaction significantly. This approximation, which treats the atom as a point-like dipole, is valid when the wavelength of the light is much larger than the atomic dimension and when the light is not of high intensity.

 The classical Hamiltonian for the interaction between a dipole with dipole moment $\vek{D}$ and a homogeneous electric field is
\begin{equation}
H_{I}=\vek{D}.\vek{E}(t)\ .
\end{equation}

This Hamiltonian can be cast in quantum mechanical form by replacing $\vek{D}$ and $\vek{E}$ with the corresponding quantum mechanical operator. The quantum electric field operator $\hat{\vek{E}}$ has been derived above.
We simplify the analysis further by adopting a \emph{two-level approximation} for the atom, where we  assume that the atom has only  two internal states, $\ket{A}$ and $\ket{B}$ with  energy difference $\hbar \omega_{AB}$.
A justification for this approximation is that, as we shall see explicitly below, energy exchange between the atom and cavity only occurs when the atomic transition frequency is close to \emph{resonance} with the mode frequency. Interaction with non-resonant modes leads to energy shifts, which can be  accounted for in the way we define the energies of levels $\ket{A}$ and $\ket{B}$. In cavity QED experiments, the atomic species are chosen such that only the desired transitions fulfill the resonance requirements and thus the two-level approximation is valid\footnote{Sometimes  more complicated  situations arise where interactions between  three, four or more  levels must be taken into account. As long as each transition is resonant with just one mode then this can be treated as a sum of  individual two-level interaction Hamiltonians, as is the case in chapter~\ref{c-5}.}.

In this two-level picture , the most general form that the dipole operator $\hat{\vek{D}}$ can take is 
\begin{equation}
\hat{\vek{D}}=\vek{D}_{AA}\ket{A}\bra{A}+\vek{D}_{BB}\ket{B}\bra{B}+\vek{D}_{AB}\ket{A}\bra{B}+\vek{D}_{BA}\ket{B}\bra{A}
\end{equation}
where $\vek{D}_{ij}=\langle i |\hat{\vek{D}}|j\rangle$.

The dipole moment operator is proportional to  the relative position vector between the nucleus and the interacting electron $\hat{\vek{D}}=-e\hat{\vek{r}}$ and thus has odd parity
 (it changes sign when $\vek{r}$ is replaced by $-\vek{r}$). Since diagonal components of the interaction operator merely shift the energies of the system's basis states  we can neglect the diagonal components and set $\vek{D}_{AA}$ and $\vek{D}_{BB}$ to zero. Furthermore, since $\hat{\vek{D}}$ is a Hermitian operator,  $\vek{D}_{AB}=\vek{D}_{BA}^*$. For brevity of notation, let us label  $\vek{D}_{AB}=\vek{d}$.

The interaction Hamiltonian can then be written

\begin{equation}\label{hinorwa}\begin{split}
\hat{H}_I=\sum_\vek{k}& \varepsilon_\vek{k}\biggl((\vek{d}\cdot{ \vekk{\epsilon}})u_\vek{k}(\vek{r})\hat{a}_\vek{k}\ket{B}\bra{A}+
(\vek{d}^*\cdot{ \vekk{\epsilon}^*})u_\vek{k}^*(\vek{r}\hat{a}^\dag_\vek{k}\ket{A}\bra{B}+\\&
(\vek{d}^*\cdot{ \vekk{\epsilon}})u_\vek{k}(\vek{r})\hat{a}_\vek{k}\ket{A}\bra{B}+
(\vek{d}\cdot{ \vekk{\epsilon}^*})u^*_\vek{k}(\vek{r})\hat{a}^\dag_\vek{k}\ket{B}\bra{A}
\biggr)\ .
\end{split}
\end{equation}

The analysis of this interaction Hamiltonian can be simplified greatly by the use of the \emph{Rotating Wave Approximation} (RWA). This is another approximation which works very well for interactions with non-intense fields. Under the RWA, the latter two terms of equation \eqref{hinorwa} are omitted. These are the terms which contain products of terms which both create or both destroy quanta of energy. It can be shown that the higher order effects to which these terms contribute  are only non-negligible  in very intense laser fields (for an example of this, see \cite{brownekeitel}). The term ``Rotating Wave Approximation'' is due to the fact that, when an interaction picture (see section~\ref{intpic}) is adopted, the neglected terms rotate at a much higher frequency than the remaining terms, which  means that they  have a much smaller effect on the system's evolution.

The interaction Hamiltonian under the RWA takes the concise form
\begin{equation}\label{Jaynes-Cummings}
\hat{H}_I=\hbar\sum_\vek{k}\left(g_\vek{k}(\vek{r})\hat{a}_\vek{k}\ket{B}\bra{A}+g^*_\vek{k}(\vek{r})\hat{a}^\dag_\vek{k}\ket{A}\bra{B}\right)
\end{equation}
where $g_\vek{k}(\vek{r})=(\vek{d}\cdot{ \vekk{\epsilon}})u_\vek{k}(\vek{r})/\hbar$ contains all spatial dependence of the coupling strength. 
This is often called  the \emph{Jaynes-Cummings} Hamiltonian, after the authors\cite{jaynescummings} who first introduced it (see \cite{plkjaynes} for a review).

\section{Interaction with a single cavity mode}

To better understand how the atom will interact with the whole electro-magnetic field, let us first consider the interaction with just a single mode. We shall see that the behaviour of this interaction depends greatly on whether the energy of the transition between the atomic levels is at resonance with the mode frequency.

As we are only considering a single mode, let us drop mode labels and write the full Hamiltonian of the atom-mode system, under the approximations introduced above, 
\begin{equation}\label{intham1}
\hat{H}=\hbar\omega \hat{a}^\dag \hat{a}+\frac{\hbar{\omega_{AB}}}{2}(\ket{B}\bra{B}-\ket{A}\bra{A})+\hbar(g \hat{a}\ket{B}\bra{A} + g^* \hat{a}^\dag\ket{A}\bra{B})\ ,
\end{equation}
where $\omega$ is the mode frequency and $\hbar\omega_{AB}$ is the transition energy between levels $\ket{A}$ and $\ket{B}$.

When dealing with resonance phenomena it is useful to define  the \emph{detuning}, here $\delta=\omega-\omega_{AB}$, between the transition frequency and mode frequency. Hence, we rewrite equation \eqref{intham1}

\begin{equation}
\hat{H}=\hbar\omega \hat{a}^\dag \hat{a}+\frac{\hbar(\omega-\delta)}{2}(\ket{B}\bra{B}-\ket{A}\bra{A})+\hbar(g \hat{a}\ket{B}\bra{A} + g^* \hat{a}^\dag\ket{A}\bra{B}) \ .
\end{equation}

Due to the RWA, this Hamiltonian only couples states in pairs $\ket{B,n}$ and $\ket{A,n+1}$, and the Hamiltonian can be rewritten as a direct sum of operators $\hat{H}_n$ which act only in these sub-spaces,

\begin{equation}
\hat{H}=\sum_{n=0}^\infty \hat{H}_n\ ,
\end{equation}
where $\hat{H}_n$ can be written in matrix form as follows,
\begin{equation}\begin{split}
&\frac{\hat{H}_n}{\hbar}=\begin{array}{c}\left(\begin{array}{cc}\ket{A,n+1}&\ket{B,n}\end{array}\right)\\ \phantom{()}\end{array}\\&\qquad\qquad\qquad\mbox{}\cdot
\left(\begin{array}{cc}
\omega(n+\half)+\delta/2& g^*\sqrt{n+1}\\
g\sqrt{n+1}&\omega(n+\half)-\delta/2\end{array}\right)\left(\begin{array}{c}\bra{A,n+1}\\\bra{B,n}\end{array}\right)\ .
\end{split}
\end{equation}

Diagonalising this matrix for all $n$ will give us the eigenbasis of the full atom-cavity Hamiltonian. Let us label these eigenstates, which are often called \emph{dressed states} $\ket{\alpha_+,n}$ and $\ket{\alpha_-,n}$ \cite{jaynescummings,cohendressed}. Their respective energies are $\hbar(\omega(n+\half)\pm\hbar(\Omega_n/2)$ where $\Omega_n=\sqrt{\delta^2+4|g|^2(n+1)}$ is known as the \emph{Rabi frequency}. 

The states themselves can be written

\begin{equation}\label{dressedplus}
\ket{\alpha_-,n}=\sin(\theta)\ket{A,n+1}+e^{i\arg(g)}\cos(\theta)\ket{B,n}
\end{equation}

\begin{equation}\label{dressedminus}
\ket{\alpha_+,n}=-\cos(\theta)\ket{A,n+1}+e^{i\arg(g)}\sin(\theta)\ket{B,n}
\end{equation}
where 
\begin{equation}
\tan(\theta)=\frac{\delta-\Omega_n}{\sqrt{\Omega_n^2-\delta^2}}\ .
\end{equation}

\subsection{Far-detuned behaviour}
Let us first consider the case of far detuning, i.e. when $|\delta|\gg g$. This means that  $\tan(\theta)$ is approximately zero and therefore $\ket{\alpha_-,n}\approx\ket{B,n}$ and $\ket{\alpha_+,n}\approx\ket{A,n+1}$. This can be interpreted as follows; the product basis states $\ket{B,n}$ and $\ket{A,n+1}$ are approximate eigenstates of the complete Hamiltonian, therefore a system  prepared in one of these states will remain there. In other words, at far off-resonance, no energy exchange occurs between the atom and the cavity mode. However,  states  $\ket{\alpha\pm,n}$ do not have the same energy as $\ket{A,n+1}$ and $\ket{B,n}$. Thus, the observed energy levels of the atom appear to have shifted from their respective natural values by $\mp\hbar(\delta-\Omega_n)/2\approx\mp\hbar g^2(n+1)/\delta$. This effect is known as the \emph{light shift}.

\subsection{Near resonance}

Now let us consider the opposite extreme, when $|\delta|\ll g$. In this limit  $\tan(\theta)$ is approximately unity and the dressed states are approximate equal superpositions of the states $\ket{A,n+1}$ and $\ket{B,n}$.

Consider a system prepared in the product state $\ket{B,n}$. When we expand this state in terms of the dressed states its time evolution is easy to calculate, 
\begin{equation}
\begin{split}
\ket{\psi_{B,n}(t)}&=e^{-i \arg(g)}e^{-i\omega(n+\frac{1}{2})t}\left( \sin(\theta) e^{-i\frac{\Omega_n}{2}t}\ket{\alpha_{+,n}}+\cos(\theta) e^{+i\frac{\Omega_n}{2}t}\ket{\alpha_{-,n}}\right)\\
&=e^{-i\omega(n+\frac{1}{2})t}\biggl(-i\frac{2g^*}{\Omega_n}\sin\left(\frac{\Omega_n}{2} t\right)\ket{A,n+1}\\ &\qquad\qquad\qquad\mbox{}+\left[\cos\left(\frac{\Omega_n}{2} t\right)+i\frac{\delta}{\Omega_n}\sin\left(\frac{\Omega_n}{2} t\right)\right]\ket{B,n}\biggr)\ .
\end{split}
\end{equation}

Likewise, the evolution of a system  prepared in state $\ket{A,n+1}$ is
\begin{equation}\begin{split}
\ket{\psi_{A,n+1}(t)}=&e^{-i\omega(n+\frac{1}{2})t}\biggl\{\left[\cos\left(\frac{\Omega_n}{2} t\right)-i\frac{\delta}{\Omega_n}\sin\left(\frac{\Omega_n}{2} t\right)\right]\ket{A,n+1}\\&\mbox{}\qquad
-i\frac{2g}{\Omega_n}\sin\left(\frac{\Omega_n}{2} t\right)\ket{B,n}\biggr\}\ .
\end{split}
\end{equation}

In the limit that $|\delta|/g$ tends to zero, these become

\begin{equation}\label{rabi}
\ket{\psi_{B,n}(t)}=e^{-i\omega(n+\frac{1}{2})t}\left(-i e^{-i \arg(g)}\sin\left( \frac{\Omega_n}{2}t\right) \ket{A,n+1}+\cos\left( \frac{\Omega_n}{2}t\right) \ket{B,n}\right)\ ,
\end{equation}

\begin{equation}\label{rabi2}
\ket{\psi_{A,n+1}(t)}=e^{-i\omega(n+\frac{1}{2})t}\left(\cos\left( \frac{\Omega_n}{2}t\right) \ket{A,n+1}-i e^{i \arg{g}}\sin\left( \frac{\Omega_n}{2}t\right) \ket{B,n}\right)\ .
\end{equation}

The atomic population oscillates between ground and excited states with frequency $\Omega_n$. This phenomenon is known as \emph{Rabi oscillation}, which is why $\Omega_n$ is called the Rabi frequency.

\section{Multi-mode case - interaction with a resonant  mode}\label{multimode}
In a standard atom cavity experiment, an atom is prepared in a state with a transition close to resonance with a particular cavity mode. Typically, the Rabi frequency will be on the order of kHz, while the difference between mode frequencies will be on the order of GHz or THz. This means that the far-detuned limit is valid for all modes except those near to resonance.

Thus, while a full treatment  of  the evolution of the system would require the full multi-mode Hamiltonian, it is usually a very good approximation to neglect the interaction with the non-resonant modes, since they only contribute  level shifts to the system's evolution. The single-mode Hamiltonian, with the off-resonant energy shifts included in the state energies,  describes the evolution of the system very well and  is the starting point for many research papers on cavity QED.

\section{Interaction with a classical field}\label{sec:intclassical}

When a laser mode interacting with an atom with a Jaynes-Cummings Hamiltonian is in a coherent state with large amplitude $|\alpha|$, solely quantum effects, such as  entanglement between the mode and the atom become negligible, and the state of the mode can be treated ``classically'' to a good approximation. As discussed in section~\ref{coherentstates}, this involves assuming that the field is in a coherent state $\ket{\alpha}$ with $|\alpha|\gg1$. One then replaces the mode operators in the Hamiltonian by their time-dependent expectation values, neglecting any effects on the time evolution from the interaction, e.g., operator $\hat{a}$ is replaced by the expectation value $\bra{\alpha e^{-i \omega t}} \hat a \ket{\alpha e^{-i \omega t}}=\alpha e^{-i \omega t}$. The free field energy in the Hamiltonian is now just a number, and can be neglected, as it just imparts a non-detectable global phase to the systems evolution. The  \emph{semi-classical} (as the atom is still treated quantum mechanically) atom-mode interaction Hamiltonian is
\begin{equation}
\hat{H}_\textrm{int: s.c}=g\alpha e^{-i \omega t} \ket{B}\bra{A} + h.c.\ .
\end{equation}
Typically, the phase of $g\alpha$ is unimportant, so $g$ is assumed to be real, and the classical Rabi frequency $\Omega_c=2 g\alpha$ is introduced, to give the standard semi-classical interaction Hamiltonian under the rotating wave approximation,
\begin{equation}
\hat{H}_\textrm{int: s.c}=\frac{\Omega}{2} e^{-i \omega t}  \ket{B}\bra{A} + h.c. \ .
\end{equation}

\section{Simplifying the Hamiltonian: the interaction picture}\label{intpic}
In quantum optics, the most interesting part of a systems evolution is often the part  caused  by its interactions. The  evolution of the system due to the free energies of the sub-systems is usually of little interest, and only complicates calculations. For this reason, one often employs the \emph{interaction picture}.

Let us define $\hat{H}_\textrm{int}$ as that part of  the Hamiltonian due to the interaction one is interested and $H_0=\hat H-\hat{H}_\textrm{int}$ as the non-interaction part. Without the interaction, the system would evolve via the unitary  $U(t)=\exp[-i\hat{H}_0  t/\hbar]$. In the interaction picture, one works with the state without this free evolution $\ket{\psi_\textrm{I.P}}=U^\dag(t)\ket{\psi}$ and operators  are correspondingly obtained as $\hat{O}_{I.P}=U(t){O}_{I.P}U^\dag(t)$.

Since the Hamiltonian determines the time evolution of the system and $U(t)$ is time-dependent, the interaction picture Hamiltonian must  be obtained in a different way. By inspecting the Schr\"odinger equation, one sees that the interaction picture  Hamiltonian must have the form
\begin{equation}\begin{split}
\hat{H}_\textrm{I.P.}&=U^\dag(t) \hat{H}U(t) +i\hbar \frac{\partial U^\dag(t)}{\partial t} U\\&=U^\dag(t) \hat{H}_{int} U(t)\ .
\end{split}  
\end{equation}

We shall use the interaction picture to simplify the calculations in chapter~\ref{c-5}.







\section{Decoherence and dissipation - the ``quantum jump'' approach}\label{quantjump}

\begin{figure}
\psfrag{g}{$\Gamma$}
\psfrag{k}{$\kappa$}
\begin{center}
\includegraphics[scale=0.5]{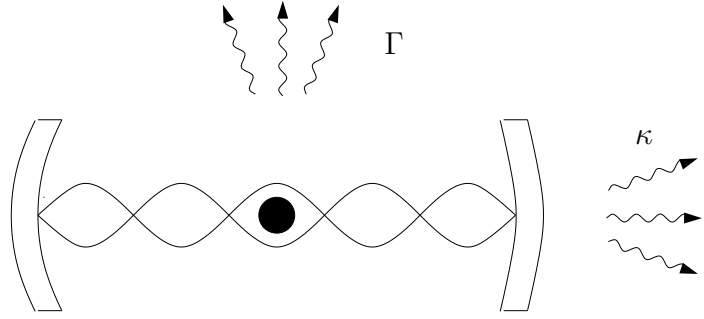}\caption{In a real cavity, the mirrors will not be perfect, photons will leak through the mirrors with rate $2\kappa$. Also, the cavity will not be totally sealed from the continuum of vacuum modes of the surrounding environment. An atom in the cavity will undergo spontaneous emission into these modes with rate $2\Gamma$.}\label{cavityfig}
\end{center}
\end{figure}

In a real experiment, the cavity will not be perfect, photons in the cavity will escape by leakage through the mirrors, and the atom will spontaneously emit photons out of the cavity due to the coupling of the atom to the modes of the electro-magnetic field of the free space on either side, as illustrated in figure \ref{cavityfig}.

The dynamics of the atom-cavity system under these conditions can be described by a Master equation. This is an evolution equation for the reduced density operator for the atom-cavity system. The dynamics of the system under these forms of decoherence are Markovian, i.e., the environment does not retain a memory of the decay events. This means that the  Master equation 
 can be written in Lindblad form \cite{lindbladform},
\begin{equation}
\dot \rho=- \frac{\im}{\hbar}[\hat{H}_{\rm eff}\rho-\rho \hat{H}_{\rm eff}^\dag]+\sum_i \hat{J_i} \rho \hat{J_i}^\dag
\end{equation}
where $\hat{J_i}$ are  ``quantum jump'' operators, and 
$\hat{H}_{\rm eff}$ is called an ``effective Hamiltonian'' for the system whose Hamiltonian without dissipation and decoherence would be $\hat{H}_c$:
\begin{equation}
\hat{H}_{\rm eff}=\hat{H}_c-i\sum_i\frac{\hat J_i^\dag \hat J_i}{2}\ .
\end{equation}

This decomposition of the Master equation has a pleasingly intuitive physical interpretation. The loss of energy from the system is described in terms of quantum jumps, which describe the loss of  a single quantum of energy from the system. Each operator $\hat{J_i}$ embodies one type of jump. The non-Hermitian effective Hamiltonian describes the continuous Schr\"odinger-like evolution under the condition that no  jump takes place. Its non-Hermitian character means that the norm of the state under this evolution will decrease, and this corresponds exactly to the probability that no jump has occurred.
Thus the system evolution can be thought of as an average over many trajectories, each of which consists of continuous Schr\"odinger-like evolution and random quantum jumps\footnote{It must be stressed however, that no physical reality can be ascribed to these trajectories. There are infinitely many other equally valid possible ``unravellings'' of the Master equation.}. This ``quantum jump approach''\cite{zollermartewalls,dalibardcastinmoelmer} (see \cite{plenioquantj} and \cite{gardinerzoller} for introductory reviews) allows one to numerically simulate the Master equation using Monte-Carlo methods.


For a two-level atom in a cavity, there are two major sources of energy loss, illustrated in figure~\ref{cavityfig}; leakage of the photons from the cavity at rate $2\kappa$, and spontaneous emission at rate $2\gamma$. The jump operator for the photon leakage is $\sqrt{2\kappa} \hat{a}$, and for the spontaneous emission is $\sqrt{2\gamma}\ket{A}\bra{B}$.
Although the construction of the Master equation in this way seems purely phenomenological, it can, in fact be derived from the physical decay processes by a fuller treatment (see, for example, \cite{meystrebook}, 3rd edition).

\section{Generation of entanglement in cavity QED}

Cavity QED has proven itself to be an excellent testing ground for the validity of quantum mechanics with a high degree of precision. Many beautiful experiments have been carried out, and in recent years, entangled states have been created and verified. There are two principle approaches for the creation of entanglement in these systems. The first is a precise control of the coherent dynamics, such that the desired entangled state evolves through Rabi oscillation. This was the first approach to be implemented experimentally. 
Rauschenbeutel and co-workers
 generated Bell states between two modes in a single cavity
using a Rydberg atom \cite{rauschen01} coherently interacting with
each mode in turn. Their scheme could be adapted in a
straightforward way to generate such states in spatially separated
cavity modes. The same Paris group, which is  led by Serge Haroche, have  performed several further such experiments summarised in the following review articles \cite{harochermp,harochereview2}  including the creation of an entangled state between two atoms and a cavity.

In  the second approach, measurement plays a key role. The first proposals of this kind were by Cabrillo and co-workers \cite{cabrillo99} and Plenio et al \cite{pleniobeige}. They showed that  Bell states could be generated between atoms conditionally, by
driving them with a weak laser pulse, and subsequently detecting a
photon spontaneously emitted by one of the atoms. The key to their scheme is that the measurement cannot in principle reveal from which cavity the photon originated, and the measurement thus projects the atoms onto the desired entangled state. Many refinements of this idea have been proposed, including the scheme by Bose et al. \cite{bose99tele} which is more relevant to cavity QED experiments will be discussed in detail in chapter~\ref{c-5}. 
Several enhanced versions of these schemes have been proposed recently,  \cite{protsenko,feng,simon,duan}, which aim to combat some of their inherent drawbacks, namely that
the schemes typically have a finite success probability, often around 50\%, and often do not work well when realistic imperfections, such as spontaneous emission and inefficient detection, are taken into account. 
%
%
It has recently been pointed out \cite{brownerudolph} that such entangling operations can be used to generate  entangled multi-qubit states known as cluster states\cite{briegelcluster}, which are a resource for universal quantum computation via single qubit measurements alone\cite{raussenbrowne}.

In the following two chapters, we describe two novel schemes for the generation of maximally entangled two qubit states between spatially separated cavities. Both of these schemes fall into the second of the above categories, and rely on the outcome of a measurement for the generation of the desired entangled state.
In chapter~\ref{ch:4}, we introduce a method of generating a Bell state between modes of two spatially separated cavities, whose robustness against experimental error stems from the fact that it relies on very short interactions between the cavities and a mediating atom.

In chapter~\ref{c-5} this principle is taken a step further, where a single brief interaction is replaced by continuous weak driving, to produce a scheme for entangling two ions or atoms trapped in spatially separated cavities. This  can, in contrast to previous schemes, achieve success probabilities arbitrarily close to 100\%, and is robust against experimental imperfections.


%% file: c-4.tex
\chapter{ Generation of Entanglement between two Optical Cavities}
\label{ch:4}
\section{Introduction}

In this chapter, we  describe a novel method \cite{brownebrief} of generating entanglement between two spatially separated
optical  cavities. 
The goal is the creation of a maximally entangled Bell state in the photon numbers of a  mode in two spatially separated cavities. Let us label the two cavities $\mathcal{A}$ and $\mathcal{B}$ and label the Fock states of the chosen mode $\ket{n}_\mathcal{A}$ and $\ket{n}_\mathcal{B}$ respectively. The state which is to be generated can thus be written
\begin{equation}
\ket{\Phi^+}=\sqrt{\frac{1}{2}}(\ket{0}_\mathcal{A}\ \ket{1}_\mathcal{B}+\ket{1}_\mathcal{A}\ \ket{0}_\mathcal{B})\ .
\end{equation}

\begin{figure}
\includegraphics[scale=0.6]{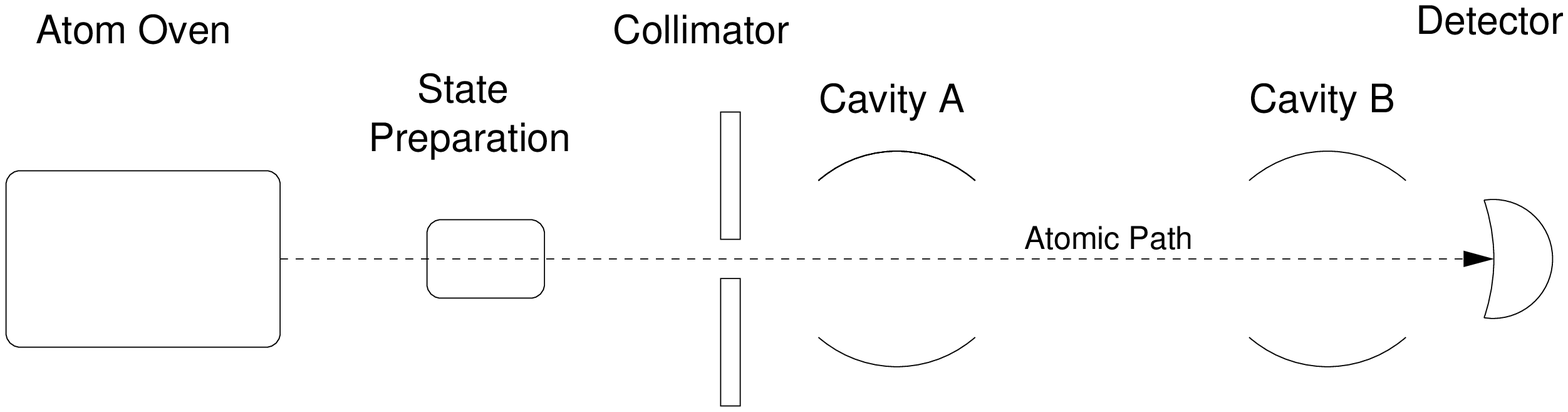}
\caption{\label{fig:layout} A schematic diagram of the layout of the scheme.}
\end{figure}

The scheme can be summarised as follows. A beam of atoms is employed, which have a transition resonant with modes in two identical  cavities.
The two cavities are positioned as in
figure \ref{fig:layout}, such that their axes are parallel, and
their centres in alignment. The cavities must be prepared in their vacuum state. Due to the high energies of their photons,  cavities with optical resonant frequencies, known as optical cavities,  are naturally very close to this state at room temperature. In the case of   microwave cavities, on the other hand, this requires cooling to low temperatures. 

The first atom of the beam  is prepared in the excited state of the resonant  transition, which we denote $\ket{B}$, as above,  and then passes along a path through the centres of each cavity. Upon leaving the second cavity the atom's state is immediately measured. If the atom is now in state $\ket{A}$, the lower state of the transition,  and if the  conditions explained below are satisfied, the entangled state $\ket{\Phi^+}$ has been generated in the cavities. If the atom is measured to have remained in  state $\ket{B}$, the procedure must be repeated, with a further atom.
Let us  examine the scheme in more detail, and explain how it works.  A useful concept in the  description of the atom-cavity systems after the atom's passage is the \emph{total effective interaction time} between the atom and each cavity.

\section{Effective interaction time}

Since the interaction Hamiltonian of an atom interacting with a cavity
\begin{equation}
H_{int}=g(\vek{r}) \hat a \ket{B}\bra{A} + h.c.
\end{equation}
is dependent on the position $\vek{r}$ of the atom, when the atom is in motion through the cavity  this Hamiltonian becomes \emph{time-dependent}. The Schr\"odinger equation for a time-dependent Hamiltonian can often not be solved analytically, and this is especially true in a situation like this, where the time-dependence is a functional of a yet unspecified path $\vek{r}(t)$. Nevertheless, in the special case that the detuning between the atomic transition and the cavity is zero, the Schr\"odinger equation can still be integrated analytically since the dressed states of the system do not depend on $|g(\vek{r})|$ and thus remain time-independent (see equations~(\ref{dressedplus}) and ~(\ref{dressedminus})). Since the mode-functions of the standing  cavity-modes we are considering are real,  the phase of $g(\vek{r})$ remains constant along its path. Thus, for simplicity, and without loss of generality, we assume that $g(\vek{r})$ is real.

If we look, for example at  the equation of motion for the system initially in the product state $\ket{B,n}$,  we see that is has a similar  form to equation~(\ref{rabi})
\begin{equation}\begin{split}
\ket{\psi(t)}&=e^{-i\omega(n+\frac{1}{2})t}\biggl(-i\sin\left(\sqrt{n+1} {\int_0^t |g(t')|dt'}\right) \ket{A,n+1}\\&\qquad\qquad\mbox{}+\cos\left(\sqrt{n+1} {\int_0^t |g(t')|dt'}\right) \ket{B,n}\biggr)
\end{split}
\end{equation}
except that the simple product $\Omega_n t = \sqrt{n+1} g t$ is replaced by an integral over time, $\sqrt{n+1} \int_0^t |g(t')|dt'$. That this equation of motion satisfies the appropriate  Schr\"odinger equation can be verified easily by differentiation.
If we define the \emph{effective interaction time} $\tau(t)=\int_0^t g(t')dt' / g_0$, where $g_0=g(\vek{r}=0)$ is the coupling strength when the atom is in the centre of the cavity, then the  equation of motion 
\begin{equation}
\ket{\psi(t)}=e^{-i\omega(n+\frac{1}{2})t}\biggl(-i\sin\left(g_0 \tau(t)\right) \ket{A,n+1}+\cos\left(g_0 \tau(t)\right) \ket{B,n}\biggr)
\end{equation}
is now equivalent to that of a stationary atom, sitting in the centre of the cavity for duration $\tau(t)$. By calculating the equation of motion for a system initially in state $\ket{A,n+1}$ as well, one sees that the unitary evolution operator for an atom interacting resonantly with a cavity with effective interaction time $\tau$ is

\begin{equation}
U=\exp\left[-i g_0 \tau(\hat a \ket{B}\bra{A} + \hat{a}^\dag \ket{A}\bra{B})\right] \ .
\end{equation}

If an atom passes through the cavity on a straight path from some point much further away from the centre of the cavity than waist of the cavity mode, to some far point on the other side of the cavity, one can characterise the  complete interaction by the \emph{total effective interaction time}.

\section{Outline of the scheme under ideal conditions}

Using the effective interaction time, we can now show very simply how the scheme works. On its path through both cavities the atom interacts resonantly with each. Let us label the total effective interaction times with each cavity $\tau_\mathcal{A}$ and  $\tau_\mathcal{B}$ respectively.
The evolution of quantum state of the system after the two atom-cavity interactions can be written 

\begin{equation}
U=\exp\left[-i g_0 \tau_\mathcal{B}(\hat a_\mathcal{B} \ket{B}\bra{A} +h.c.)\right] \exp\left[-i g_0 \tau_\mathcal{A}(\hat a_\mathcal{A} \ket{B}\bra{A} +h.c.)\right] 
\end{equation}

where we assume that the coupling strength $g_0$ is the same in each cavity and where we label mode operators for each cavity $\hat{a}_\mathcal{A}$ and  $\hat{a}_\mathcal{B}$ respectively.

In the scheme we propose, the cavities are aligned as shown in
figure \ref{fig:layout} and the atom passes along the straight
line through the centre of both cavities at constant velocity.
This means that the effective interaction times between the atom
and each cavity will be equal. It may, of course, be difficult to
control the path of the atom with sufficient accuracy that the
interaction times are exactly equal, and the effect of this is
discussed in section \ref{collim}. For now, however, let us assume
that $\tau_\mathcal{A}$ and $\tau_\mathcal{B}$ are both equal to some value $\tau$.
In the limit when the effective interaction times are very small, i.e. when $g_0\tau\ll1$, $U_{AB}$ can be expanded to the first order in $g_0\tau$, and takes the following form
\begin{equation}
U\approx \iden - \im g_0\tau\biggl[(\hat{a}_\mathcal{A}+\hat{a}_\mathcal{B}) \ket{B}\bra{A}+(\hat{a}^\dag_\mathcal{A}+\hat{a}^\dag_\mathcal{B}) \ket{A}\bra{B}\biggr].
\end{equation}

Thus, the state of the system after the atom has left the second cavity,  $\ket{\psi}=U_{AB}\ket{\psi_{\text{init}}}$, can be written in this limit
\begin{equation}\label{gencavoutcome}
\ket{\psi}\approx \ket{B}\ket{\psi_{\text{cav}}}-\im g_0\tau\ket{A}(a^\dag_{A}+a^\dag_\mathcal{B})\ket{\psi_{\text{cav}}}.
\end{equation}

When the state of the atom is now measured, the cavity modes are projected into one of two states, depending on the measurement outcome.
If $\ket{B}$ is detected, the cavity returns to its initial state.
This is important for a non-deterministic process, because it
means that it can be repeated immediately from the same starting
conditions. If the ground state $\ket{A}$ is detected, the cavity
modes are now in the (un-normalised) state 
$(a^\dag_{A}+a^\dag_\mathcal{B})\ket{\psi_{\text{cav}}}$.
Thus, if the cavities are initially in the vacuum state, the state generated in the cavities would be $(a^\dag_{A}+a^\dag_\mathcal{B})\ket{0}_\mathcal{A}\ket{0}_\mathcal{B}=\ket{1}_\mathcal{A}\ket{0}_\mathcal{B}+\ket{0}_\mathcal{A}\ket{1}_\mathcal{B}$, which when normalised is the Bell state $\ket{\Psi^+}$ introduced above.

If, following a successful run, one were to repeat the scheme
immediately and carry on until $n$ atoms had been detected in the
ground state, in the limit that  $g_0\tau$ is small, the state
generated would have the form
$(a^\dag_{A}+a^\dag_\mathcal{B})^n\ket{\psi_{\text{cav}}}$. More specifically, if the
cavities are initially in their vacuum state, the state would be 
$(a^\dag_{A}+a^\dag_\mathcal{B})^n\ket{0}_\mathcal{A}\ket{0}_\mathcal{B}$. This is equivalent
to the state produced when a $n$-photon Fock state and a vacuum
state are incident together on a 50:50 beam splitter. However,
the numerical simulations we have undertaken have suggested that the fidelity of states
produced via this method would decrease swiftly with increasing
$n$. This is due to two reasons. Firstly, the short interaction
time approximation becomes worse when higher photon numbers are
present in the cavities since the time scale of the interactions
is faster (the Rabi frequency scales with $\sqrt{n+1}$). Secondly,
when more than one photon is in the cavities, a failed attempt does not reset the state of the cavities to the state
before the run, so, as more repetitions are made, the fidelity of
the final state gets worse and worse. For these reasons, this
 does not appear to be  a good scheme for
the generation of such states.

However, as we will show below, the generation of single-photon
Bell states is not affected by these problems. Firstly, we find
that the fidelity of the generated state remains close to unity
for values of $g_0\tau$ much greater than the above approximation
would be valid. Secondly,  ``failures'' reset the state
of the cavities to the vacuum state exactly, so the fidelity of
the state generated is unaffected by the number of runs which were
required.
In the limit that $g_0\tau$ tends to zero, the probability of the
detector measuring a ground state is approximately $2(g_0\tau)^2$,
which, in this limit, is vanishingly small. In a practical scheme,
one would need to work in a parameter range where the probability
of success was high enough that few repetitions would be required
to achieve a successful run. However, this is the case for higher values of
$g_0\tau$, where the above short-time  approximation would no longer be valid.
Fortunately, starting with the simple pure state
$\ket{B}\ket{0}_\mathcal{A}\ket{0}_\mathcal{B}$, it is straightforward to solve the
Schr\"odinger equation and calculate exactly the state of the
system after the interactions have taken place. This state
$\ket{\psi}$, under the assumption that the interaction times
$\tau$ are exactly equal is,
\begin{equation}
\begin{split}
    \ket{\psi}=&\cos^2(g_0\tau)\ket{B}\ket{0}_\mathcal{A}\ket{0}_\mathcal{B} -
    \im\cos(g_0\tau)\sin(g_0 \tau)\ket{A}\ket{0}_\mathcal{A}\ket{1}_\mathcal{B}\\
&\qquad -\im\sin(g_0\tau)\ket{A}\ket{1}_\mathcal{A}\ket{0}_\mathcal{B}.
\end{split}
\end{equation}

\begin{figure}\begin{center}
\includegraphics[scale=1.4]{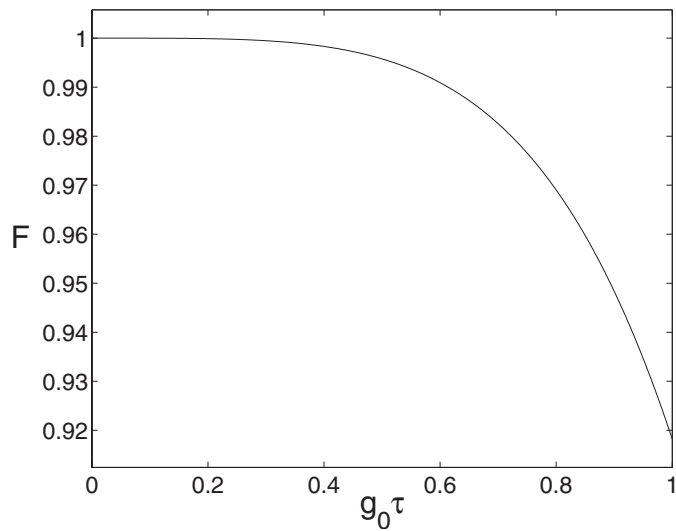}
\end{center}
\caption{\label{fig:idealfid} The fidelity of the Bell state
generated in the cavity modes after a successful run, plotted
against $g_0\tau$ with values from 0 to 1. Note that the values of
$g_0\tau$ corresponding to a full Rabi oscillation in each cavity
is $g_0\tau=\pi$.}
\end{figure}

Let us consider a measurement of the atom's state. If
the excited state is detected the
state of the cavities is projected back to the vacuum state,
independent of $\tau$, as in the approximate case. 
If the ground state is detected, the following entangled state is
generated in the cavity,
\begin{equation}
\ket{\psi_{\text{cav}}}=\cos(g_0 \tau)\ket{0}_\mathcal{A}\ket{1}_\mathcal{B}+\ket{1}_\mathcal{A}\ket{0}_\mathcal{B},
\end{equation}
 where normalisation has been omitted. As $g_0\tau$
approaches zero, this tends to the desired state $\ket{\Psi^+}$.
One can quantify how close this state generated is to
$\ket{\Psi^+}$ in terms of the fidelity
$F=\abs{\braket{\Psi^+}{\psi_{\text{cav}}}}^2$,
\begin{figure}\begin{center}
\includegraphics[scale=1.4]{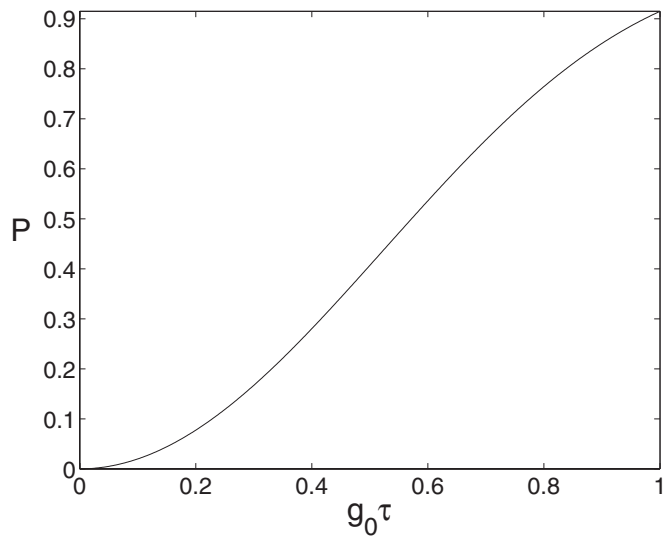}
\end{center}
\caption{\label{fig:idealprob} The probability that a single run will lead to the detection of the atom in its ground state, and thus the successful generation of a Bell state, plotted for values of $g_0\tau$ from 0 to 1.}
\end{figure}
\begin{equation}\label{eq:idealfid}
F=\frac{1}{2}+\frac{\cos(g_0\tau)}
{\cos^2(g_0\tau)+1}=1-\frac{(g_0\tau)^4}{16}+
O\bigl((g_0\tau)^6\bigr)\ .
\end{equation}
Note that there is no term proportional to $(g_0\tau)^2$ since the series expansions of $\cos(g_0\tau)$ and $(\cos^2(g_0\tau)+1)$ are proportional up to this power of $(g_0\tau)$.
The expression is plotted in figure \ref{fig:idealfid}. We see that the fidelity remains very close to unity for a surprisingly large range of $g_0\tau$. For example, the fidelity remains above $1-4\times10^{-3}$ for $g_0\tau=0.5$. Additionally, the function is extremely flat in the wide range of values between $g_0\tau=0$ and $g_0\tau=0.5$. Therefore the scheme would be insensitive to variations in the interaction times within this range.

For the scheme to be useful, the probability of a single run leading to the successful generation of an entangled state, $P_{\text{success}}$, needs to be high enough that prohibitively many repetitions are not required.
\begin{equation}
P_{\text{success}}=1-\cos^4(g_0\tau)=2(g_0\tau)^2+O\bigl((g_0\tau)^4\bigr)
\end{equation}
The success probability is plotted in figure \ref{fig:idealprob}. For $g_0\tau=0.5$, this probability is approximately 0.4, so a successful run would probably be achieved in 2 or 3 repetitions.
The optimal parameter range for the scheme depends upon the fidelity of the state required. The higher the value of $g_0\tau$ chosen the higher the success probability will be but the lower the fidelity of the state generated. If, for example, fidelities of $0.95$ were acceptable, and a minimum success probability of 0.5 were desired, the scheme could operate between approximately $g_0\tau=0.6$ and $g_0\tau=0.9$. The exact value of $g_0\tau$, however, can lie anywhere in this range, so no fine tuning of the interaction time is required.
Note that even if extremely high fidelities such as 0.999 are required, the necessary parameters (up to $g_0\tau=0.35$), still allow a success probability of up to 0.22,  meaning that a successful run would probably be reached after 4 or 5 repetitions.

\section{Practical considerations}\label{practical}


The above description is quite general and  no
particular type of atom or cavity has been specified.
A number of aspects must be considered in the choice of a physical
system to implement the scheme.
Firstly, it would be desirable that the entangled states, once
generated, would be as long-lived as possible. This favours
microwave cavities over optical cavities, since the lifetime of a
photon in an optical cavity is currently at the very most one
microsecond \cite{WolfgangLange}, whereas microwave cavities with
a photon lifetime of a millisecond have been made
\cite{harochermp}.
It takes an atom travelling at 500ms$^{-1}$ around 20$\mu$s to
traverse a typical microwave cavity, which is much smaller than
this lifetime, so there would be sufficient time for further atoms
to probe and interact with the cavity mode before the entangled
state has dissipated.
If a microwave cavity is used, an atom with a microwave transition
is then needed. Microwave transitions occur in atomic fine and
hyperfine structure, however dipole transitions between these
states are forbidden, and their interaction with the cavity mode
would be much too weak to implement this scheme -- the interaction
times that would be required would be much greater than the photon
lifetime of the cavity.
Rydberg atoms, on the other hand, although more difficult to
prepare, have large dipole moments and thus would interact
strongly with cavity modes.

Experiments have been carried out, for example by the Paris group
of Haroche and coworkers \cite{harochermp}, which have parameters
close to that required in this scheme. Recall that  our scheme
requires that the product of parameters $g_0$ and $\tau$ is at the
minimum 0.2 and maximally 0.5 to 0.8, depending on the fidelity of
Bell state one wants to generate.
In the Paris experiments, a Rydberg atom interacts with a
microwave cavity Rydberg atoms interact resonantly with microwave
cavity modes with a Rabi frequency of approximately 47kHz. This
means that  $g_0=47000\pi s^{-1}=1.48\times10^{5}
s^{-1}$.
Atoms in an atomic beam from an oven source travel at speeds of
the order of hundreds of meters per second. In the Paris
experiment atoms with a speed of 500ms$^{-1}$ are selected. This
means that the effective interaction time is such that a single
Rabi oscillation is performed, i.e. $g_0\tau=\pi$.
This is a factor of 4 to 8 lower than the parameter range for our
scheme. It would be difficult to lower $\tau$ by using faster
atoms, since the velocity of the atoms scales with the square root
of the atom oven temperature, so a lower $g_0$ would be required.
This could be obtained be using a larger cavity, and since $g_0$
scales with $1/\sqrt{V}=1/L^{\frac{3}{2}}$, so a cavity of mirror
separation 3 or 4 times as great as in the Paris experiments would
lead to a effective interaction times in the required range. Thus
Rydberg atoms and microwave cavities could be employed to
implement the scheme.

The disadvantage in using Rydberg atoms is that, at present, the efficiency of state detection schemes is much lower than unity. In  \cite{harochermp}, for example, they report a detection efficiency of 40\%. We will discuss the implications this has for the scheme in subsection \ref{detection}.
First, however, we consider the effect on the fidelity of the
entangled states produced in the scheme if the path of the atom
through the cavities is not well controlled and deviates from the
line through the centre of the cavities.

\subsection{The atomic path}\label{collim}

If the effective interaction times with both cavities are not exactly the same, this can reduce the fidelity of the entangled state produced by a successful run of the scheme.
 In our discussion above, we assumed that the two interaction times were exactly equal. In practice, however, it could be difficult to control the path of the atom so precisely.
Let us consider first the effect that differing interaction times
would have in the fidelity of the Bell state generated by a
successful run of the scheme. Let the effective interaction
between the atom and cavity $\mathcal{A}$ and the atom and cavity $\mathcal{B}$ be $\tau$
and $\tau(1-\epsilon)$, respectively. We can rewrite equation
\eqref{eq:idealfid}, to take the differing interaction times into
account, and find the following expression for the fidelity,
\begin{equation}
F=\frac{1}{2}+\frac{\cos(g_0\tau)\sin(g_0\tau)\sin\bigl(g_0\tau(1-\epsilon)\bigr)}
{\cos^2(g_0\tau)\sin^2\bigl(g_0\tau(1-\epsilon)\bigr)+\sin^2(g_0\tau)}\ .
\end{equation}

\begin{figure}\begin{center}
\includegraphics[scale=1.4]{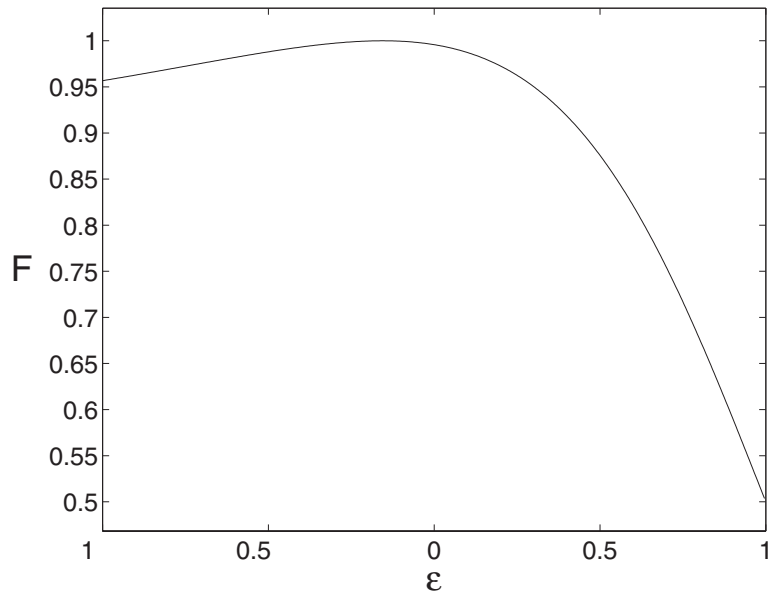}
\end{center}
\caption{\label{fig:epsilon} The fidelity of the Bell state generated in the cavity, if the effective interaction times are $\tau$ and $\tau(1-\epsilon)$ for the interactions with cavities A and B respectively, is plotted here as a function of $\epsilon$ for $g_0\tau=0.5$. }
\end{figure}

\begin{figure}\begin{center}
\includegraphics[scale=1.4]{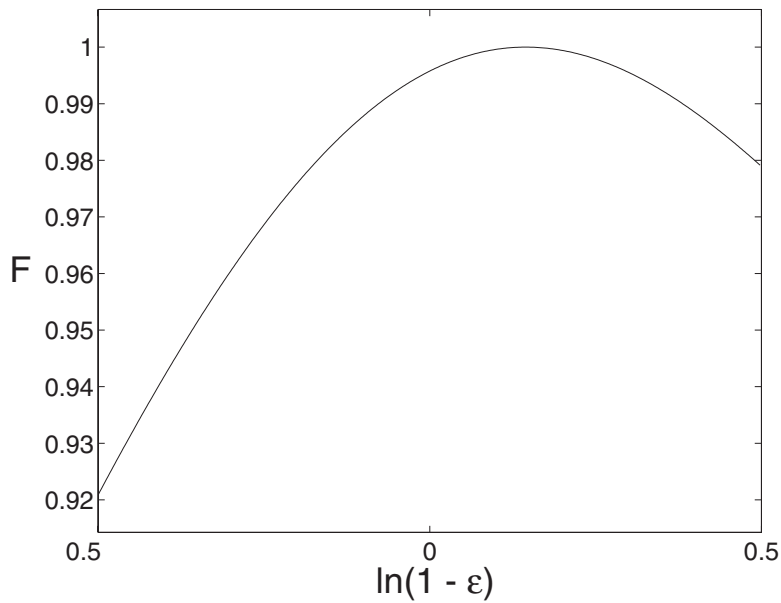}
\end{center}
\caption{\label{fig:logepsilon} The fidelity of the Bell state generated in the cavity plotted again as a function of $\epsilon$ for $g_0\tau=0.5$, this time plotted against $\ln(1-\epsilon)$. }
\end{figure}

The fidelity is plotted for $g_0\tau=0.5$ as a function of $\epsilon$ in figure \ref{fig:epsilon}.
The asymmetry of the plot is partly an artifact of the choice of parameterisation, but, if we take this into account, by plotting $F$ against $\ln(1-\epsilon)$, as in figure \ref{fig:logepsilon}, we see that the asymmetry remains. This is due to the asymmetry in the scheme itself regarding the interactions with the two cavities. When the atom enters the first cavity it is always in the product state $\ket{B}$, whereas, when entering the second it is always entangled with the first cavity. This makes the interaction with cavity $\mathcal{B}$ slightly weaker, and is the reason that the maximal value of $F$ occurs when $\epsilon$ has a small negative value. The slightly longer interaction time compensates for the interaction being slightly weaker.

Since it is reasonably easy to position the cavities in the
desired place to very high precision, let us assume that they are
perfectly aligned in the layout illustrated in figure
\ref{fig:layout}.
\begin{figure}\begin{center}
\includegraphics[scale=0.6]{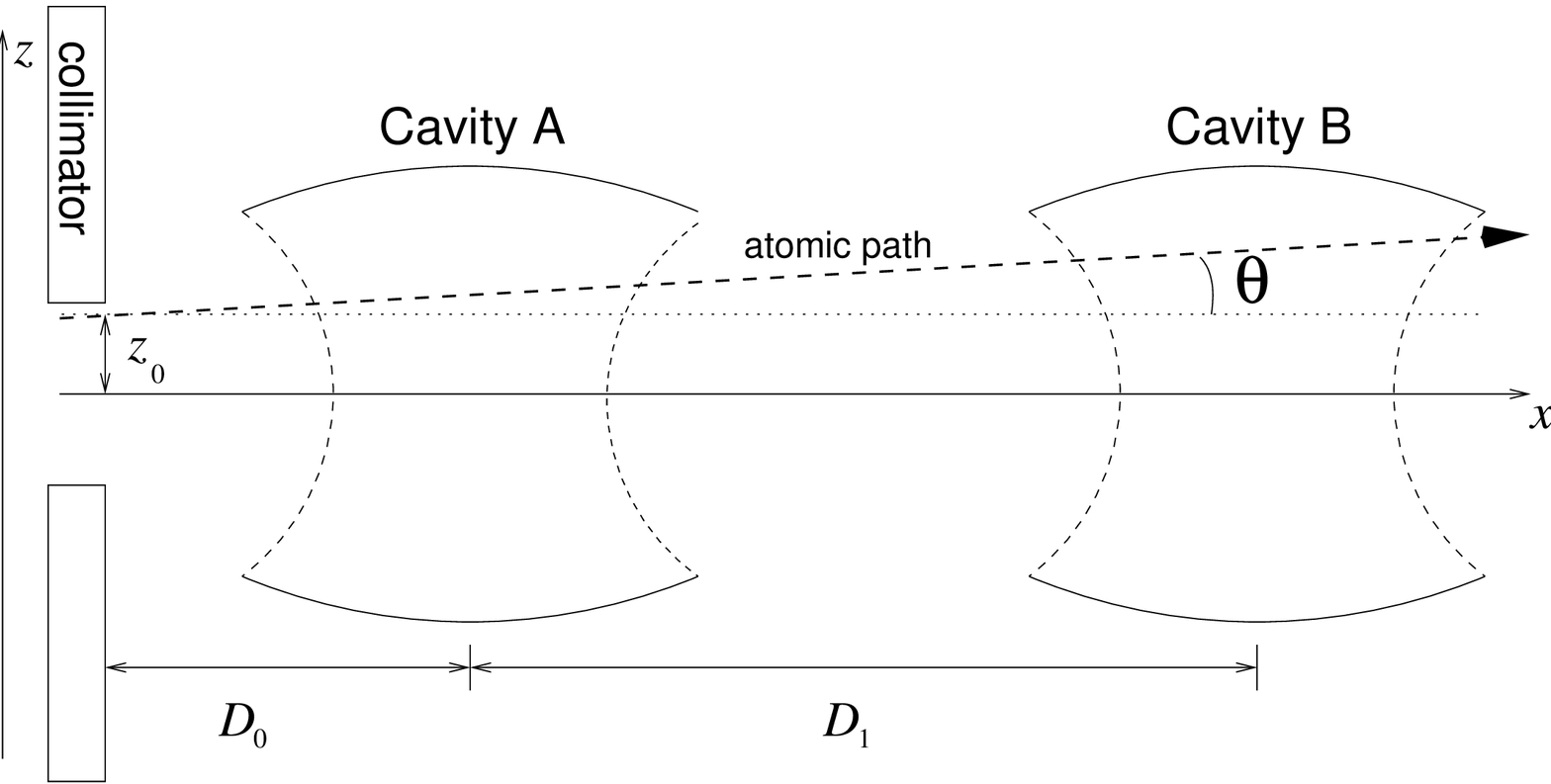}
\end{center}
\caption{\label{fig:zdiag} A general straight line path through the cavities can be defined in terms of the parameters $y_0$, $z_0$, $\phi$ and $\theta$, where  distances are measured in meters and  angles in radians. This figure illustrates $z_0$ and $\theta$, the other two parameters, $y_0$ and $\phi$, are equivalently defined in the perpendicular $x-y$-plane.
The distance between the exit of the collimator and the centre of cavity $\mathcal{A}$ is $D_0$. The distance between the centres of the cavities is $D_1$.
}
\end{figure}
The factor which would be harder to control would be the atom's path through the apparatus, since it would be travelling on a ballistic path after being ejected from a heat oven, and although such atomic beams can be highly collimated, there will generally be a small amount of spread in the transverse direction, meaning that the atom's path will diverge slightly from the central path.

To calculate how much this would affect the interaction times we consider the cavity geometry. We assume that the cavities are in a Fabry-Perot configuration as discussed in section~\ref{sec:cavmodes}. Since the atoms will be assumed to remain close to the $z$-axis of each cavity, we may use the approximate mode function
\begin{equation}
u(x,y,z)=e^{-\frac{x^2+y^2}{w_0^2}}\cos{\frac{2\pi z}{\lambda}},
\end{equation}
where $w_0$ is the mode waist. The z-axis lies along the line connecting the centres of the two mirrors and the origin is in the centre of the cavity. The waist of a Gaussian mode, $w_0$, is a function of the cavity geometry and the field wavelength,

\begin{equation}
w_0^2=\frac{\lambda\sqrt{L(2R-L)}}{2 \pi},
\end{equation}

where, as above,  $L$ is the separation between and $R$ the radius of curvature of the mirrors.
If the atom travels along the central axis of the cavity with constant speed $v$, the effective interaction time is $\sqrt{\pi}w_0/v$.
Let us consider a general straight path through the system, which can be defined in terms of four parameters, $y_0$ and $z_0$, the initial displacements from the central line in the $y$ and $z$ directions, and $\phi$ and $\theta$ the angles between the atomic path and the central line in the  $y$ and $z$ directions, illustrated for $z_0$ and $\theta$ in figure \ref{fig:zdiag}.
We can calculate the effective interaction times between the atom and cavity as it travels along this path with speed $v$ and find, that the interaction time with each cavity, in terms of displacements $\delta_y$ and $\delta_z$ which differ for each cavity, is

\begin{equation}\label{effectint}
\begin{split}
\tau_{{\rm
eff}}=&\frac{\sqrt{\pi}\omega_0}{v\cos{\theta}}\exp\left[-\frac{\delta_y^2}{\omega_0^2}\left(1-\frac{\sin^2\theta}{\cos^2\phi}\right)\right]\exp\left[-\frac{k^2\omega_0^2\tan^2\theta}{4}\right]\\
&\cos\left[k\delta_z-k\delta_y\left(\frac{\sin\theta\sin\phi}{\cos^2\theta}\right)\right]
.
\end{split}
\end{equation}
For the interaction with cavity $\mathcal{A}$, $\delta_y=y_0+\cos(\phi)D_0$
and $\delta_z=z_0+\cos(\theta)D_0$, and for the interaction with
cavity $\mathcal{B}$ $\delta_y=y_0+\cos(\phi)(D_0+D_1)$ and
$\delta_z=z_0+\cos(\theta)(D_0+D_1)$.
We can use these expressions to calculate $\epsilon$ in terms of
these parameters, written here to the  second order in $y_0$,
$z_0$, $\phi$ and $\theta$ and their products,

\begin{equation}\label{eq:epsilon}
\begin{split}
\epsilon\approx&\frac{1}{w_0^2}\left[(D_1\phi)^2+(D_1\phi)(2D_0\phi)+(2y_0)(D_1\phi)\right]\\&\mbox{}+\frac{2\pi^2}{\lambda^2}\left[(D_1\theta)^2+(D_1\theta)(2 D_0\theta)+(2z_0)(D_1\theta)\right].
\end{split}
\end{equation}
Let us discuss the constraints this would have on the
collimation of a typical experiment. In the cavity QED experiments
of Haroche and coworkers \cite{harocheprl761800,harochermp},
Rydberg atoms interact resonantly with microwave cavities. In a
typical experiment, a cavity mode with waist $w_0=5.97$mm and
wavelength $\lambda=5.87$mm is employed. For $\epsilon$ to be
small, the quantities in parentheses in equation
\eqref{eq:epsilon} must be much smaller than $w_0$ and
$\lambda/\sqrt{2}\pi=1.32$mm. This means that the atomic beam must
be collimated so that the effective beam radius is much smaller
than this distance. In \cite{harocheprl761800} an effective  beam
radius of $0.25$mm is reported. If we assume from this that, in
the worst case, this would mean that $y_0,z_0\approx$0.25mm and
$D_1\phi,D_1\theta\approx$0.25mm, we can estimate that $\epsilon$
would be less than $0.2$. If $g_0\tau=0.8$ and $\epsilon=0.2$,
this would corresponds to a reduction of the fidelity of the
entangled state produced by a successful run from 0.96 if both
interaction times are exactly equal to 0.93.
%
Equation \eqref{eq:epsilon} also provides another reason why
optical cavities would be unsuitable for the scheme. The typical
waist of an optical cavity  tends to be much smaller
than that of a microwave cavity, so the demands on
the atomic beam collimation required if optical cavities were used
would be extremely high.

\subsection{Detector efficiency}\label{detection}
Our analysis in the previous section assumes that the atomic state
detector has perfect efficiency. In practice, the detection
efficiency will be less than unity. Indeed, as mentioned above,
current state detection methods for Rydberg atoms have an
efficiency of just 40\% \cite{harochermp}. The detection process
ionises the atom, destroying the state, so increased efficiency
cannot be obtained by placing detectors in series.
A single detection failure will disrupt the scheme, since it will
cause a mixed state to be created in the cavities. The scheme must
then be halted, and one would then have to wait until this state
would have dissipated from the cavities, and the cavities have
returned to the vacuum state. Otherwise, if further atoms are sent
through the cavities immediately, they will interact with the
mixed state, and any ``successful'' run, will have generated a
mixed state with much reduced fidelity. Rather than halting the
flow of atoms through the cavities, their interaction with the
cavities could be prevented for the cavity dissipation time by the
application of an electric field to the system, to create a Stark
shift in the atoms such that they are no longer resonant with the
cavities.
Therefore, one would like a detection efficiency  high enough that the probability of a detection failure, during the typical number of runs needed before a successful ground state measurement is made, is low.
The mean number of runs to generate the entangled state in the cavity, if the detector were ideal, is simply the inverse of the success probability. The probability that the detector works every time during the process is therefore simply $D^{1/P_{\text{success}}}$, which for all values of $P_{\text{success}}$ is less than or equal to $D$. Therefore, for a reliable scheme, a high detection efficiency would be desirable.

For example, let us consider an implementation of this scheme with Rydberg atoms travelling through the cavities such that $g_0\tau=0.5$. The success probability  $P_{\text{success}}$ is 40.7\% and with current detectors with efficiency 40\%, $P_{\text{det}}$ would be around 10\%. This would  mean, typically one would have to repeat the whole process, including preparation of the cavities, 10 times before it could reach its successful conclusion. However, in light of the fact that the cavity dissipation time is of the order of milliseconds, and the time taken for each run much less than this, even in this case, the time needed to repeat the scheme enough times to generate the Bell state would be a fraction of a second.



\section{Conclusion}

We have described  a scheme for the generation of high-fidelity Bell
states between two spatially separated cavity modes. The scheme is
non-deterministic, but we have shown that within the range of
parameters $g_0\tau$ between approximately 0.3 and 0.9, fidelities
higher than 0.95 are obtained, with success probabilities for a
single run greater than 1/5 for this entire range.

The most appropriate physical system to implement this scheme
would be a combination of microwave cavities and Rydberg atoms.
The low detection efficiency for Rydberg states would increase the
number of times which the scheme would need to be repeated before
a successful run, and require extra time after each detection
failure to allow the mixed state produced in the cavities to
dissipate. Nevertheless, the scheme would still be successful
within a reasonable number of repetitions. The scheme requires
that the atomic beam used is highly collimated otherwise the
fidelity of the states produced may be degraded, but the
collimation which has already been achieved in current experiments
is high enough that this effect would be small.

There are some disadvantages in generating entangled states in cavity modes. Firstly,  they have a  short life time which means that either the state would have to be utilised within a short time-window, or it must be mapped onto a fresh atom. This mapping can be achieved, however  in the same experimental framework, except that the atom must be introduced in the lower state $\ket{A}$ and remain in the cavity for an effective Rabi oscillation. An alternative approach would be to work with atoms or ions trapped within the cavities via external fields. In the next chapter we consider such a situation, and demonstrate  again the advantages that an approach based on weak interactions combined with measurements can have.


%% file: c-5.tex
\chapter{Robust Creation of Entanglement between Ions in Spatially Separate Cavities}
\label{c-5}

\section{Introduction}

In the previous chapter, we introduced a scheme for entanglement generation, where high fidelities are obtained, because we work in a regime where the interactions involved are very brief. However, an unavoidable side-effect of the use of this technique is that the probability of success for a single run is low. In this chapter, we use similar ideas to greatly enhance a previous proposal for the generation of a maximally entangled two-qubit state between ions trapped two spatially separated cavities. Various schemes for trapping ions and atoms have been proposed and experimentally implemented \cite{paul,liebfried} including trapping inside cavities\cite{kimbletrap,walthertrap}. Our proposal uses once again the principle of using a slight interaction plus a  measurement to achieve experimental robustness. This time, however, by applying a driving field continuously, but weakly, this  robust scheme also exhibits arbitrarily high success probabilities.

The structure of this chapter is as follows. First we will briefly summarise the proposal of Sougato Bose and co-authors \cite{bose99tele} and discuss its advantages and drawbacks. Then we will introduce and discuss our improved version of the scheme \cite{PRLbrowneplenio} and demonstrate it's robustness against experimental imperfections.

\section{System Hamiltonian}

Before describing this proposal in detail we first derive a Hamiltonian which describes the evolution of the the ion-cavity system, which is also applicable to our modified scheme.

\begin{figure}
\psfrag{gg}{$g$}
\psfrag{AA}{$\ket{A}$}
\psfrag{BB}{$\ket{B}$}
\psfrag{CC}{$\ket{C}$}
\psfrag{ga}{$\gamma_A$}
\psfrag{gb}{$\gamma_B$}
\psfrag{om}{$\Omega$}
\psfrag{DD}{$\Delta$}
\psfrag{wl}{$\omega_l$}
\psfrag{wc}{$\omega_c$}
\begin{center}
\includegraphics[scale=1.2]{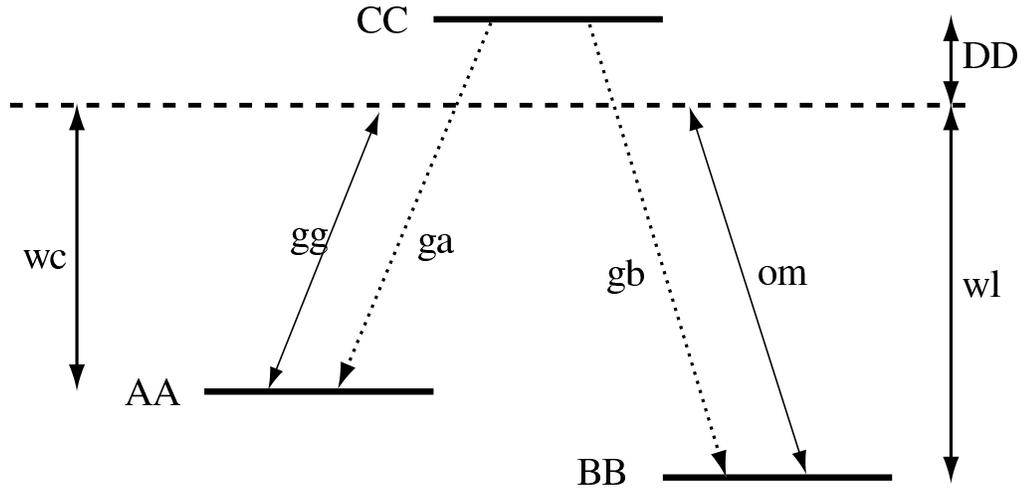}
\end{center}
\caption{\label{levelstructure} Internal level scheme of the ions. Levels $\ket{A}$ and $\ket{C}$ are coupled via a non-resonantly interaction with a cavity mode. Levels $\ket{B}$ and $\ket{C}$ are coupled by an external ``classical'' driving field. Spontaneous decay is indicated by arrows from $\ket{C}$ to $\ket{A}$ and $\ket{B}$.}
\end{figure}

We consider atoms (or ions) with  three energy levels $\ket{A}$, $\ket{B}$ and $\ket{C}$, sitting in a cavity such that a single cavity mode of frequency $\omega_c$ couples with the transition between levels $\ket{A}$ and $\ket{C}$ with detuning $\Delta$. The transition between  levels $\ket{B}$ and $\ket{C}$ is driven by an external field with frequency $\omega_l$ which we can treat classically.
Neglecting decay processes, we can write down a full Hamiltonian from the system, using the results from chapter~\ref{ch:3}

\begin{equation}\label{hnoadiabat}\begin{split}
\hat{H}=&\hbar\Delta \ket{C}\bra{C}-\hbar\omega_c\ket{A}\bra{A}-\hbar\omega_l\ket{B}\bra{B}+\hbar\omega_c \hat{a}^\dag \hat{a}
\\&+\hbar \bigl( g \ket{C}\bra{A}\hat{a}+h.c.\bigr)+ \hbar\frac{\Omega}{2}\bigl( e^{-i\omega_l t}\ket{C}\bra{B}+h.c.\bigr )\ ,
\end{split}
\end{equation}

where $g$ is the atom-cavity mode coupling constant, and $\Omega$ is the classical Rabi frequency of the interaction with the external laser field. 
Let us simplify this by transforming to an interaction picture with $\hat{H}_\textrm{int}=\hbar\Delta \ket{C}\bra{C}+\hbar \bigl( g \ket{C}\bra{A}\hat{a}+h.c.\bigr)+ \hbar(\Omega/2)\bigl( e^{-i\omega_l t}\ket{C}\bra{B}+h.c.\bigr )$
\begin{equation}\begin{split}
\hat{H}_{I.P.}&=e^{+i(\hat{H}-\hat{H}_\textrm{int})/\hbar}(\hat{H}-\hat{H}_0)e^{-i(\hat{H}-\hat{H}_\textrm{int})/\hbar}\\
&=\hbar\Delta \ket{C}\bra{C}+\hbar \left( g \ket{C}\bra{A}\hat{a}+h.c.\right)+ \hbar\frac{\Omega}{2}\left( \ket{C}\bra{B}+h.c.\right)\ .
\end{split}
\end{equation}

The schemes below operate in the high detuning regime, where $\Delta$ is much greater than the Rabi frequencies of each interaction. We may therefore simplify the Hamiltonian  further by \emph{adiabatic elimination} of  level $\ket{C}$.
To perform this, first we diagonalise the  Hamiltonian, as we did in chapter~\ref{ch:3} for the two-level system. We write Hamiltonian is rewritten as a sum of operators $\hat{H}_n$  each acting on the sub-space spanned by $\ket{A,n+1}$, $\ket{B,n}$ and $\ket{C,n}$, and write  $\hat{H}_n$ in matrix form as
\begin{equation}
\hat{H}_n=\hbar\left( \begin{array}{ccc}
0&0&g^* \sqrt{n+1}\\
0&0&\Omega/2\\
g  \sqrt{n+1} &\Omega/2&\Delta
\end{array}
\right)\ .
\end{equation}
In the limit that $\Delta\gg g,\Omega$, the eigenvalues of  this matrix become approximately the following values, $E_1=0$, $E_2= \hbar(4|g|^2(n+1)+\Omega^2)/(4\Delta)$ and $E_3\approx\Delta$, with corresponding eigenvectors
\begin{equation}
\ket{\psi_1}=\sqrt{\frac{1}{4|g|^2(n+1)+\Omega^2}}\left(-\Omega\ket{A,n+1}+2 g \sqrt{n+1}\ket{B,n}\right)\ , 
\end{equation}
\begin{equation}
\ket{\psi_2}\approx \sqrt{\frac{1}{4|g|^2(n+1)+\Omega^2}} \left( 2 g^*\sqrt{n+1}\ket{A,n+1}+\Omega\ket{B,n}\right)\ ,
\end{equation}
\begin{equation}
\ket{\psi_3}\approx\ket{C,n}\ .
\end{equation}
Thus, in this limit, $\ket{C,n}$ is an approximate eigenstate of the full Hamiltonian, and does not couple to the \{$\ket{A,n+1}$, $\ket{B,n}$\} subspace. Since the initial state of the system in our scheme will be the in latter sub-space, we therefore simplify the Hamiltonian by restricting it to this sub-space, \emph{adiabatically eliminating} $\ket{C,n}$. Casting the restricted diagonalised Hamiltonian back in terms of the product state basis vectors, we obtain
\begin{equation}\label{Had}
\begin{split}
\hat{H}_{ad}&=\hbar\sum_n \frac{4|g|^2(n+1)+\Omega^2}{4\Delta}\ket{{\psi_2}}\bra{{\psi_2}}\\
&=\hbar\frac{|g|^2}{\Delta} \ket{A}\bra{A} \hat{a}^\dag \hat a+\hbar\frac{\Omega^2}{4\Delta}\ket{B}\bra{B}+\left( \hbar\frac{\Omega g}{2\Delta} \ket{B}\bra{A}\hat{a} +h.c. \right)\ .
\end{split}
\end{equation}

Note that if system is prepared in the \{$\ket{A,1}$,$\ket{B,0}$\} sub-space and the classical field strength is chosen such that $\Omega=2g$ then this Hamiltonian is exactly like a resonant two-level Jaynes-Cummings Hamiltonian with Rabi Frequency $2 g^2/\Delta$.

\section{An earlier proposal}

\begin{figure}
\begin{center}
\includegraphics[width=14cm]{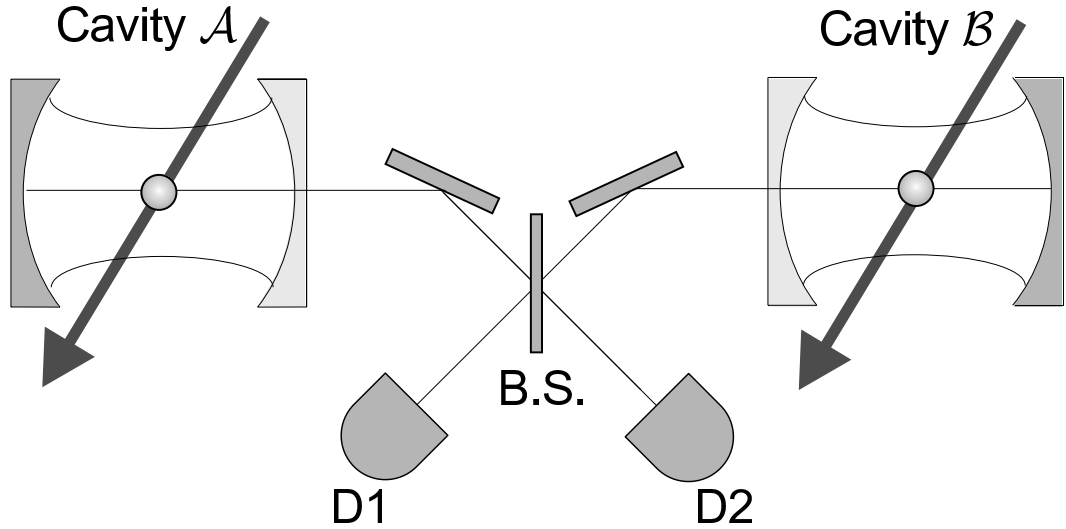}
\end{center}
\caption{\label{figurecavity} We consider a set-up in which
individual ions are trapped inside two spatially separated optical
cavities $\mathcal{A}$ and $\mathcal{B}$ where they are also driven by an external classical field. Photons can leak out of the cavities and are
then mixed on a beam splitter B.S. and subsequently detected by
photo-detectors $D_1$ and $D_2$.}
\end{figure}

Here, we introduce a proposal \cite{bose99tele} for the generation of a maximally entangled state between two ions trapped in spatially separated cavities, on which our improved scheme is based. Although this proposal was primarily presented  as a teleportation scheme by Bose and co-workers, we focus here on the author's modified version in which a maximally entangled state is prepared between two ions (or atoms) trapped in two spatially separated cavities. Bose and co-authors considered the setup illustrated in figure~\ref{figurecavity}. Two cavities each contain a trapped ion (or atom) driven by an external field, as described in figure~\ref{levelstructure} and in the previous section. One mirror in each cavity is semi-silvered such that photons leak out at rate $2\kappa$.

Each atom is prepared in state $\ket{B}$ and the cavities are cooled to their vacuum state, before the classical driving fields are applied. Driving fields with strength $\Omega=2g$ are applied for exactly one quarter a Rabi cycle, that is for duration $\pi/4 g^2$, and then the driving field is switched off. This drives each ion-cavity sub-system through a quarter of a Rabi oscillation into the state,
\begin{equation}
\ket{\psi}=\sqrt{\frac{1}{2}}\left(\ket{B}\ket{0}-i\ket{A}\ket{1}\right)
\end{equation}
where $\ket{0}$ and $\ket{1}$ are the zero and one photon states of the cavity field. Adding subscripts $\mathcal{A}$ and  $\mathcal{B}$ to denote the cavity, the combined state of both ion-cavity systems is
\begin{equation}
\begin{split}
\ket{\psi}=\frac{1}{2}\biggl(&\ket{B}_\mathcal{A}\ket{B}_\mathcal{B}\ket{0}_\mathcal{A}\ket{0}_\mathcal{B}-\ket{A}_\mathcal{A}\ket{A}_\mathcal{B}\ket{1}_\mathcal{A}\ket{1}_\mathcal{B}\\
&\mbox{}-i\bigl[\ket{B}_\mathcal{A}\ket{A}_\mathcal{B}\ket{0}_\mathcal{A}\ket{1}_\mathcal{B}+\ket{A}_\mathcal{A}\ket{B}_\mathcal{B}\ket{1}_\mathcal{A}\ket{0}_\mathcal{B}\bigr]
\biggr)\ .
\end{split}
\end{equation}
 For simplicity here we assume that the leakage rate is much slower than the Rabi frequency of the state preparation step above, and neglect it (a fuller treatment may be found in \cite{bose99tele}).

One can model this mathematically using  a Master equation in the quantum jump approach, as described in section~\ref{quantjump}, however, we can explain Bose et al's scheme qualitatively without resorting to this level of analysis. Let us simply assume that the photons will leak out of the cavity at some time, where they will mix on a beam splitter and be detected by detectors D1 and D2. Since the cavities leak at the same rate, when a single photon is detected, one cannot distinguish from which cavity it originated, and this lies at the heart of the scheme's ability to create entanglement.

The action of the 50:50 beam splitter on the four possible input basis states $\ket{0}\ket{0}$, $\ket{0}\ket{1}$, $\ket{1}\ket{0}$ and $\ket{1}\ket{1}$ is the following transformation
\begin{align}
\ket{0}\ket{0}\rightarrow&\ket{0}\ket{0}\\
\ket{0}\ket{1}\rightarrow&\sqrt{\frac{1}{2}}(\ket{0}\ket{1}+i\ket{1}\ket{0})\\
\ket{1}\ket{0}\rightarrow&\sqrt{\frac{1}{2}}(i\ket{0}\ket{1}+\ket{1}\ket{0})\\
\ket{1}\ket{1}\rightarrow&i\sqrt{\frac{1}{2}}(\ket{0}\ket{2}+\ket{2}\ket{0})\ . 
\end{align}

One waits for a length of time such that it is very likely that all photons in the cavities have leaked out and, if present, been detected. One can interpret the detection outcome as a measurement on the initial state of the cavity. If one and only one photon is detected, then this is equivalent to having performed a joint projection on the states of the two cavities onto  $\ket{0}_\mathcal{A}\ket{1}_\mathcal{B}+i\ket{1}_\mathcal{A}\ket{0}_\mathcal{B}$ or  $i\ket{0}_\mathcal{A}\ket{1}_\mathcal{B}+\ket{1}_\mathcal{A}\ket{0}_\mathcal{B}$ depending on which detector clicked. This projection leaves the two ions in a maximally entangled state $\ket{A}_\mathcal{A}\ket{B}_\mathcal{A}-i\ket{B}_\mathcal{A}\ket{A}_\mathcal{B}$ or  $-i\ket{A}_\mathcal{A}\ket{B}_\mathcal{A}+\ket{B}_\mathcal{A}\ket{A}_\mathcal{B}$. Neglecting imperfections in the state preparation due to the cavity leakage, and assuming perfectly efficient detection, the success probability is 50\%. Half the time either zero or two photons will be detected, which leave the ions in a product state. The scheme must then be repeated. Note that a detector which can discriminate between different one or two photons is needed here, a great technological challenge.

In \cite{bose99tele}, the authors simulate the performance of the scheme with a full treatment of the cavity decay and consider inefficient detectors. They find, in their more thorough treatment, that the success probability is lowered by the small chance of photon leakage during the preparation stage. They also find that low detection efficiencies are particularly damaging, as they mean that states which should have led to two photons being detected generating product states are wrongly identified as successes, which badly degrades the fidelity and the degree of entanglement of the states generated.

Using the same basic setup as this scheme, we now  use the principle of employing weak or brief interactions and measurement to give a new proposal\cite{PRLbrowneplenio} which,
 (i) succeeds
with $100\%$ probability under ideal conditions, (ii) allows the
achievement of high fidelity entanglement outside the strong
coupling regime upon the detection of a photon, (iii) is robust
against detector inefficiencies and absorption losses in the
cavity mirrors and (iv) can be adapted, with the same efficiency,
to entangle ions trapped in a single optical cavity.

\section{Weakly driven approach}
We begin the discussion of our new scheme by first analysing in more detail the evolution of each of the driven ion-cavity sub-systems including  cavity leakage. Our initial analytic treatment will be valid for the so-called ``strong-coupling regime'' where spontaneous emission from the ions is neglected. After this we will consider a more experimentally realistic ``weak-coupling regime'' with significant spontaneous emission and also include finite detection efficiency in our model.

First let us concentrate on a single ion-cavity sub-system. We assume here that light leaks through one of the cavity mirrors, with rate $2 \kappa$.
Adopting a quantum-jump approach, we write the Master equation for the ion-cavity system, 

\begin{equation}
    \dot\rho = -i(\hat{H}_{\text{eff}}\rho -
    \rho \hat{H}_{\text{eff}}^{\dagger}) + 2\kappa  \hat{a}\rho
    \hat{a}^{\dagger}
\end{equation}
where we have defined the  non-Hermitian effective  Hamiltonian
\begin{eqnarray}\label{nonherm}
\hat{H}_{\textrm{eff}} = \hat{H}_{ad} - i\kappa  \hat{a}^{\dagger}\hat{a}
\end{eqnarray}
 where $\hat{H}_{ad}$ is defined above in equation~(\ref{Had}).

In contrast to \cite{bose99tele}, our proposal operates in what we call the 
 ``weak driving limit'' such  that the condition
$\frac{g\Omega}{2\Delta} \ll \kappa$ is satisfied. Intuitively
this implies that the rate of transitions between levels $|A\rangle$
and $|B\rangle$ of the ions will be slower than the rate of cavity decay. Since our scheme is initiated with both ion-cavity sub-systems in the product state $\ket{B,0}$,
this 
implies that the population in level $|A\rangle$ of the atoms
will be small, unless a photon is detected. 
More precisely,  in the weak driving limit, under the condition that no photon is detected,  the state of the sub-system will
quickly approach the following state

\begin{equation}\label{steadyst}
    |e_1\rangle=\sqrt{\frac{1}{1+|x|^2}}(x\ket{A,1}+\ket{B,0})
\end{equation}
where $x=-i\frac{g\Omega}{2\Delta\kappa}$. 

To see this, we diagonalise the non-Hermitian Hamiltonian in equation~(\ref{nonherm}). If the sub-system begins with the ion in state  $\ket{B}$ and both cavity and external modes in the vacuum state, it remains in the sub-space $\{\ket{B,0},\ket{A,1}\}$. In the weak-driving limit we obtain the following eigenvalue, eigenvector pairs,

\begin{align}
E_1\approx& \frac{\Omega^2}{4\Delta}-i\kappa(|x|^2)\qquad\ket{e_1}=\sqrt{\frac{1}{1+|x|^2}}(x\ket{A,1}+\ket{B,0})\\
E_2\approx& \frac{|g|^2}{\Delta}-i\kappa\qquad\ket{e_2}=\sqrt{\frac{1}{1+|x|^2}}(\ket{A,1}-x\ket{B,0})\end{align}

where $x$ is defined above.

Now, since this Hamiltonian is non-Hermitian, the eigenvalues are not all real and the eigenvectors are not orthogonal. Nevertheless, since the eigenvectors are still linearly independent, one can still represent an arbitrary state $\ket{\psi}=\alpha \ket{A,1}+\beta\ket{B,0}$ as a unique superposition of these (normalised) eigenstates\footnote{Note, however, the weights associated with each eigenvector are no longer simple overlaps with $\ket{\psi}$ but have a more complicated form.}. Once decomposed in this form, the time evolution, under the condition that no photon leaks from the cavity, may be simply calculated as above by associating the factor $e^{-i E_i t /\hbar}$ with each eigenstate, where $E_i$ is the associated eigenvalue. Since $E_i$ may be complex, as well as the usual oscillatory part of this factor, the imaginary part of $E_i$ contributes to an exponential decay factor. 

Now, in the weak-driving limit of our non-Hermitian Hamiltonian, the imaginary part of $E_1$ is tiny, but  $E_2$ has the large imaginary part $-i\kappa$. Thus the state will converge towards eigenvalue $\ket{e_1}$ at rate $1/\kappa$, with a success probability dependent on the overlap between the starting state and  $\ket{e_1}$. Since, in the weak-coupling regime,  the overlap between $\ket{e_1}$ is close to unity, the norm of the conditional state remains very close to unity during this time scale, so the chance of this state preparation stage of the protocol failing via the leakage of a photon is negligible. 

Thus, after a short period, both ion-cavity sub-systems have attained the state $\ket{e_1}$. We now recall that the light escaping from the two cavities is mixed at a 50:50 beam-splitter, and then passes into the detectors $D1$ and $D2$. Now that we are considering both atom-cavity subsystems, let us give operators the suffices $\mathcal{A}$ or $\mathcal{B}$ according to which subsystem they belong. Due to the linear transformation imposed by the beam splitter, it is impossible for the detectors to distinguish from which cavity the detected photon originated and a detection at detector $D1$ will now correspond to the jump operator $\hat{J}_1=\sqrt{1/2}(i \hat{a}_\mathcal{A}+\hat{a}_\mathcal{B})$ and a detection at $D2$ to $\hat{J}_2=\sqrt{1/2}( \hat{a}_\mathcal{A}+i \hat{a}_\mathcal{B})$. 
We write the Master equation for the combined state of both atom-cavity systems as follows,

\begin{equation}
\dot\rho=-i\left[\hat{H}_{\mathcal{A},\mathcal{B}}\ \rho-\rho\hat{H}_{\mathcal{A},\mathcal{B}}^\dag\right]+2 \kappa \hat{a}_\mathcal{A}\rho \hat{a}_\mathcal{A}^\dag+2 \kappa \hat{a}_\mathcal{B}\rho \hat{a}_\mathcal{B}^\dag
\end{equation}
where $\hat{H}_{\mathcal{A},\mathcal{B}}=(\hat{H}_\textrm{eff})_\mathcal{A}+(\hat{H}_\textrm{eff})_\mathcal{B})$. One can rewrite the jump part of the equation in a form more suitable to the detection scenario by using the identity $2 \kappa \hat{a}_\mathcal{A}\rho \hat{a}_\mathcal{A}^\dag+2 \kappa \hat{a}_\mathcal{B}\rho \hat{a}_\mathcal{B}^\dag=2 \kappa \hat{J}_1\rho \hat{J}_1^\dag+2 \kappa \hat{J}_2\rho \hat{J}_2^\dag$.

Let us assume that each sub-system begins in state $\ket{e_1}$. The state of the system, after a very short time $\Delta t$ will be
\begin{equation}
\rho(\Delta t)=\rho(0)+\Delta t \left(-i\left[\hat{H}_{\mathcal{A},\mathcal{B}}\ \rho(0)-\rho(0)\hat{H}_{\mathcal{A},\mathcal{B}}^\dag\right]+2 \kappa \hat{J}_1\rho(0)\hat{J}_1^\dag+2 \kappa \hat{J}_2\rho(0)\hat{J}_2^\dag\right)
\end{equation}
which in the weak coupling limit, where $x\ll 1$, gives to a close approximation,
\begin{equation}
\rho(\Delta t)\approx(1-4\kappa|x|^2\Delta t)\ket{\psi_0}\bra{\psi_0}+2\kappa|x|^2 \Delta t \ket{\psi_1}\bra{\psi_1}+2\kappa|x|^2 \Delta t \ket{\psi_2}\bra{\psi_2}
\end{equation}
where $\ket{\psi_0}=\ket{e_1}_\mathcal{A}\ket{e_1}_\mathcal{B}$ and
\begin{equation}
\begin{split}
\ket{\psi_1}&=\frac{\hat{J}_1}{x}\ket{\psi_0}=\frac{1}{\sqrt{2}(1+|x|^2)}\biggl[\ket{B,0}_\mathcal{A}\ket{A,0}_\mathcal{B}+i\ket{A,0}_\mathcal{A}\ket{B,0}_\mathcal{B}\\&\mbox{}+x\left(i\ket{A,0}_\mathcal{A}\ket{A,1}_\mathcal{B}+\ket{A,1}_\mathcal{A}\ket{A,0}_\mathcal{B}\right)\biggr]
\ ,
\end{split}
\end{equation}
and likewise
\begin{equation}
\begin{split}
\ket{\psi_2}&=\frac{\hat{J}_2}{x}\ket{\psi_0}=\frac{1}{\sqrt{2}(1+|x|^2)}\biggl[i\ket{B,0}_\mathcal{A}\ket{A,0}_\mathcal{B}+\ket{A,0}_\mathcal{A}\ket{B,0}_\mathcal{B}\\&\mbox{}+x\left(\ket{A,0}_\mathcal{A}\ket{A,1}_\mathcal{B}+i\ket{A,1}_\mathcal{A}\ket{A,0}_\mathcal{B}\right)\biggr]
\ ,
\end{split}
\end{equation}
which are normalised in the limit that $x$ is small, when both of these  become very close to a maximally entangled state in the ions' internal degrees of freedom.

Thus with probability $4\kappa|x|^2 \Delta t$ one of the detectors clicks, and a state very close to a maximally entangled state is created between the ions. Otherwise, there is no click and the state is reset to $\ket{\psi_0}$. If one models the detectors as taking repeated measurements at some short time $\Delta t$ apart, then in each time window the probability of a click is $4\kappa|x|^2 \Delta t$ multiplied by the probability that no detection has occurred so far. Taking $\Delta t$ to zero, the probability of the first click happening at time t is  $P(t)=4\kappa|x|^2 \exp[-4\kappa|x|^2 t]$ and thus the mean time before the first
detection event will be
\begin{equation}\label{tav}
T_{\text{av}}=\int_0^\infty t P(t) dt=\frac{1}{4\kappa|x|^2}=\frac{\kappa\Delta^2}{|g|^2\Omega^2}\ .\end{equation}



In the weak driving limit, $x$ is much smaller than 1 and can be made arbitrarily small by increasing the detuning and weakening the driving fields. Thus a maximally entangled state of arbitrarily high fidelity can be generated with unit success probability.

However, this result is still only valid in the strong coupling
limit as we have so far neglected the effect spontaneous emission. We have also assumed perfectly efficient detectors which are not experimentally available. We can investigate how our scheme performs taking these effects into account by numerical simulation.
\subsection{Spontaneous emission}

Let us first consider the effect of  spontaneous emission. Since spontaneous decay will principally occur from state $\ket{C}$, we can no longer use the adiabatically eliminated Hamiltonian, and return to the  Hamiltonian in equation~\eqref{hnoadiabat} for each cavity. We  add spontaneous emission jump operators to the Master equation as described in section~\ref{quantjump}, assuming that two decay channels exist, from $\ket{C}$ to $\ket{A}$ with rate $\gamma_A$ and from $\ket{C}$ to $\ket{B}$ with rate $\gamma_B$.
We can now use the quantum jump approach to calculate the evolution numerically using a Monte Carlo simulation.
Since now sometimes the photons will be lost by spontaneous emission, sometimes neither detector will click. Thus, one must choose a ``waiting time'', $T_w$, the length of time the detectors will be monitored before the attempt is judged to have failed, and the system re-initiated. This in turn, gives us a meaningful concept of success probability for the scheme.
There are also two parameters relevant to our simulation technique. The first is the simulation time step. This is the length of time between the ``coin tosses'' which choose whether a quantum-jump has taken place. For a good simulation this time must be small compared to the decay times of the norms under the effective Hamiltonian evolution. We find that these remain on the order of $T_\textrm{av}=1/4\kappa|x|^2$, even for large spontaneous decay rates, as expected and thus  choose time steps of $T_\textrm{av}/10^5$. Secondly, we need to repeat the simulations enough time such that a fair average is obtained. Clearly, there is a trade-off between accuracy of simulation and calculation time. Since we are more interested here in the qualitative behaviour of changing parameters than an exact quantitative calculation, we choose $10^4$ repetitions, which gives sufficient accuracy and reasonable calculation times.

\begin{figure}
\begin{center}
\includegraphics[width=14cm]{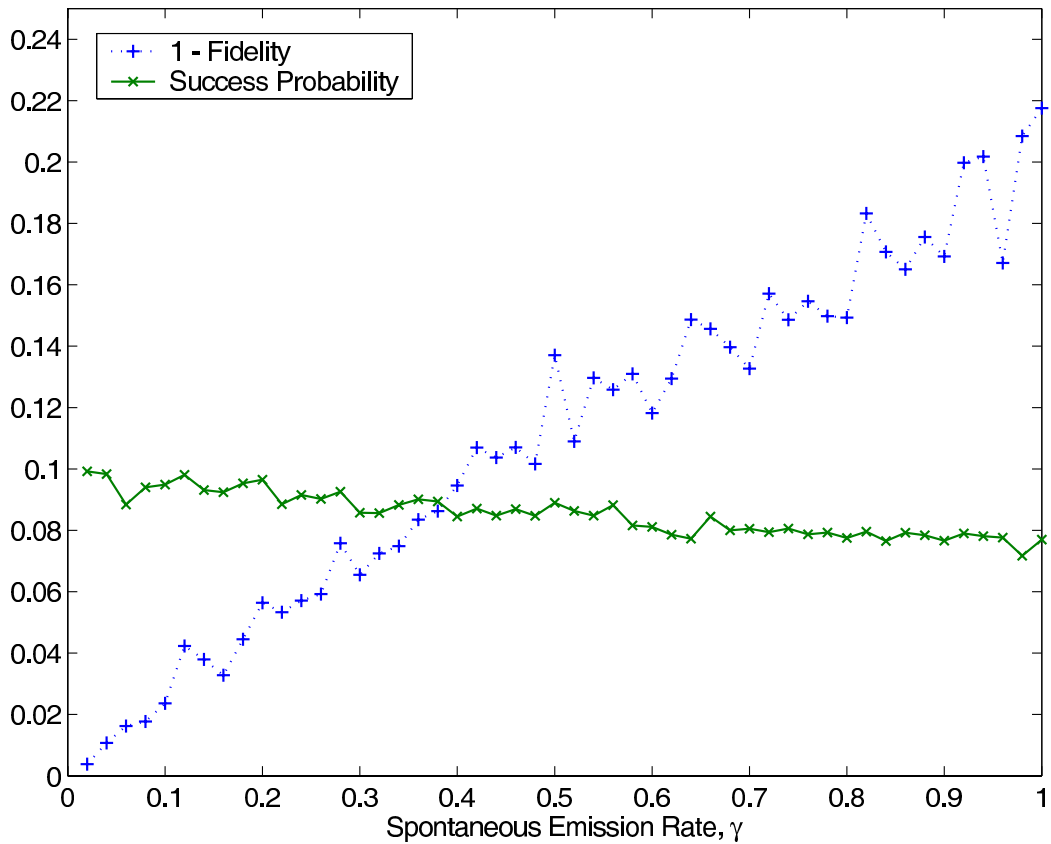}
\end{center}
\caption{\label{effspon} The success probability (solid green line) and
the deviation from unit fidelity, are plotted versus the spontaneous emission rate, which is assumed to be equal for both decay channels. The parameters chosen for the simulation were $\Omega=2g, \kappa=10g, \Delta=20g$. A time-step of $0.1g$ and a waiting
time of $T_w=100/g$ were used. The plot has been obtained from a quantum jump
simulation where each point is the
result of an average over $10^4$ runs.}
\end{figure}

The results of a simulation of the effect of spontaneous emission are plotted in figure~\ref{effspon} for the the parameters detailed in the figure caption. As expected, we see that finite spontaneous emission rates cause a linear   reduction in the fidelity and the success probability. This  corresponds to events where spontaneous decay has driven one or both of the cavities into the state $\ket{A,0}$.  This state is an eigenstate of the single cavity Hamiltonian and the system remains there. If this happens in a single cavity, any following detector click will lead to the undesired, and unheralded  production of a product state, leading to a reduction in the fidelity of the state generated, and if this happens in both cavities no detection will occur, leading to the reduction in success probability.

\begin{figure}
\begin{center}
\includegraphics[width=14cm]{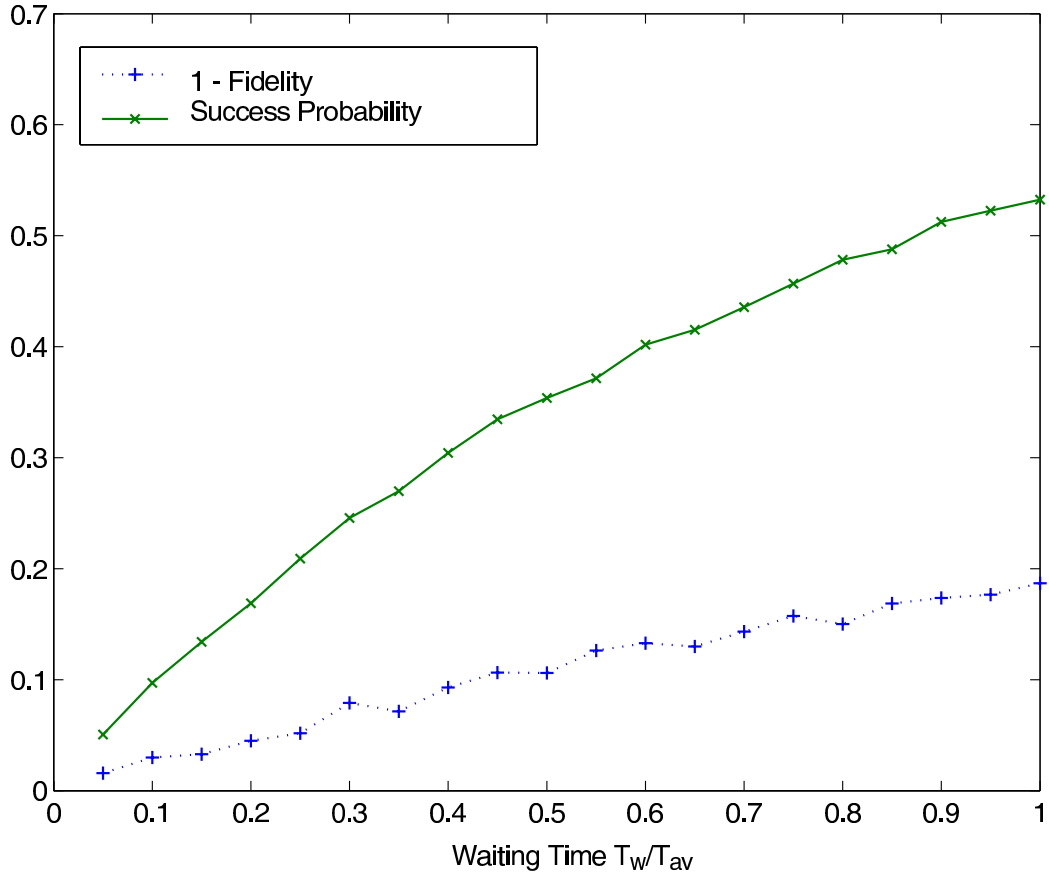}
\end{center}
\caption{\label{effwait} The success probability (solid green line) and
the deviation from unit fidelity, are plotted versus the waiting time $T_w / T_av$ for $\Omega=2g, \kappa=10g, \gamma_{31}=\gamma_{32}=0.1g$ and $\Delta=200g$.  Each point is the
result of an average over $10^4$ runs of the scheme.}
\end{figure}

Working with a weak coupled system with strong spontaneous emission, one would expect that the fidelity of the generated state be maximised by reducing the waiting time since this reduces the chance that a spontaneous decay has occured, as a trade off against success probability. This confirmed by the simulation results  plotted in figure~\ref{effwait}.

\subsection{Detection efficiency}

\begin{figure}
\begin{center}
\includegraphics[width=14cm]{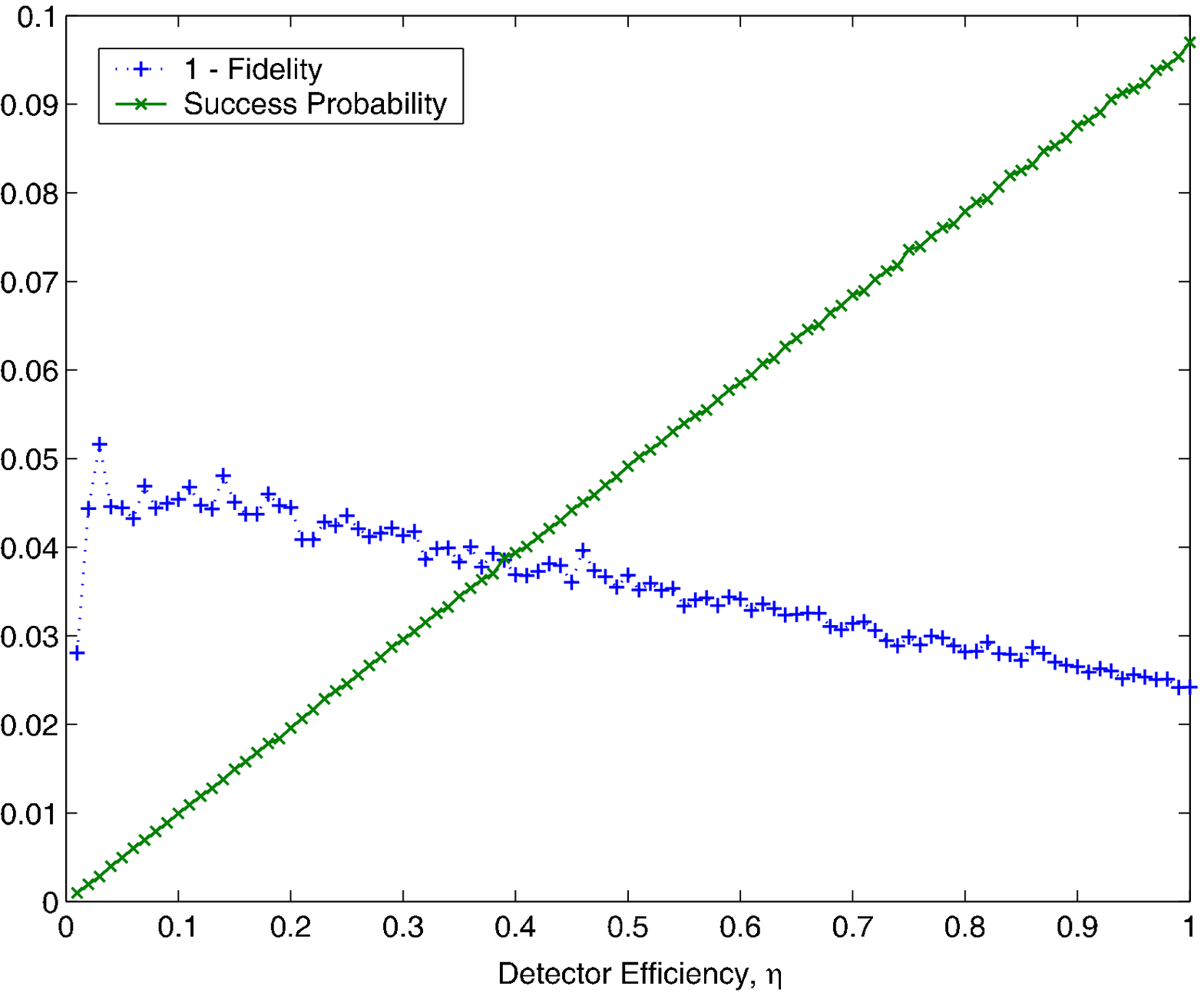}
\end{center}
\caption{\label{effprob} The success probability (solid  line) and
the deviation from unit fidelity, are plotted versus the detector
efficiency for $\Omega=2g, \kappa=10g, \gamma_{31}=\gamma_{32}=0.1g$ and $\Delta=20g$ and a waiting
time of $T=100/g$.  Each point is the
result of an average over $10^6$ runs of the scheme.}
\end{figure}
Another important experimental imperfection is non-unit detection efficiency. In fact, not only are current photo-detectors unable to detect all incident photons, but other effects, such as absorption of photons in the cavity mirrors and the photons escaping from the cavity in the wrong direction will all contribute to give an effective detector efficiency $\eta$ much lower than one. There is a simple way to model such a situation.
The case of imperfect detectors is modelled as follows, the 
usual projection operator $|n\rangle\langle n|$ is replaced by 
\begin{equation}
|n\rangle\langle n|\longmapsto 
\sum_{k=n}^\infty
{k \choose n}\eta^n
(1-\eta)^{k-n}|k\rangle\langle k|\ .
\end{equation}
Each term in the sum corresponds to $(k-n)$ photons being missed or absorbed before the detector.

It is straightforward to incorporate imperfect detectors into our Monte Carlo simulation. We expect that our scheme will exhibit robustness against such an imperfections since, in the weak-driving regime, there is only a very small probability that photons will leak from both cavities, which is the only event which will decrease our state fidelity, by contributing an admixture of the product state $\ket{A}\ket{A}$. Indeed, the simulation backs up this view. 
In Fig. \ref{effprob} we have plotted both the success probability
and the achieved error (1-fidelity) for fixed $\Omega=2g,
\kappa=10g, \gamma_{A}=\gamma_{B}=0.1g$ and
$\Delta=20g$ and a waiting time of $T=100/g$ against the detector
efficiency. We observe, as expected, a weak linear reduction in the state fidelity see that the success probability scales by a factor $\eta$.


\subsection{Dark counts}

A further experimental imperfection which must be considered is
the presence of ``dark counts'', i.e. when the detector fires
although no light is incident upon it. Clearly, this will degrade to
some extent the performance of all schemes which rely on a single
detector click to generate an entangled state, and will lead, in
general, to a loss in fidelity of the state produced. However, in
the present scheme, the time-window in which a click due to a
photon should occur is far shorter than the mean time between dark
counts. For example, in \cite{Buttler} a dark count rate of
approximately 1400 s$^{-1}$ is reported, thus the mean time
between dark counts is on the order of ms. In the optical regime,
the atom-cavity coupling $g$, detuning $\Delta$, cavity decay rate
$\kappa$ and the coupling with the classical field $\Omega$ will
all be at least on the order of MHz \cite{Kimble}. Thus, using equation~\eqref{tav} one can estimate that $T_\text{av}$, the mean time
before a proper click occurs in this scheme, is on the order of
$\mu$s. Since the time-window for detection in this scheme can
thus be made much smaller than the mean time between dark counts,
their effect  on this scheme can be made very small.

So far we have considered the case where we were faced with the
task of entangling two spatially separated ion-cavity systems. The
key ingredient in the detection scheme was the beam-splitter that
erased the which-path information from the system, so that a
photo-detection event could lead to entanglement between the
cavities. However, the above method could also be used to entangle
two ions trapped in a single cavity whose decay is monitored by a
single photo-detector if the system is set-up such that the
detection of a photon does not provide any information about which
ion the photon was emitted from.

\section{Conclusion}

In summary, we have presented here a new approach to entangling ions in spatially separated cavities that, under ideal
conditions, allows for the deterministic generation of perfect
entanglement between individual ions in distant cavities.
Employing a weak interaction as in the previous chapter  provides a robust method of  state-preparation which is resilient against many experimental imperfections. We therefore expect this scheme to work well when operated with current experimental setups. The reason why this approach is so effective can be understood as follows. Ignoring decoherence and dissipation, the evolution of the system is unitary and can be thus written as an exponential $\exp[-i \chi \hat{O}]$, where $\hat{O}$ is a Hermitian operator of unit norm. In the regime of weak or brief interactions, $\chi$ is small, and the operator may thus be approximated by $\iden-i\chi \hat{O}$. Then, a measurement takes place which projects out one of these terms. With high probability the former term is selected, and the system has returned to its previous state, but with a small probability, the latter term, which generates the desired entangled state, is realised. By keeping $\chi$ small the actual value of $\chi$ remains unimportant. While in the first scheme, presented in chapter~\ref{ch:4}, this led to a low success probability, in the proposal presented in this chapter, constant weak driving and continuous detection allow, in the ideal scenario, arbitrarily high success probabilities to be achieved. The schemes remain robust in the face of decoherence, dissipation and imperfect measurement, as long as measurements can still resolve that latter term sufficiently well. Thus, such techniques could be adapted to many other physical
systems, and may be especially useful in systems which are hard to control with precision.

%% file: c-6.tex
\chapter{Gaussian States and Operations}\label{ch:6}

In the second half of this thesis, we move from the cavity QED setting to consider the quantum states of  light pulses. As we discussed in section~\ref{sec:laserlight}, the  quantum state of a stream of  pulses generated by an ideal laser source  may be written, 

\begin{equation}
\rho_\textbf{pulses}=\frac{1}{2\pi}\int d\phi ( \ket{\lambda e^{i\phi}} \bra{\lambda e^{i\phi}})^{\otimes N}\ .
\end{equation}

As we discussed in section~\ref{sec:laserlight}, as long as all phase measurements on the light are made using reference pulses split from the same source (which is just the approach that experimentalists take) the pulses will be indistinguishable from a pure product of coherent states, $\ket{\lambda e^{i\phi}}$. In the following chapters, for mathematical and conceptual convenience we will shall use the universal approach taken in quantum optics and ignore the uncertainty in the overall phase of the generated light. This is perfectly valid as long as a beam of reference pulses split from the source as described in section~\ref{sec:laserlight} is available to the experimenter along with the pulse whose quantum state is being manipulated, and we assume the presence of such reference pulses in all the following discussion.

\section{Gaussian states}\label{sec:gausstates}
Clearly the quantum state of a light mode can be extremely complicated, as a general state of a quantum harmonic oscillator has infinitely many parameters. Fortunately, there is a class of states  which require only a small number of parameters for their specification, which also have immense experimental importance, especially in the frame-work of linear optics. These are the \emph{Gaussian States}. 
Gaussian States are states whose \emph{characteristic function}, introduced in section~\ref{sec:states}, are Gaussian \cite{simon87,arvind95,jenseqis}.

In section~\ref{sec:states}, we restricted ourselves to single-mode states, but to describe entangled states, we need a multi-mode notation. Let us label each mode of the $n$-mode system with a number $i$, and write the \emph{quadrature operators} (again introduced in section~\ref{sec:states}), for the $i$th mode $\hat X_i$ and  $\hat P_i$. It is convenient to represent these operators in terms of a column vector\footnote{As with the quadrature operators themselves, conventions for the definition of all of these multi-mode constructs, in terms of their normalisation and ordering, vary widely in the literature.}
 $\hat{\vek{R}}=(\hat{X}_1, \hat{P}_1,\ldots,\hat{X}_n, \hat{P}_n)$, and likewise form a vector of the multi-mode  phase-space variables ${\vekk{\xi}}=({X}_1, {P}_1,\ldots,{X}_n, {P}_n)$.

The canonical commutation relations can now be expressed
\begin{equation}\label{ccr}
[\hat{R}_j,\hat{R}_k]=i\Sigma_{j,k}
\end{equation}
where $\Sigma$ is the \emph{symplectic matrix}, a direct sum of $n$ anti-symmetric two-by-two matrices,
\begin{equation}\label{sympmat}
\Sigma=\bigoplus_{i=1}^n \left(\begin{array}{cc} 0 & 1\\-1 &0
  \end{array}
\right)\ .
\end{equation}
We can thus express the multi-mode Weyl operator $\hat{W}_\vekk{\xi}$ as follows

\begin{equation}
\hat{W}_\vekk{\xi}=e^{i \vekk{\xi}^T \Sigma \hat{\vek{R}}}
\end{equation}
where we have employed the \emph{symplectic product}, $\vekk{\xi}^T \Sigma \hat{\vek{R}}$ between vectors $\vekk{\xi}$ and  $\hat{\vek{R}}$.
We write the characteristic function for an $n$-mode state $\rho$,

\begin{equation}\label{charfunrho}
\chi_\rho(\vek{\xi})=\Tr[\rho \hat{W}_\vekk{\xi}]\ .
\end{equation}
The state $\rho$ may be obtained from its characteristic function as follows

\begin{equation}\label{gausstaterho}
\rho=\frac{1}{(2\pi)^n}\int \chi_\rho(-\vek{\xi})\hat{W}_\vekk{\xi}
d^{2n}\xi \ ,
\end{equation}
which can be verified using the identity $\Tr[\hat{W}_\vekk{\alpha}\hat{W}_\vekk{\beta}]=(2 \pi)^n \delta^{2n}(\vekk{\alpha}+\vekk{\beta})$ \cite{cahillglauber}.
Gaussian states are all those for which the characteristic function has a Gaussian form. This means that the characteristic function may be written\footnote{Again, the normalisations in this are a matter of convention.},
\begin{equation}\label{gauschar}
\chi_\rho=\exp[-\frac{1}{4} \vekk{\xi}^T\Sigma \Gamma \Sigma^T \vekk{\xi} + i \vek{d}^T \Sigma \vekk{\xi}]\ .
\end{equation}
The $2n\times2n$ matrix $\Gamma$ is called the \emph{covariance matrix} and $\vek{d}$ is the \emph{displacement vector}. The displacement vector describes the centre of the Gaussian in phase-space, and the \emph{covariance matrix}, naturally, the variances and co-variances. 
By equating partial derivatives of $\chi_\rho(s\vekk{\alpha}+t\vekk{\beta})$ in terms of $s$ and $t$,  in equations~(\ref{charfunrho}) and~(\ref{gauschar}), one obtains expressions for $\Gamma$ and $\vek{d}$ in terms of $\rho$.
The displacement vector consists of nothing more than the expectation values of the quadrature operators,

\begin{equation}
d_j=\Tr\bigl[\rho \hat{R}_j\bigr]=\langle  \hat{R}_j \rangle\ .
\end{equation}

The elements of the covariance matrix have the following form,
\begin{equation}\begin{split}
\Gamma_{j,k}&=\Tr\bigl[\rho\bigl\{(\hat{R}_j-\langle  \hat{R}_j \rangle),(\hat{R}_k-\langle  \hat{R}_k \rangle)\bigr\}_+\bigr]\\
&=2 \textrm{Re}\bigr(\Tr\bigl[\rho(\hat{R}_j-d_j \rangle)(\hat{R}_k-d_k \rangle)\bigr]\bigl)\\
&=2\Tr\bigl[\rho(\hat{R}_j-d_j)(\hat{R}_k-d_k)\bigr])-i\Sigma_{j,k}
\end{split}
\end{equation}
where $\{\hat{A},\hat{B}\}_+=\hat{A}\hat{B}-\hat{B}\hat{A}$ is an anti-commutator. We see from this definition that $\Gamma$ is a real symmetric $2n\times2n$ matrix.
Thus an $n$-mode state is described by $2n+(1/2)(2n)(2n+1)=3n+2n^2$ parameters. Furthermore, the displacement vector contains only quantities local to each mode, which means that all of the entanglement properties of the state must only depend on the state's covariance matrix.

It is important to mention that not all possible covariance matrices correspond to a physical state. In fact, for $\rho$ to be Hermitian and normalisable, the covariance matrix must satisfy the following inequality
\begin{equation}\label{gausineq}
\Gamma+i\Sigma\ge 0 \ ,
\end{equation}
where the matrix inequality expression means that the right hand side is positive semi-definite. This inequality also contains the Heisenberg uncertainty relations for the quadrature operators.

Not only do the Gaussian states have a useful theoretical description, they correspond to states of light of great experimental importance. These include the coherent states which we introduced in section~\ref{coherentstates}, thermal states of a radiation field and the \emph{squeezed states}\cite{stolersq,yuensq}, a very important class of non-classical states which have been heavily studied in quantum optics (see \cite{loudonknight} or \cite{teichsaleh} for a review), both as single mode and entangled multi-mode states.

\subsection{Descriptions of important Gaussian states}

We shall briefly review how these important states are represented in the covariance matrix and displacement vector formalism.The covariance matrix for a single-mode \emph{coherent state} $\ket{\alpha}$ is simply  the identity matrix $\iden$ and their displacement vector is $(\textrm{Re}(\sqrt{2}\alpha),\textrm{Im}(\sqrt{2}\alpha))^T$.
When a radiation mode is in thermal equilibrium with a heat bath at temperature $T$ its state is called a \emph{Gibbs thermal state},
\begin{equation}
\rho=(1-e^{-\kappa})\sum_n e^{-\kappa n}\ket{n}\bra{n}
\end{equation}
where $\kappa=(k_B T)^{-1}$. Thermal states are Gaussian States with covariance matrix, $
\Gamma=
\coth(\kappa/2)\iden$ and zero displacement.

The Heisenberg uncertainty relation for the canonically conjugate quadrature operators is $\Delta X \Delta P\ge 1/2$. For coherent states, both $\Delta X$ and $\Delta P$ are $\sqrt{1/2}$, so these are states of minimum uncertainty. To reduce $\Delta X$ further one must increase $\Delta P$ or vice versa. Minimum uncertainty states with $\Delta X$ or $\Delta P$ less than $\sqrt{1/2}$ are called  (single-mode) \emph{squeezed states}. 

It is convenient to express the squeezed  states in terms of the single-mode squeezing operator $\hat{S}(\zeta)$

\begin{equation}
\hat{S}(\zeta)=\exp\left[-\frac{\zeta}{2}\hat{a}^{\dag2}+\frac{\zeta^*}{2}\hat{a}^{2}\right]
\end{equation}
where $\zeta=r e^{i\phi}$ is the complex squeezing parameter. Single mode squeezed states are therefore expressed $\hat{S}(\zeta)\ket{0}$. The degree of squeezing\footnote{If one interprets the uncertainty in the squeezed quadrature measured in an experiment as ``noise'', then squeezing may be quantified in terms of  of deciBels.  In fact, this is the approach taken in most experimental papers. The conversion of a squeezing value $d$ expressed in deciBels to the squeezing parameter $r$ proceeds via the following relation $r=(\ln(10)/20)d\approx0.12 d$.} is quantified by $r$  and the angle of the quadrature in phase-space which is being squeezed is expressed by $\phi$. The uncertainty of the squeezed quadrature is  reduced by a factor $e^{-r}$ and the uncertainty of the \emph{anti-squeezed} conjugate quadrature increases by a factor $e^{r}$. This is directly captured in the form of the covariance matrix,
\begin{equation}
\Gamma= \mathcal{R}(\phi)^T \left(\begin{array}{cc} e^{-2r}& 0\\0& e^{2r}
  \end{array}
\right)  \mathcal{R}(\phi)\ ,
\end{equation}
where 
\begin{equation}
\mathcal{R}(\phi)=\left(\begin{array}{cc} \cos(\phi) & -\sin(\phi) \\ \sin(\phi) & \cos(\phi)
  \end{array}\right)
\end{equation}

is a phase-space rotation through angle $\phi$. The displacement remains unaffected by the squeezing operator and remains zero, although a more general family of squeezed states may be constructed by allowing finite displacements as well.
Experimentally, the highest degree of squeezing which can been obtained for a continuous  laser beam is 7.2dB \cite{squeezecw} corresponding to $r=0.82$. The highest degree of squeezing recorded for pulsed light is 3.1dB or $r=0.36$ \cite{squeezepulse}.


So far, I have only detailed single-mode Gaussian states. The next case is an important example of a two-mode entangled state. It is called a \emph{two-mode squeezed state} and is again best expressed in terms of a two-mode squeezing operator, as $\hat{S_2}(\zeta)\ket{0}\ket{0}$, where

\begin{equation}
\hat{S_2}(\zeta)=\exp\left[-\zeta\hat{a}_1^{\dag}\hat{a}_2^{\dag}+{\zeta^*}\hat{a}_1\hat{a}_2\right]
\end{equation}
and where  modes are numbered 1 and 2. We shall express the complex squeezing parameter $\zeta$ in terms of its modulus and argument, $\zeta=r e^{i\phi}$. For non-zero $r$ this state is entangled, and its entanglement increases with increasing $r$. Its covariance matrix is written,
\begin{equation}\label{eq:tmss}
\Gamma=[\mathcal{R}(\phi/2)\oplus\mathcal{R}(\phi/2)]^T\left(\begin{array}{cccc}
C_r &0&- S_r&0\\
0&C_r &0& S_r\\
-S_r&0&C_r &0\\
0&S_r&0&C_r
\end{array}
\right)\mathcal{R}(\phi/2)\oplus\mathcal{R}(\phi/2)
\end{equation}
where $C_r=\cosh(2r)$ and $S_r=\sinh(2r)$.
The displacement vector for this state is zero.

One of the beautiful features of the covariance matrix formalism is the ease with which one can identify the reduced state of a sub-system, one merely reads off the corresponding sub-matrix. Thus, for example, we see that the reduced state of two modes in a two-mode squeezed state is a thermal state with $\coth(\kappa/2)=\cosh(r/2)$. 
By writing out the state vector of the two-mode squeezed state, we see that it also possesses  correlations between the photon numbers of the entangled modes,
\begin{equation}
\hat{S_2}(r e^{i\phi})=\textrm{sech}(r)\sum_{n=0}^\infty [-e^{i\phi}\tanh(r)]^n \ket{n}\ket{n} \ .
\end{equation}
The ``squeezing'' in this state is an enhancement in the correlations between the quadratures of the two modes. Let us set $\phi$ to zero and consider the uncertainty in the sum of the $X$ quadratures for both modes,
\begin{equation}\begin{split}
\Delta(X_1+X_2)&=\sqrt{\langle(\hat{X}_1+\hat{X}_2)^2\rangle}\\
&=\sqrt{\frac{\Gamma_{1,1}+\Gamma_{3,3}}{2}+\Gamma_{1,3}}\\
&=e^{-r}\ .
\end{split}
\end{equation}
We see that, in this case, the sum of the quadratures is squeezed, and similarly one can show that the difference of the quadratures is anti-squeezed. When we consider the sum and difference of the conjugate quadratures $\hat{P}_1$ and $\hat{P}_2$, we find that the opposite occurs, namely the difference is squeezed and the sum is anti-squeezed.

Two-mode squeezed states may be generated by mixing two single-mode squeezed states, with orthogonal squeezed quadratures, at a 50:50 beam-splitter \cite{furusawatele,silbersq,bowersq}, or alternatively, generated directly via a non-linear process known as parametric down-conversion   \cite{ousqueeze,zhangsqueez}. A detailed treatment of the theory of squeezed states, both single and multi-mode, can be found in the literature \cite{maandrhodes,barnettradmore}.

\section{Gaussian operations}
Now that we have examined the most important Gaussian States, we turn our attention to the Gaussian operations \cite{arvind95}, namely those which map all Gaussian states onto Gaussian States. We shall see that these operations have significant experimental importance in the optical setting. We will not discuss the formal description of the Gaussian operations in great detail here  it is worth briefly describing how operations on states translate into operations on covariance matrices and displacement vectors. We will focus here on  unitary operators alone, for  an excellent review which includes more general operations and measurement, see \cite{jenseqis}.

Unitary transformations on Gaussian states correspond to \emph{symplectic} transformations on covariance matrices and displacement vectors. A symplectic transformation $\mathcal{S}$ is one which preserves the symplectic product between any pairs of vectors, and corresponds to a canonical transformation of quadrature operators \cite{plkekert}, preserving the canonical commutation relations. This leads to the condition,

\begin{equation}
\mathcal{S}^T \Sigma \mathcal{S}=\Sigma
\end{equation}
where $\Sigma$ is the symplectic matrix defined above. Symplectic transformations are represented by real matrices with unit determinant.
For a given unitary transformation on a state $U$, there is a corresponding unique real symplectic transformation such that $U \hat{W}_\vekk{\xi} U^\dag=\hat{W}_\vekk{S\xi}$. It is then straightforward to confirm that when a Gaussian state is acted on by $U$, its displacement vector transforms $\vek{d}\mapsto \mathcal{S}\vek{d}$ and its covariance matrix, $\Gamma\mapsto \mathcal{S}\Gamma \mathcal{S}^T$.
Once can show (see \cite{impnogauss}, \cite{giedkenogauss} and \cite{jaromirnogauss}), that, in the optical setting, all Gaussian operations can be implemented with \emph{beam-splitters}, \emph{phase shifters}, \emph{squeezers} and \emph{homodyne} measurement.
Apart from the detection, these are the operations which comprise \emph{linear optics}.

Linear Optics is so-called as it encompasses all operations which can be expressed as linear transformations of the creation and annihilation operations. For example, a single mode pure state
\begin{equation}
\ket{\psi}=\sum_n \alpha_n \ket{n}=\sum_n \frac{\alpha_n}{\sqrt{n!}}(\hat{a}^\dag)^n\ket{0}
\end{equation}
is transformed by a single mode linear transformation $L[\cdot]$ to 
\begin{equation}
\ket{\psi'}=\sum_n \frac{\alpha_n}{\sqrt{n!}}(L[\hat{a}^\dag])^n\ket{0}\ .
\end{equation}
These operations may be divided into two categories, \emph{passive} linear operations, which preserve the mean photon number  of the state, and \emph{active} linear operations, which do not.

\subsection{Passive linear optics}\label{sec:passive}
All passive linear operations can be implemented via networks of phase shifters and beam splitters. Phase shifters are single mode operations which transform $\hat{a}^\dag$ to $e^{i\phi}\hat{a}^dag$, where $\phi$ is a particular phase. Thus, for example, the coherent state $\ket{\alpha}$ is transformed to the state $\ket{e^{i\phi}\alpha}$. Phase shifts are often implemented by passing the laser pulse through a crystal with a certain length with a refractive index higher than air.
The symplectic transformation which  the phase shift generates is, as one would expect, a rotation in phase space
\begin{equation}
\mathcal{S}=\left( \begin{array}{cc} 
\cos(\phi) & -\sin(\phi) \\
\sin(\phi) & \cos(\phi) 
\end{array}
\right)\ .\end{equation}

The beam-splitter takes two-modes as input and outputs two-modes. It implements a 
linear two-mode transformation given here in its most general form \cite{prasad}

\begin{equation}
\left(\begin{array}{c} \hat{a}^\dag_1 \\ \hat{a}^\dag_2
  \end{array} \right) \mapsto \left(\begin{array}{cc} T & R \\ -R^* & T^*
  \end{array} \right) \left(\begin{array}{c} \hat{a}^\dag_1 \\ \hat{a}^\dag_2
  \end{array} \right)
\end{equation}
where $|T|^2+|R|^2=1$. $T$ (or sometimes $|T|^2$) is referred to as the transmittivity and $R$  (or sometimes $|R|^2$) as the reflectivity. Thus for a 50:50 beam splitter $|T|^2=|R|^2=1/2$. 

Beam-splitter operations are typically implemented by allowing the optical pulses to meet on a semi-silvered mirror. The phases of $T$ and $R$ may then be chosen by appending extra phase shifters to the inputs and/or outputs of the beam splitter, as appropriate. In this thesis, we shall assume that $T$ and $R$ are always real, and always specify any extra phase shifts. In that case, the symplectic beam splitter transformation is

\begin{equation}
\mathcal{S}=\left( \begin{array}{cc} 
T\iden_2 & -R\iden_2 \\
R\iden_2 & T\iden_2
\end{array}
\right)\end{equation}
where $\iden_2$ is a two by two identity matrix.

Approximate displacement operations may be achieved with linear optics by mixing the mode to be displaced with an intense coherent state on a beam-splitter with near 100\% transmittivity.

\subsection{Active linear optics}
In contrast  to the passive linear operations ``squeezing'',  the physical realisation of the single mode squeezing operator $\hat{S}(\zeta)$, changes the energy of the mode and thus requires driving by an external field. This field $\hat{b}$, couples to the mode which is to be squeezed $\hat{a}$ inside a non-linear crystal with the interaction Hamiltonian,

\begin{equation}
\hat{H}=i \left(\chi (\hat{a})^2\hat{b}^\dag-\chi^*(\hat{a}^\dag)^2\hat{b} \right)\ .
\end{equation}
The driving field is an intense ``classical'' coherent state $\ket{\beta}$, and thus, taking the classical approximation introduced in section~\ref{sec:intclassical}, this implements the  squeezing unitary
\begin{equation}
\hat{S}(2\chi^* \beta)=\exp\left[-\chi^*\beta  (\hat{a}^\dag)^2+\chi\beta^* (\hat{a})^2\right]\ .
\end{equation}
The symplectic transformation for a squeezing operation with real $\zeta=r$ is,
\begin{equation}
\mathcal{S}=\left( \begin{array}{cc} 
e^{r} & 0 \\
0 & e^{-r} 
\end{array}
\right)\ .\end{equation}
A complex squeezing parameter corresponds to extra phase rotations, as above.
Of course, many approximations have gone into the analysis here, and in physical realisations there is both a limit in the degree of squeezing which is attainable, and, typically, extra unwanted noise will be added to the state.

\subsection{Homodyne measurement}

The statistics of the quadrature operator on a mode may be probed by homodyne detection \cite{personickthesis,yuenshapiro,banahomo}. This is implemented by mixing the mode with a phase coherent reference pulse (i.e. from the same source) in an intense coherent state on a beam-splitter, and then measuring the difference in intensity of the two output ports. The measurement corresponds to a projection onto the eigenstate of $\hat{X}$ with the measured eigenvalue. By adding phase shifts to the reference pulse, other quadratures may be measured. A more elaborate setup using four beam splitters, known as eight-port homodyning, leads to projections onto coherent states.

\subsection{Non-unitary Gaussian maps}
If one further ingredient is added, the above operations include all possible Gaussian operations. This is to go beyond merely unitary operations and projections, by adding ancilla modes prepared in some fiducial state, allowing joint operations between our system and these ancilla modes, and tracing out the ancilla modes afterwards. The general completely positive Gaussian maps which this leads to are characterised in \cite{impnogauss,giedkenogauss,jaromirnogauss}.

\section{Entanglement in Gaussian states}\label{sec:entgausss}

As one would expect for a system with infinitely many  degrees of freedom,  in general, quantifying entanglement in infinite-dimensional systems can be difficult. In particular, many standard entanglement measures show undesirable characteristics such as discontinuities  \cite{jensentinfinitedim}, and one can show that any state is arbitrarily close to a state of infinite entanglement. However, as Eisert and co-authors show, if one makes the reasonable demand that all states must be finitely bounded in energy, then 
these problems do not arise, and entanglement measures are much better behaved.

The matter is simpler still for Gaussian states, where, due to their small number of parameters, particular entanglement measures can be calculated directly. Additionally, since the displacement vector contains purely local properties, the entanglement of a state depends on its covariance matrix alone.

The question of separability of Gaussian states was first considered independently in \cite{duancriterion} and  \cite{simoncriterion}. In particular, Simon showed that the Positivity of Partial Transpose (PPT) criterion is both necessary and sufficient for bi-partite Gaussian states.
In phase-space, transposition corresponds to time-reversal,  momentum is reversed, position remains unaltered. Consider a Gaussian state $\rho$ with covariance matrix $\Gamma$ in a system of $n\times m$ modes. The entanglement between the former $n$ and latter $m$ modes may be investigated by considering the partially transposed covariance matrix $\Gamma'$ where the momenta of the latter $m$ modes have been reversed. Thus the PPT criterion becomes the condition that  $\Gamma'$ fulfills the inequality
\begin{equation}
\Gamma'+i\Sigma\ge 0\ .
\end{equation}

If this condition is not fulfilled, the state $\rho$ must be entangled. 
Werner extended this result to show than if entanglement between a single mode in a  multi-mode Gaussian state and the remaining modes are consider this criterion is also necessary and sufficient \cite{werneronetimesn}.

As we described in section~\ref{sec:entanglement}, one can  derive a measure of entanglement from this condition, the ``logarithmic negativity'', which expresses to how great a degree the PPT inequality is violated. For Gaussian States, this quantity can be simply expressed in terms of the \emph{symplectic eigenvalues} of the covariance matrix. The symplectic eigenvalues \cite{simoncovmatrix} are the absolute eigenvalues of the matrix $\Sigma\Gamma$, which always occur in pairs. Any covariance matrix can be brought into a diagonal form with these entries by symplectic operations \cite{williamsonnormalform}. For a state $\rho$, with covariance matrix $\Gamma$, if $\Gamma'$ is the partially transposed covariance matrix, the logarithmic negativity is defined to be,

\begin{equation}
E_{\mathcal{N}}=\sum_k \textrm{Max}[-\log_2(\gamma'_k),0]
\end{equation}
where $\gamma'_k$ are the symplectic eigenvalues of $\Gamma'$ and only one member of each pair of eigenvalues is included in the sum.

Two further important quantities can be expressed concisely in terms of the symplectic eigenvalues, the Von Neumann entropy $S_\textrm{vN}=-\textrm{Tr}[\rho \ln \rho]$ and linear entropy  $S_\textrm{lin}=1-\textrm{Tr}[\rho^2]$,

\begin{equation}
S_\textrm{vN}=\sum_k \left[ \frac{\gamma_k+1}{2}\log\left(\frac{\gamma_k+1}{2}\right)-\frac{\gamma_k-1}{2}\log\left(\frac{\gamma_k-1}{2}\right)\right]\ ,
\end{equation}

and
\begin{equation}
S_\textrm{lin}=1-\prod_k\frac{1}{\gamma_k}=1-|\Gamma|^{-\frac{1}{2}}\ ,
\end{equation}

where $\gamma_k$ the symplectic eigenvalues of $\Gamma$ and,  as above, each eigenvalue pair is counted only once. The determinant of $\Gamma$ is written $|\Gamma|$. This is equal to the square of the product of the symplectic eigenvalues since the determinant is unchanged by symplectic (unit determinant) operations. A corollary of this statement is that pure states always have covariance matrices with unit determinants.
For a full derivation of these equations see, for example, \cite{krugerthesis}.
The expression for the von Neumann entropy allows us, of course, to calculate the entanglement of pure Gaussian states.

\section{Gaussian states in quantum information}\label{sec:gaussinqip}

The experimental feasibility of Gaussian operations makes Gaussian states an attractive setting for implementation of quantum information science protocols. These protocols have in the main been proposed for qubit systems, however many may be translated to the continuous variable domain.

Braunstein and Kimble showed how quantum teleportation could be implemented with linear optics using two-mode squeezed states as an entanglement resource instead of Bell states \cite{braunsteinkimble} and their proposal was implemented soon afterwards\cite{furusawatele}. 
Quantum key distribution (QKD) also translates readily to the continuous variable regime. Optical states are particularly suitable for this task as they can be easily transported over long distances. QKD schemes based on the transmission of coherent states  \cite{grosshans}, squeezed states \cite{ralphqkd} and shared two-mode squeezed states \cite{ralphqkd} have been proposed and realised \cite{grosshansnature,silberqkd1,silberqkd2}. So far only the latter two have been proven absolutely secure against all possible eavesdropping attacks\cite{gottesmannpreskill}.

Other qubit-based protocols which have been translated into linear optical proposals on Gaussian states include dense-coding \cite{samdense}, entanglement swapping \cite{samswap} and error-correction \cite{braunsteinerrorcorlinop}.
An important protocol which had not, however, been developed for continuous variable states is entanglement distillation. In the following chapters we shall describe a new protocol for the distillation of entangled Gaussian states.


%% file: c-7.tex
\chapter{Distillation of Continuous Variable Entanglement}\label{ch:7}
\section{Distribution of entanglement}\label{sec:entdistrib}

As discussed in chapter~\ref{ch:1}, many of the most intriguing features and useful applications of entanglement are encountered when the entangled sub-systems are spatially separated. However, entangled states can only be created by global operations on the sub-systems, which requires either that the sub-systems interact directly with one another, or each interacts directly with some  mediating third party, as in the protocols described in chapters~\ref{ch:4} and~\ref{c-5}. In either case, for entanglement to be established between two distant laboratories, quantum systems must physically traverse  the distance between them.

It is unavoidable that during this transmission, some interaction with the ``environment'', which is how we shall describe systems outside of the experimenter's control, will occur. This usually leads to the state of the system changing,  typically becoming  more mixed due to entanglement with the environment. This in turn degrades the entangled state which is supposed to be being transmitted.

As an example of this, let us consider a two-mode squeezed state, as introduced in section~\ref{sec:gausstates}, with real squeezing parameter $\lambda$, generated at a central source of pulsed radiation. The two pulses are transmitted separately through  optical fibres  to two parties, who we shall name Alice and Bob\footnote{The use of the names Alice and Bob, instead of the dry labelling  $A$ and $B$, has been adopted from classical communication theory to become ubiquitous in quantum information science. Indeed, it is hard to find a paper on quantum key distribution or entanglement distillation where they do not appear!}.

To a first approximation, an optical fibre acts as an \emph{absorbing channel}. This is a transmission line  which absorbs some of the radiation which passes through it. A thorough analysis \cite{stefanchannelpaper} has shown that such a channel may be accurately modelled in a very simple way. We pretend that the light pulse has been mixed with a vacuum mode on a beam splitter of transmittivity $T=\sqrt{\tau}$. The parameter $\tau$ then corresponds to the \emph{transmission coefficient} of the channel, expressing the ratio between the intensities of transmitted and incident light. Since the beam splitter is a Gaussian operation, the absorbing channel preserves the Gaussian nature of input states. 
In the phase-space formalism it is easy to calculate the effect of such a channel. The displacement vector is reduced by a factor of $\sqrt{\tau}$, and the covariance matrix $\Gamma$ transforms to $\tau \Gamma +(1-\tau)\iden$.

Let us consider the effect of passing both pulses of a two-mode squeezed state with real squeezing parameter $r$ through two channels, each with equal transmission coefficients. The covariance matrix is transformed, from the form expressed in equation~(\ref{eq:tmss}) to

\begin{equation}\label{eq:noisytmss}
\left(\begin{array}{cccc}
1+\tau(C_r-1) &0&- \tau S_r&0\\
0&1+\tau(C_r-1) &0& \tau S_r\\
-\tau S_r&0&1+\tau(C_r-1) &0\\
0&\tau S_r&0&1+\tau(C_r-1) 
\end{array}
\right)
\end{equation}
where $C_r=\cosh(2r)$ and $S_r=\sinh(2r)$ as before.
We can calculate the logarithmic negativity of this state using the method described in the previous section, and find,
\begin{equation}
E_\mathcal{N}=-\log_2[1-\tau(1-e^{-2r})]\ .
\end{equation}
Thus, as expected, while entanglement still increases with increasing $r$, the absorption $\tau<1$ reduces its value. Furthermore, while the entanglement of the pure two-mode squeezed state is unbounded, the entanglement of the state after the absorbing channels, is bounded by $-\log_2[1-\tau]$.

\section{Entanglement distillation}\label{sec:distillsec}

This degradation in the entanglement of the states at their source is a generic feature of the decoherence processes which are unavoidable in the  transmission of quantum states. For the applications of spatially distributed entangled Gaussian states discussed above in section~\ref{sec:gausstates}, highly entangled states are required. Since, by definition, the entanglement of a spatially distributed state cannot be increased by local operations and classical communication, another approach is required.  This problem was first considered for qubit systems, and Bennett and co-workers proposed a scheme \cite{bennettdistillation} whereby if the two spatially separated parties, Alice and Bob, share a supply of many copies of a mixed and weakly entangled state, which had a fidelity greater than $1/2$ with respect to a maximally entangled state, they could
 \emph{distill} a small number of copies of states arbitrarily close to a maximally entangled state  by two-qubit operations, measurement and classical communication. 

Clearly, the quantum systems most appropriate for transmission over long distances are light pulses. Light pulses can embody qubits in, for example, their polarisation, and schemes for the implementation of the quantum gates needed for entanglement distillation have been proposed \cite{klm}. However, these schemes are very technically challenging and in particular, a full scheme would require detectors with close to unit efficiency, which are so far unavailable. So far, therefore, only a few steps of such a scheme have been implemented \cite{pandistillation}. 

Of course, light pulses can also distribute entangled Gaussian states, which we have seen in the previous chapter, have many applications in quantum information science. It is thus natural to ask whether one can distill Gaussian states. In particular, since Gaussian operations are so experimentally accessible, it would be ideal to have an entanglement distillation protocol for Gaussian states which employed solely these operations.

\section{Distillation with Gaussian operations}

With the formal description of Gaussian operations, which we have outlined above, the question of finding a Gaussian entanglement distillation procedure becomes an matter of matrix analysis. This  was carried out in \cite{impnogauss}, \cite{jaromirnogauss} and \cite{giedkenogauss}. Their proofs are too technical to sketch here, but, in light of what \emph{is} possible with Gaussian operations alone, the result is surprising. If one acts with solely Gaussian operations and Gaussian measurements on a supply of Gaussian states it is impossible to distill any greater entanglement than the initial states already possess. Any joint operations, measurements and classical communication which Alice and Bob implement will leave them with a state with either the same or less entanglement than  the initial states.

This seems a disappointing result, but there is a loophole, that means that optical distillation of Gaussian states is not entirely ruled out. There is one operation available to the linear optics lab, that is not a Gaussian operation. This is photo-detection. In the next section, we will show that even very simple protocols using passive linear optics and photo-detection can allow Alice and Bob to generate states more highly entangled than their initial supply, but the states will no longer be Gaussian.

\section{Procrustean entanglement enhancement with linear optics and photo-detection}\label{sec:procmeth}

\begin{figure}
\vspace{1cm}
\begin{center}
\includegraphics[width=10cm]{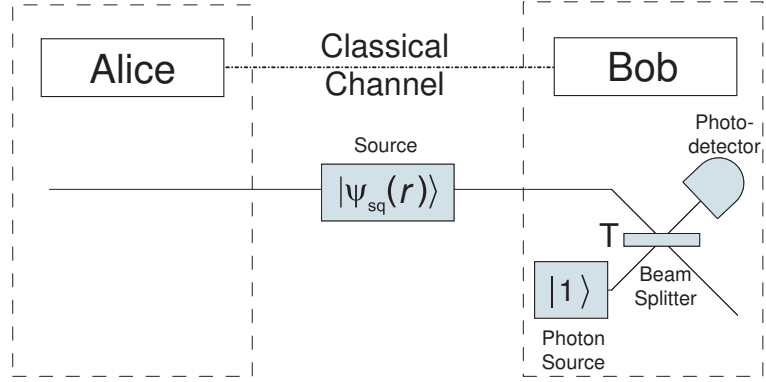}
\end{center}
\caption{\label{fig:proc1} A ``Procrustean'' linear optical entanglement enhancement protocol. Bob mixes his half of a two-mode squeezed state with a single photon on a beam-splitter of transmittivity $T$ and places a photon counter at one of its output ports. }
\end{figure}
Imagine Alice and Bob share a pure two-mode squeezed state with real squeezing parameter $r$ and Bob performs the following operation. He mixes his entangled pulse with a single photon on a beam splitter with transmittivity $T$ as illustrated in figure~\ref{fig:proc1}, and measures with a photo-detector one of the beam splitter outputs. Let us assume for now that the photo-detector works with 100\% efficiency and can resolve any number of photons in the measured mode. It is then a matter of algebra to show that after the measurement, if $m$ photons are registered Alice and Bob's state is collapsed into the following,

\begin{equation}
\ket{\psi}=\sqrt{\frac{1}{\sum_{n=m-1}^\infty|\alpha_n(m)|^2}}\sum_{n=m-1}^\infty \alpha_n \ket{n}_A\ket{n+1-m}
\end{equation}
where
\begin{equation}
\alpha_n(m)=(-\tanh(r))^n T^{n-m}R^{m-1}\left[-R^2\sqrt{{n \choose m}(n+1-m)}+T^2\sqrt{{n \choose m-1}m}\right]
\end{equation}
and we have assumed for simplicity that $T$ and $R$ are real, and we define ${n \choose k}=0$ if $k>n$.

\begin{figure}
\includegraphics[width=14cm]{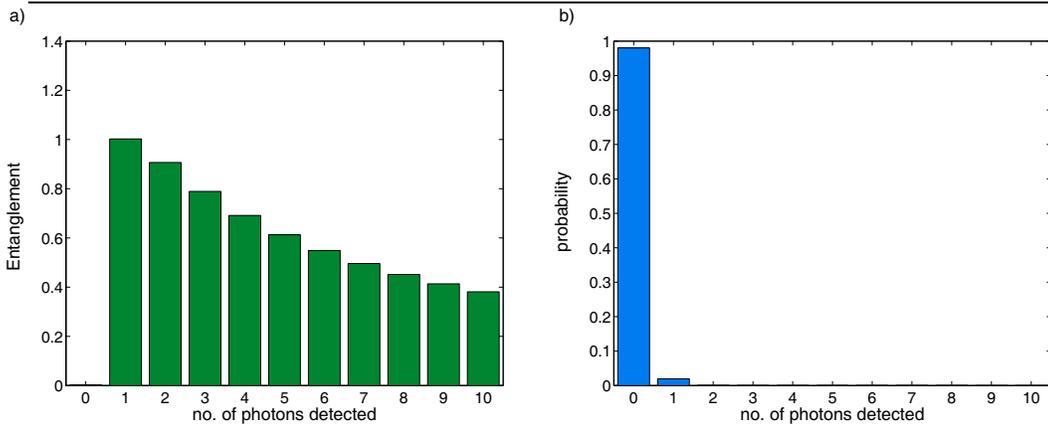}
\caption{\label{fig:procrus1}The pure state entanglement (a) of the state generated and  the event's probability (b), when a certain photon-number is detected in the simple protocol described in section~\ref{sec:procmeth}. Alice and Bob share on a initial two-mode squeezed state with squeezing parameter $\tanh(r)=0.1$ and a beam-splitter of transmittivity $T=\tanh(r)=0.1$ is used. The probabilities which are close to zero and can thus not be read clearly on the graph decrease exponentially, e.g. for 2-photons the probability is $3\times10^{-4}$ and for 3-photons the probability is $4\times10^{-6}$. For comparison, the pure state entanglement of the input state is $0.08$.}
\end{figure}

As this is a pure state, one can immediately calculate the entanglement shared by Alice and Bob in terms of the von Neumann entropy of the reduced density matrix $\rho_A$, $E_{vN}=-\textrm{Tr}[\rho_A \log_2 \rho_A]$. The analytic form of this does not give us much insight, so we shall plot the entanglement of the state generated when $m$ photons are detected, and the probability that this will occur (given a perfect detector) in figure~\ref{fig:procrus1}.

This shows that with this very simple protocol employing the non-Gaussian operation of photo-detection, a huge gain in entanglement is achievable, if at a very low probability. 
The effect of the beam-splitter and the photo-detector is to single out a particular  decomposition of the input state, of which the part corresponding to zero-photons in the detector has a high amplitude and  very weakly entanglement, and the 1-photon part has low amplitude and high entanglement. 
For the low values of $r$ which are experimentally attainable, the probability of a two-or-more photon detection event is low, however, the beam splitter parameters can always be chosen such that, as in figure~\ref{fig:procrus1} the one-photon detection event corresponds to the most highly entangled state.

The fact that a degree of entanglement of almost unity is obtained in the above example is not accidental, since the value of $T$ was been chosen to ensure that the state generated in close to a  maximally entangled two qubit state in the $\ket{0}$ and $\ket{1}$ photon basis. If one chooses $T\approx \tanh(r)/\lambda$ for some value $\lambda$, then the state generated conditional on one photon being detected is approximately $\ket{0}\ket{0}-\lambda\ket{1}\ket{1}$, the approximation being more and more valid as the original squeezing $r$ decreases. Thus the above method generates approximate Bell states, conditionally on a particular measurement outcome being obtained. In the next section we shall see an important and relevant use for such states.

So far we have assumed perfect conditions, which we shall now relax. Firstly, we have assumed that the detector is both 100\% efficient and resolves different photon numbers. However, since the probability of a two-or-more photon detection event occurring is so small, when the detector does click it can be assumed that only one photon was present, thus the use of an inefficient detector which merely registers the presence or absence of photons will only minimally affect the state produced, leading to a tiny admixture of the states from these multi-photon detection events. 

Secondly, if Alice also possesses such a photo-detector, Bob does not need a single-photon source, which are a great technological challenge. If the pair is willing to sacrifice some entangled pairs, Bob need only wait till Alice detects a single photon on her side of a pair and this will project Bob's counterpart into a single photon which he can use.

\begin{figure}
\includegraphics[width=15cm]{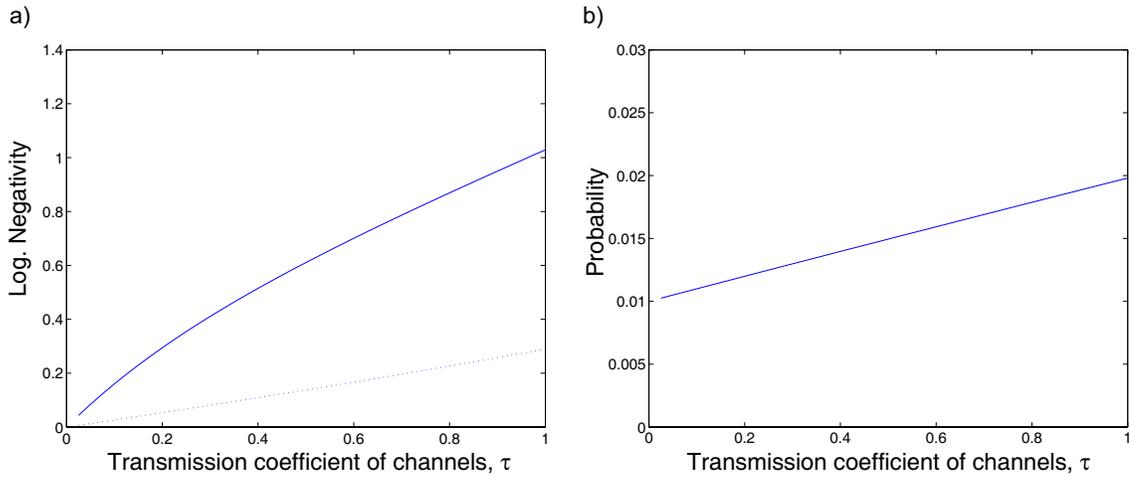}
\caption{\label{noisyproc} In this figure we see the result of the procrustean entanglement concentration  protocol on a mixed input state which is the result of a two-mode squeezed state with squeezing $\tanh(r)=0.1$  being passed through absorbing channels with transmission coefficient $\tau$. The beam splitter transmittivity is chosen to be $T=0.1$,  the same as the plot in figure~\ref{fig:procrus1}. In plot (a) the entanglement of the generated state is plotted alongside the entanglement of the input state (dotted line) as a function of $\tau$. In plot (b) we see that the success probability has decreases weakly and linearly with decreasing $\tau$.}
\end{figure}

Most importantly, if the protocol were realised between distant laboratories, the light pulses will have passed to Alice and Bob via a channel which will at the very least absorb some of the signal decreasing the entanglement and increasing the entropy of the initial state available to Alice and Bob. Thus it is natural to ask whether this protocol can give a similar boost in entanglement for  a mixed input state. We consider here the state described by the covariance matrix in equation~(\ref{eq:noisytmss}). The matrix elements of the state generated by the Procrustean protocol, in the case when one and only one photon is incident on the detector may be calculated analytically, and the entanglement of the state, in terms of its logarithmic negativity thus calculated.  In figure~\ref{noisyproc} we plot the entanglement of the state before and after a successful implementation of the protocol. We see that even for these mixed states, this extremely simple procedure can provide a significant increase in the entanglement.

It will be convenient for the next chapter to write out the zero-photon and one-photon matrix elements of the (unnormalised) state $\rho$ generated here.
\begin{equation}
\rho_{0,0,0,0}=1
\end{equation}
can be set for convenience since $\rho_{0,0,0,0}>0$.
\begin{equation}
\rho_{1,1,0,0}=\rho_{0,0,1,1}=\lambda\rho_{0,0,0,0}
\end{equation}
\begin{equation}
\rho_{1,1,1,1}=\lambda^2\rho_{0,0,0,0}\left(1+(\tau-1)\tanh^2(r)\right)
\end{equation}
\begin{equation}
\rho_{0,1,0,1}=\epsilon\lambda^2\rho_{0,0,0,0}\left(1-(\tau-1)\tanh^2(r)\right)
\end{equation}
where 
\begin{equation}
\lambda=\frac{(2T^2-1)\tanh(r)\tau}{T(\tanh^2(r)(\tau-1)^2)-1}
\end{equation}
and $\epsilon=(1-\tau)/\tau$. All other zero and one photon matrix elements are zero.
Note that when $r$ is very small, these matrix elements depend only on the two parameters $\epsilon$ and $\lambda$, and furthermore for any given values of $r$ and $\tau$, $\lambda$ may be set by choosing the appropriate value of $T$.

The first investigation of local measurement-based protocols for probabilistically increasing the entanglement of a state can be found in \cite{bennettprocust} who coined the term ``Procrustean'' for simple measurement-based protocols on a single copy of the state\footnote{Procrustes was a villain in Greek legend, who would invite travellers to sleep in his iron bed, promising that it would be a perfect size for them. However, this perfect fit was achieved by  shorter guests being stretched or chopping off the legs of taller ones! The entanglement concentration methods here are Procrustean, in the sense that the projection, corresponding to the successful measurement outcome, ``chops off'' the more weakly entangled portion of the state leaving a more highly entangled state behind.}.
The protocols described here were introduced by us in \cite{brownegauss} and \cite{jensgauss}, but a more comprehensive investigation of similar ideas can be found in \cite{billproc1,billproc2}. 
Since any kind of entanglement concentration or distillation is impossible with Gaussian operations, it is perhaps surprising how well these very simple Procrustean protocols with a single photo-detection as their only non-Gaussian element can enhance the entanglement of their input states, both pure and mixed.
Theses states however, are manifestly non-Gaussian and not useful for the applications described in section~\ref{sec:gaussinqip}. It would therefore appear, at first sight, that these protocols still do not aid us in the search for an entanglement distillation procedure for Gaussian states. In the next chapter, however, we introduce a simple linear optical protocol called ``Gaussification''. We will show that this allows Alice and Bob, via local linear optics operations, measurements and classical communication only, to convert entangled non-Gaussian states into Gaussian states and also, often, further increase the entanglement in the process.


%% file: c-8.tex
\chapter{``Gaussification'' with Linear Optics}\label{c-8}

\section{Introduction}

In this section we introduce an iterative protocol, that we shall call ``Gaussification'' \cite{brownegauss,jensgauss}. This protocol acts on a supply of  many copies of a non-Gaussian two-mode entangled state, under  conditions which we describe below. After several non-deterministic iterations, it produces a small number of states that are arbitrarily close to Gaussian states and which are often more entangled than the input. 
First we will outline the protocol and then, under the assumption that all operations are ideal, we will consider analytically all its important properties. In the final section of this chapter we will  model how the procedure would be affected by the most important experimental imperfections of a real implementation of the scheme.

\begin{figure}
\includegraphics[width=14cm]{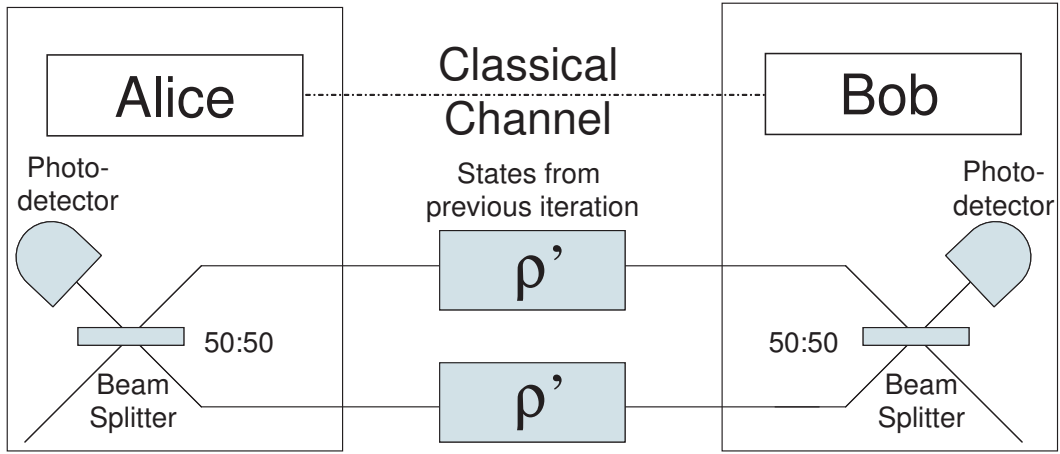}
\caption{\label{gausfig}A single iteration of the ``Gaussification'' protocol.  }
\end{figure}

 Each step of the protocol proceeds as illustrated in figure~\ref{gausfig}. It is an iterative procedure. Each iteration consists of
\begin{enumerate}
\item In the first instance, Alice and Bob act on shared two-mode entangled states supplied from a source. In later iterations they use pairs of states generated by a previous successful iteration.
\item They each mix their half of two copies of the state on a 50:50 beam splitter.
\item They measure one of the outputs with a detector which  distinguishes the presence from the absence of photons.
\item If no photons are registered, the iteration was a success, and the resultant state is saved for the next iteration.
\end{enumerate}

\begin{figure}
\psfrag{en}{\large $E_\mathcal{N}$}
\psfrag{lam}{\large $\eta$}
\begin{center}
\includegraphics{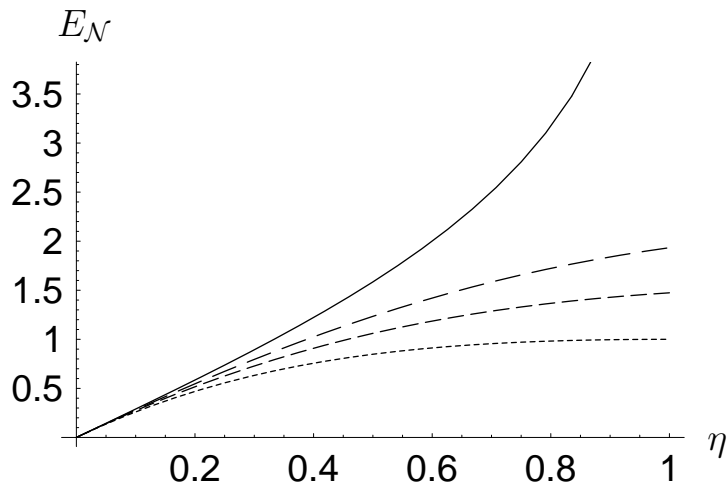}
\end{center}
\caption{\label{fig:pureentt}This figure depicts the logarithmic negativity $E_\mathcal{N}$ as a function of $\eta$ for, i) the state $\ket{\psi}=\sqrt{1/(1+\eta^2)}(|0,0\rangle+\eta|1,1\rangle)$ used as input supply for the Gaussification (dotted line), ii) the state generated by a single (dashed) and two iterations (wider dashing) of the protocol and iii) the limiting two-mode squeezed state which, in the ideal case, will be reached after many iterations.
We use the logarithmic negativity here for ease of comparison with the mixed state cases later in the chapter. }
\end{figure}

\begin{figure}
\psfrag{pr}{\large $P$}
\psfrag{lam}{\large $\eta$}
\begin{center}
\includegraphics{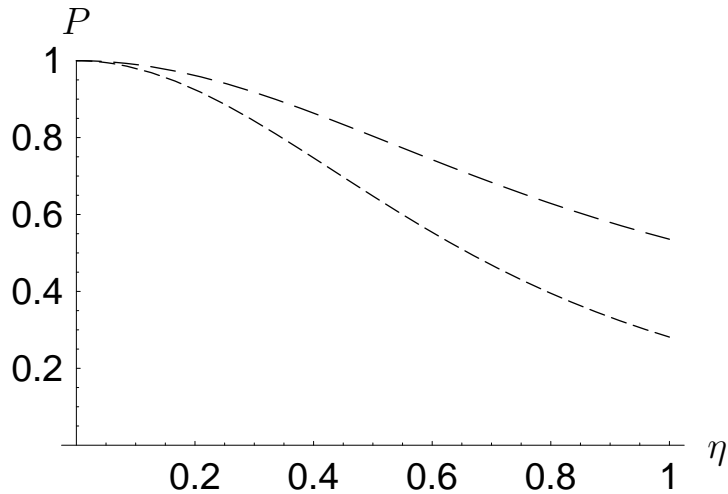}
\end{center}
\caption{\label{fig:pureprob} Here we see the success probability $P$ for the first (dashed) and second (wider dashing) iterations of the protocol with input supply $\ket{\psi}=\sqrt{1/(1+\eta^2)}(|0,0\rangle+\eta|1,1\rangle)$.}
\end{figure}

Before we investigate the properties  of this protocol for  arbitrary input states, let us illustrate it using the simple input state $\ket{\psi^{(0)}}=\ket{00}+\eta\ket{11}$. We label the two entangled modes 1 and 2 and use the labels $A$ and $B$ to denote Alice and Bob. Thus the initial state of the two copies of $\ket{\psi^{(0)}}$ is
\begin{equation}\begin{split}
\ket{\psi^{(0)}}\otimes\ket{\psi^{(0)}}=&\ket{0}_{A1}\ket{0}_{B1}\ket{0}_{A2}\ket{0}_{B2}+\eta \ket{1}_{A1}\ket{1}_{B1}\ket{0}_{A2}\ket{0}_{B2}\\&\mbox{}+\eta\ket{0}_{A1}\ket{0}_{B1}\ket{1}_{A2}\ket{1}_{B2}+\eta^2\ket{1}_{A1}\ket{1}_{B1}\ket{1}_{A2}\ket{1}_{B2}\ .
\end{split}
\end{equation}
The 50:50 beam splitter transformation transforms the state,
\begin{equation}\begin{split}
\ket{\psi^{(0)}}\rightarrow&\ket{0}_{A1}\ket{0}_{A2}\ket{0}_{B1}\ket{0}_{B2}+\frac{\eta}{2} \bigl(\ket{1}_{A1}\ket{0}_{A2}+\ket{0}_{A1}\ket{1}_{A2}\bigr)\bigl(\ket{1}_{B1}\ket{0}_{B2}+\ket{0}_{B1}\ket{1}_{B2}\bigr)\\&\mbox{}+\frac{\eta}{2} \bigl(-\ket{1}_{A1}\ket{0}_{A2}+\ket{0}_{A1}\ket{1}_{A2}\bigr)\bigl(-\ket{1}_{B1}\ket{0}_{B2}+\ket{0}_{B1}\ket{1}_{B2}\bigr)\\&\qquad\mbox{}+\frac{\eta^2}{2} \bigl(-\ket{2}_{A1}\ket{0}_{A2}+\ket{0}_{A1}\ket{2}_{A2}\bigr)\bigl(-\ket{2}_{B1}\ket{0}_{B2}+\ket{0}_{B1}\ket{2}_{B2}\bigr)
\end{split}
\end{equation}
so that, conditional on a measurement of zero photons in modes $A2$ and $B2$, the state is transformed to

\begin{equation}
\ket{\psi^{(1)}}=\ket{0}_{A1}\ket{0}_{B1}+\eta\ket{1}_{A1}\ket{1}_{B1}+\frac{\eta^2}{2}\ket{2}_{A1}\ket{2}_{B1}\ .
\end{equation}
The probability of obtaining these measurement outcomes is $(1+\eta^2+\eta^4/4)/(1+\eta^2)^2$, which for $\eta=0.5$ is approximately  0.8. We observe that the state remains a superposition of states of equal photon number and also that the ratio between the coefficients of $\ket{00}$ and $\ket{11}$ has remained constant. It is simple to calculate the state produced after a further iteration. In figures~\ref{fig:pureentt} and~\ref{fig:pureprob} we plot the entanglement of the state produced after the 1st and 2nd iterations, and the success probability each time. We see that for all values of $\eta$ each iteration generates a state with a higher degree of entanglement. Furthermore, the success probability for each iteration for all values of $\eta$ is high, staying between 25\% and 100\%.


To see if these features apply to other input states, we now consider inputs of the more  general form $\ket{\psi^{(0)}}=\sum_n \alpha_n^{(0)} \ket{n,n}$. States generated by successful iterations retain this form. Let us   introduce the notation $\ket{\psi^{(i)}}=\sum_n \alpha_n^{(i)} \ket{n,n}$ for the state generated after $i$ successful iterations of the Gaussification protocol. By writing out the action of the beam splitter and measurement on this general state one finds a recurrence relation relating the coefficients of the $(i+1)$th states to the $i$th,

\begin{equation}
\label{eq:recursion}
    \alpha_{n}^{(i+1)}= 2^{-n}\sum_{r=0}^{n} {n \choose r}
    \alpha_{r}^{(i)} \alpha_{n-r}^{(i)} \ .
\end{equation}

This relation allows us to investigate the effect of many iterations of the protocol. First, we can ask which states remain unchanged by the scheme. Let us label such a state $\ket{\psi}^{(\infty)}$ and its coefficients  $\alpha_{n}^{(\infty)}$. These must be solutions to the equations
\begin{equation}\label{eq:fixedschmidt}
    \alpha_{n}^{\infty}= 2^{-n}\sum_{r=0}^{n} {n \choose r}
    \alpha_{r}^{(\infty)} \alpha_{n-r}^{(\infty)} 
\end{equation}
for all $n$. We solve these equations  inductively. The 0th equation gives us $\alpha_0^{(\infty)}=(\alpha_0^{(\infty)})^2$, therefore $\alpha_0^{(\infty)}=1$. Of course, since the states are not normalised, this alone does not specify the state at all. Setting $\alpha_0^{(\infty)}=1$, the $n=1$ equation gives us $\alpha_1^{(\infty)}=\alpha_1^{(\infty)}$. We therefore have a free choice for this parameter, and set it to $\zeta$. Proceeding we find that $\alpha_2^{(\infty)}=\zeta^2$ and  $\alpha_3^{(\infty)}=\zeta^3$, and this pattern continues. In fact the general solution to equations~\eqref{eq:fixedschmidt} is $\alpha_n^{(\infty)}=\zeta^n$, which can be verified immediately by substitution. This state is nothing other than the two-mode squeezed state with squeezing $\tanh(r)=-\zeta$.

Since this family of states are the only fixed points of the iteration,  starting with a supply of states in the form, $\ket{\psi^{(0)}}=\sum_n \alpha_n^{(0)} \ket{n,n}$ will lead after many iterations to a two-mode squeezed state, a Gaussian state. We can investigate the convergence via recurrence relations~\eqref{eq:recursion}. For convenience, let us adopt a normalisation convention where $\alpha^{(0)}_0=1$. As we have already seen, this means that $\alpha_0^{(i)}$ remains at this value. Similarly, in this normalisation convention $\alpha_1^{(i)}$ also remains at its initial value. We prove the convergence of the remaining coefficients by induction. We first postulate that each coefficient will converge to the value $\alpha_n^{(\infty)}$, and for that aim introduce deviations form this final value $\delta_n^{(i)}=\alpha_n^{(i)}-\alpha_n^{(\infty)}$. We can now rewrite equation~(\ref{eq:recursion})
\begin{equation}
\delta_n^{(i+1)}=2^{-n}\sum_{r=0}^n{n \choose r}\left( \delta_r^{(i)} \delta_{n-r}^{(i)}- \delta_r^{(i)}\alpha_{n-r}^{(\infty)}- \delta_{n-r}^{(i)}\alpha_{r}^{(\infty)}\right)\ .
\end{equation}
Let us take the limit that $i$ goes to infinity and assume that all coefficients $\alpha_m^{(i)}$ for $m<n$ converge to $\alpha_m^{(\infty)}$ in this limit.

\begin{equation}
\lim_{i\to\infty}\left(\delta_{n}^{(i+1)}\right)=2^{(1-n)}\lim_{i\to\infty}\left(
\delta_{n}^{(i)}\right)
\end{equation}
Thus as long as $2^{1-n}<1$, that is as long as $n>1$, if all coefficients $\alpha_m^{(i)}$ converge to $\alpha_m^{(\infty)}$ for $m<n$,  $\alpha_m^{(i)}$ will converge to  $\alpha_m^{(\infty)}$ in the limit that $i$ tends to infinity. Since we have already shown that $\alpha_0^{(i)}$ and $\alpha_1^{(i)}$ converge, the convergence of the other coefficients follows by induction. This means that formal  convergence in the coefficients occurs for all input states. However not all these final states correspond to physical states. In particular, when $\zeta\ge1$, the state $\sum_n\zeta^n\ket{n,n}$ is unnormalisable, and therefore not a physical state. Thus we conclude the following; for an large input supply of states in the form $\ket{\psi^{(0)}}=\sum_n \alpha_n^{(0)} \ket{n,n}$, repeated iteration of successful runs of the Gaussification protocol lead to the generation of a small number of two-mode states with squeezing $\tanh(r)=-\alpha_1^{(i)}/\alpha_0^{(i)}$ as long as $\alpha_1^{(i)}/\alpha_0^{(i)}<1$, otherwise the procedure does not converge.

\begin{figure}[th]
\vspace{0.5cm}

\hspace{2cm}(a)

\centerline{
\includegraphics[width=6.5cm]{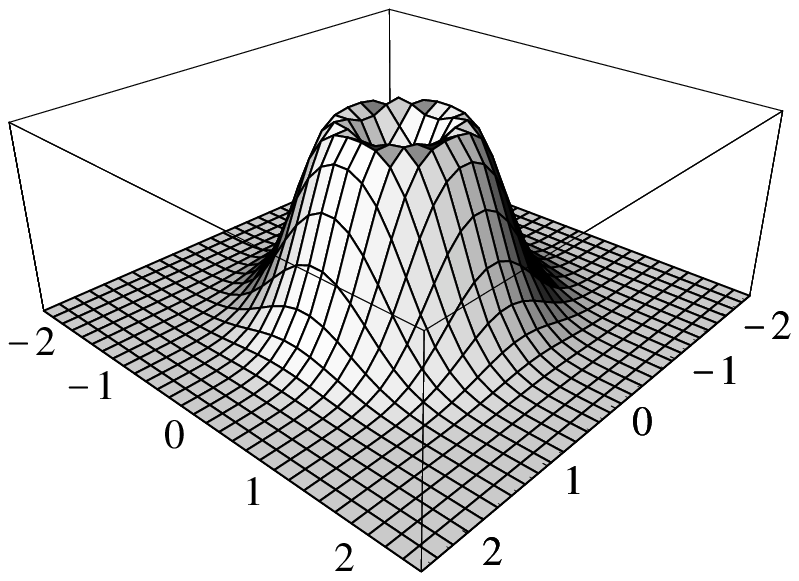}
}
\hspace{2cm}(b)

\centerline{
\includegraphics[width=6.5cm]{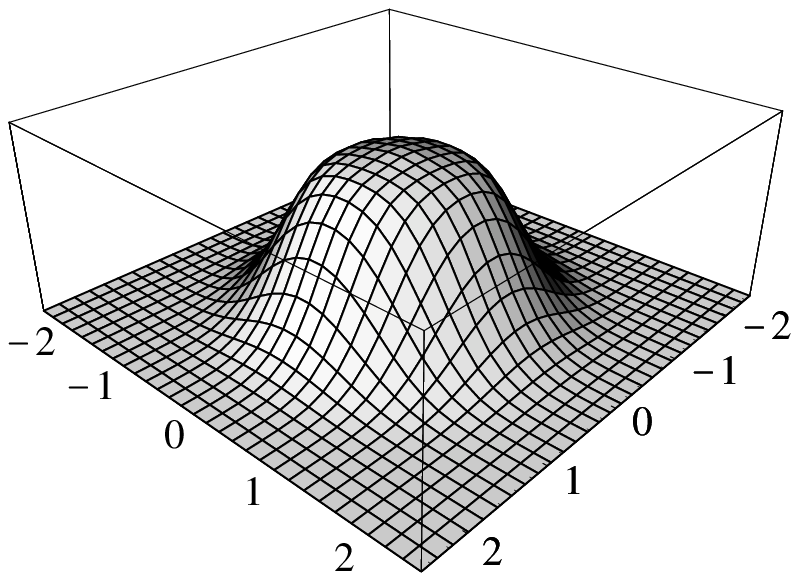}
}
\hspace{2cm}(c)

\centerline{
\includegraphics[width=6.5cm]{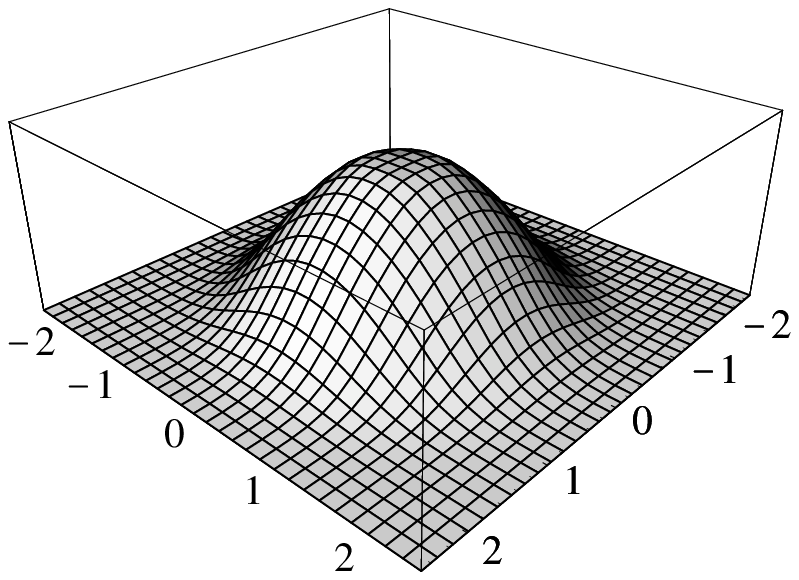}
}

\vspace{.2cm}
\caption{\label{fig:Wigner} The Wigner function of the reduced
state of  one system (a) initially and after (b) one and (c) two steps for the
initial pure state $\ket{\psi}\propto\ket{0,0}+0.6\ket{1,1}$.}
\end{figure}

We can illustrate the way the state converges toward Gaussian form by plotting the Wigner function of the single-mode states of the reduced
state of one mode alone for the initial state with parameter $\eta$ set to 0.6, and the state
after one and two steps.
The Wigner function 
is the Fourier
transform of the characteristic function,
\begin{equation}
        W(\xi) = \frac{1}{(2\pi)^2} \int e^{i \xi^T \Sigma \eta} \chi(\eta) d^2 \eta\ .
\end{equation}
Figure~\ref{fig:Wigner} shows the Wigner function of the Bob's reduced state initially and after one and two iterations of the protocol. While initially, the Wigner function is
far from being a Gaussian in phase space, its non-Gaussian features
disappear quickly.

\section{General input states}\label{sec:geninput}

In the previous section we  illustrated the Gaussification procedure with  simple pure states as input. Now we turn to input states which are completely general  pure or mixed states $\rho^{(0)}$. We will proceed in the same line of  analysis which was applied in the previous section. We will thus prove that;
\begin{enumerate}
\item All two-mode Gaussian states with zero displacement vector are invariant under the Gaussification iteration.
\item Beginning with an arbitrary input supply, the protocol will either lead to convergence to a particular Gaussian state or it will not converge.
\item There are analytic forms for the convergence and the limiting states for a given input $\rho^{(0)}$ in terms of its matrix elements.
\end{enumerate}

The proof that all two-mode Gaussian states are invariant can be achieved in a straight-forward manner, using the covariance matrix formalism. Let the covariance matrix of $\rho^{(0)}$ be $\Gamma$. The covariance matrix of Alice and Bob's pair of two-mode states will be $\Gamma\oplus\Gamma$, where the modes are ordered $A1$, $B1$, $A2$, $B2$. The symplectic 50:50 beam splitter operation $S$ is
\begin{equation}
S=\sqrt{\frac{1}{2}}\left(\begin{array}{cc} \iden_2 & -\iden_2\\ \iden_2 & \iden_2
  \end{array}\right)\ .
\end{equation}

Since the beam splitter operation will be applied to modes in pairs $A1$, $A2$ and  $B1$, $B2$ (not the above order) we also need a swap operation $\Lambda$ 
\begin{equation}
\Lambda=\left(\begin{array}{cc} 0 & \iden_2\\ \iden_2 & 0
  \end{array}\right)\ .
\end{equation}
Thus the application of beam splitters to $\rho^{(0)}$ can be expressed
\begin{equation}\begin{split}
&\Gamma\oplus\Gamma \rightarrow(\iden_2\oplus\Lambda\oplus \iden_2)(S\oplus S)(\iden_2\oplus\Lambda\oplus \iden_2)\Gamma\oplus\Gamma(\iden_2\oplus\Lambda\oplus \iden_2)(S\oplus S)^T(\iden_2\oplus\Lambda\oplus \iden_2)\\&\qquad\qquad=\Gamma\oplus\Gamma\ .
\end{split}
\end{equation}
Thus the action of simultaneously mixing modes $A1$ and $A2$, and $B1$ and $B2$ leaves the covariance matrix of the states invariant. 

Now let us look at how displacement vector is transformed. Let $\vek{d}$ be the displacement vector of $\rho^{(0)}$. The beam splitters transform $\vek{d}$ thus,
\begin{equation}
\vek{d}\oplus\vek{d} \rightarrow (\iden_2\oplus\Lambda\oplus \iden_2)(S\oplus S)(\iden_2\oplus\Lambda\oplus \iden_2)\vek{d}\oplus\vek{d}=\vekk{0}\oplus\sqrt{2}\vek{d}
\end{equation}
where $\vekk{0}$ is the zero vector.
After the measurement of modes $A2$ and $B2$, the resultant state is a Gaussian state with the same covariance matrix as $\rho^{(0)}$ and zero displacement vector. This proves the first of our assertions above.

To analyse the convergence properties of the protocol for a general input we proceed as above, constructing a recurrence relation for the matrix elements of the state $\rho^{(i)}$ produced after $i$ successful iterations;
\begin{equation}\label{recurrence1}
        \rho_{a,b;c,d}^{(i+1)} =
        \sum_{s=0}^a
        \sum_{t=0}^b
        \sum_{n=0}^c
        \sum_{m=0}^d
        M_{a,b;c,d}^{s,t;n,m} 
        \rho^{(i)}_{s,t;n,m} \rho_{a-s,b-t;c-n,d-m}^{(i)}\ ,
\end{equation}
with coefficients $M_{a,b;c,d}^{s,t;n,m}$ given by
\begin{equation}\label{nl1}
        M_{a,b;c,d}^{s,t;n,m}  =
        2^{-(a+b +c+d)/2} (-1)^{(a+b+c+d)-(s+t+n+m)}
        \left[
        \binom{a}{s}
        \binom{b}{t}
        \binom{c}{n}
        \binom{d}{m}
        \right]^{1/2}\ .
\end{equation}
We first observe that all matrix elements $\rho_{a,b,c,d}^{(i)}$ where $(a+b+c+d)$ is odd will vanish after the first iteration, since the terms in the sum of equal amplitude pick up opposite signs and cancel out.
Following the notation which we used above, we label a state which is invariant under this map $\rho^{(\infty)}$. We find that either  $\rho^{(\infty)}_{0,0,0,0}=1$ or $\rho^{(\infty)}_{0,0,0,0}=0$. In this latter case, all other matrix elements are also zero, so we ignore this trivial non physical case.  If we  normalise the state such that $\rho^{(\infty)}_{0,0,0,0}=1$, the following six coefficients of $\rho^{(\infty)}$ are completely free; $\rho_{1,0,1,0}^{(\infty)}$,  $\rho_{0,1,0,1}^{(\infty)}$, $\rho_{1,0,0,1}^{(\infty)}$, $\rho_{2,0,0,0}^{(\infty)}$, $\rho_{1,1,0,0}^{(\infty)}$. One can then show by induction that once these 10 parameters (since two of the elements are real) are chosen, all other matrix elements are already determined. In fact $\rho_{a,b,c,d}$ is a polynomial in these elements of degree $D\le \textrm{Max}\{a,b,c,d\}$.

We now need to show that all of these fixed points correspond, if they correspond to physical states, to zero-displacement Gaussian states whose covariance matrices are uniquely specified by these 10 parameters.
To do this, we calculate the relevant matrix elements for a general zero-displacement Gaussian state. We start with equation~(\ref{gausstaterho}), substituting the relevant characteristic function,

\begin{equation}
\rho=\frac{1}{(2\pi)^2}\int\exp[-\frac{1}{4} \vekk{\xi}^T\Sigma \Gamma \Sigma^T \vekk{\xi}]
\hat{W}_\vekk{\xi}
d^{2n}\xi \ .
\end{equation}
The matrix elements of the Weyl operator can be calculated via its normal ordered form \cite{cahillglauber}. This gives the following,
\begin{equation}
\begin{split}
&\langle m|W(\vekk{\xi})|n\rangle=\\&\left(\frac{n!}{m!}\right)^{\frac{1}{2}}e^{-\frac{(\xi_1^2+\xi_2^2)}{4}}\left(\frac{\xi_1+i\xi_2}{\sqrt{2}}\right)^{m-n}L^{(m-n)}_{n}\left[\frac{\xi_1^2+\xi_2^2}{2}\right]
\end{split}\end{equation}
where $L^{(i)}_{j}[x]$ are associated Laguerre polynomials.

We can thus write the matrix element $\rho_{a,b,c,d}=\langle a,b|\rho|c,d\rangle$

\begin{equation}
\begin{split}
\rho_{a,b,c,d}=&\frac{1}{(2\pi)^2}\left(\frac{c!d!}{ a!b!}\right)^{\frac{1}{2}}\int d^{4}\xi\exp\left[-\frac{1}{4} \vekk{\xi}^T\Sigma (\Gamma+\iden) \Sigma^T \vekk{\xi}\right]
\left(\frac{\xi_1+i\xi_2}{\sqrt{2}}\right)^{a-c}\\&\left(\frac{\xi_3+i\xi_4}{\sqrt{2}}\right)^{b-d}L^{(a-c)}_{c}\left[\frac{\xi_1^2+\xi_2^2}{2}\right]L^{(b-d)}_{d}\left[\frac{\xi_3^2+\xi_4^2}{2}\right]
 \ .
\end{split}
\end{equation}

We first note that all matrix elements $\rho_{a,b,c,d}$, where $(a-c)+(a-d)$ is odd (or equivalently $(a+b+c+d)$ is odd) must vanish as the integrand is then an odd function. The element $\rho_{0,0,0,0}$ may be found by Gaussian integration,
\begin{equation}
\rho_{0,0,0,0}=\frac{1}{(2\pi)^2}\int d^4\xi \exp\left[-\frac{1}{4} \vekk{\xi}^T\Sigma (\Gamma+\iden) \Sigma^T \vekk{\xi}\right]\ .
\end{equation}

The integral converges if $\Gamma+\iden$ is positive, and  by recognising that the integral is invariant under symplectic transformations of $\Gamma+\iden$, in particular symplectic diagonalisation, one can reduce it to a product of standard Gaussian integrals to give,
\begin{equation}
\rho_{0,0,0,0}=\frac{4}{\gamma_1\gamma_1}=\frac{4}{|\Gamma+\iden|^\frac{1}{2}}
\end{equation}
where $\gamma_1$ and  $\gamma_2$ are symplectic eigenvalues of $\Gamma+\iden$ and $|\Gamma+\iden|$ is its determinant. We have used here the fact that the determinant of a matrix is invariant under symplectic operations.
To calculate the remaining matrix elements, 
we multiply out the multinomials in the expression and integrate each individually. Using the identity
\begin{equation}
x^ke^{-ax^2}=\left.\frac{\partial^k }{\partial j^k}e^{-ax^2+jx}\right|_{j=0}
\end{equation}
each term can be re-written as a differential.
We use the following identity to ``complete the square''
\begin{equation}
-\frac{1}{4}\vekk{\xi}^T\Sigma(\Gamma+\iden)\Sigma^T\vekk{\xi}+\vek{j}^T\vekk{\xi}=-\frac{1}{4}\vekk{\xi}'^T\Sigma(\Gamma+\iden)\Sigma^T\vekk{\xi}'+\vek{j}^T\Sigma(\Gamma+\iden)^{-1}\Sigma^T\vek{j}
\end{equation}
where $\vekk{\xi}'=(\vekk{\xi}-2\Sigma(\Gamma+\iden)^{-1}\Sigma^T\vek{j})$
and thus each multinomial term takes the following form,
\begin{equation}
\begin{split}
&\frac{1}{(2\pi)^2}\int d^4\xi \exp[-\frac{1}{4}\vekk{\xi}^T\Sigma(\Gamma+\iden)\Sigma^T\vekk{\xi}]\xi_1^{k_1}\xi_2^{k_2}\xi_3^{k_3}\xi_4^{k_4}\\&=\left[\frac{\partial^{k_1}}{\partial j_1^{k_1}}\frac{\partial^{k_2}}{\partial j_2^{k_2}}\frac{\partial^{k_3}}{\partial j_3^{k_3}}\frac{\partial^{k_4}}{\partial j_4^{k_4}}\right]\int d^4\xi
\exp[-\frac{1}{4}\vekk{\xi}^T\Sigma(\Gamma+\iden)\Sigma^T\vekk{\xi}+\vek{j}^T\vekk{\xi}]\\
&=\frac{4}{|\Gamma+\iden|^{\frac{1}{2}}}\left[\frac{\partial^{k_1}}{\partial j_1^{k_1}}\frac{\partial^{k_2}}{\partial j_2^{k_2}}\frac{\partial^{k_3}}{\partial j_3^{k_3}}\frac{\partial^{k_4}}{\partial j_4^{k_4}}\right]\exp[\vek{j}^T\Sigma(\Gamma+\iden)^{-1}\Sigma^T\vek{j}]\ .
\end{split}
\end{equation}

 Each matrix element can now be calculated by expanding its multinomials and differentiating as appropriate. Introducing for convenience the matrix $B=(\Gamma+\iden)^{-1}$, we find that
\begin{eqnarray}
        \sigma_{1,0,1,0}&=&1-B_{1,1}-B_{2,2}, \nonumber\\
        \sigma_{0,1,0,1}&=& 1-B_{3,3}- B_{4,4}, \nonumber\\
        \sigma_{1,0,0,1}&=& - B_{1,3} - B_{2,4} + i(B_{1,4}- B_{2,3}), \nonumber\\
        \sigma_{2,0,0,0}&=& 2^{-1/2} (-B_{1,1} +B_{2,2}-2 i B_{1,2}), \nonumber\\
        \sigma_{0,2,0,0}&=& 2^{-1/2} (-B_{3,3} + B_{4,4} - 2 i B_{3,4}), \nonumber\\
        \sigma_{1,1,0,0}&=& -B_{1,3} + B_{2,4}-
        i(B_{1,4}+B_{2,3}).\label{en}
\end{eqnarray}
where $\sigma_{a,b,c,d}=\rho_{a,b,c,d}/\rho_{0,0,0,0}$.

Our analysis is now almost complete. We showed above that the most general states which are invariant under an iteration of the Gaussification protocol have null matrix elements $\rho_{a,b,c,d}$  when $(a+b+c+d)$ and are completely specified by ten free parameters, corresponding to the six matrix elements highlighted. We now see that for any choice of these parameters there exists a ``state'' with a Gaussian covariance matrix $\Gamma=B^{-1}-\iden$ (although it may not be a physical state satisfying the condition $\Gamma\ge -i\Sigma$), as long as $B^{-1}$ exists and is positive. This is the case as long as $|B|\neq0$. However, if we recall that $\rho_{0,0,0,0}=4|B|^{1/2}$, we see that $|B|$ is only zero when $\rho_{0,0,0,0}$ is zero, which is only the case when $\rho^{(\infty)}$ is the unphysical null state, so we can discount this.
Thus we have shown that the only states invariant under an iteration of the protocol  are Gaussian states with zero displacement. 

We now consider the criteria for convergence. First we ask, given a particular starting state to what state will it converge? Labelling the matrix elements of the starting state $\rho^{(0)}_{a,b,c,d}$ and choosing a normalisation such that $\rho^{(0)}_{0,0,0,0}=1$, we find from equation~(\ref{eq:recursion}) that the six matrix elements highlighted above converge to their final value after a single iteration;
\begin{eqnarray}
        \rho^{(i)}_{1,0,1,0}&=&\rho^{(1)}_{1,0,1,0}=\rho^{(0)}_{1,0,1,0}-|\rho^{(0)}_{1,0,0,0}|^2 \nonumber\\
        \rho^{(i)}_{0,1,0,1}&=&\rho^{(1)}_{0,1,0,1} = \rho^{(0)}_{0,1,0,1} -|\rho^{(0)}_{0,1,0,0}|^2\nonumber\\
        \rho^{(i)}_{1,0,0,1}&=& \rho^{(1)}_{1,0,0,1}= \rho^{(0)}_{1,0,0,1} -\rho^{(0)}_{1,0,0,0}\rho^{(0)}_{0,0,0,1}\nonumber\\
        \rho^{(i)}_{2,0,0,0}&=&  \rho^{(1)}_{2,0,0,0}= \rho^{(0)}_{2,0,0,0}-(1/\sqrt{2})(\rho^{(0)}_{1,0,0,0})^2 \nonumber\\
        \rho^{(i)}_{0,2,0,0}&=& \rho^{(1)}_{0,2,0,0}=\rho^{(0)}_{0,2,0,0}-(1/\sqrt{2})(\rho^{(0)}_{0,1,0,0})^2 
 \nonumber\\
        \rho^{(i)}_{1,1,0,0}&=& \rho^{(1)}_{1,1,0,0}=\rho^{(0)}_{1,1,0,0}-\rho^{(0)}_{1,0,0,0}\rho^{(0)}_{0,1,0,0}
    \ .\label{firstiter}
\end{eqnarray}

One can show by induction in an analogous way as above for the simple pure state case that the states converge matrix-elementwise after many iterations to the Gaussian state with covariance matrix $B^{-1}-\iden$, where $B$ is calculated from the matrix elements of $\rho^{(1)}$ via the inverse of the relations in equations~(\ref{en}). Although the matrix elements always converge to their respective final values, the state as a whole can only converge if $B^{-1}-\iden$ corresponds to a physical covariance matrix, that is $B^{-1}-\iden\ge -i\Sigma$.

Thus we have proved all the assertions made at the start of this section. Namely that the only states invariant under the protocol are Gaussian states centred in phase space and that for any given starting supply $\rho^{(0)}$,  whether or not, and to which state it will converge  can be calculated analytically from just a small number of matrix elements of $\rho^{(0)}$.

\section{Examples}

\subsection{An example mixed input}
In chapter~\ref{ch:7}, we considered the states generated by a simple optical Procrustean entanglement concentration protocol. In particular we considered the states generated  when the input state was a two-mode squeezed state where both modes had passed through absorbing channels with the same transmission coefficient $\tau$. In the limit that the initial squeezing was small, this process generates a family of entangled mixed states described by two real parameters $\lambda$, which can be take any real value between -1 and 1, by setting the beam splitter transmittivity in the Procrustean protocol to an appropriate value, and $\tau$ with the following matrix elements

\begin{equation}\label{mixedinputrhost}
\rho_{0,0,0,0}=\frac{1}{1+\lambda^2+\lambda^2(1-\tau)/\tau}
\end{equation}
\begin{equation}
\rho_{1,1,0,0}=\rho_{0,0,1,1}=\lambda\rho_{0,0,0,0}
\end{equation}
\begin{equation}
\rho_{1,1,1,1}=\lambda^2\rho_{0,0,0,0}
\end{equation}
\begin{equation}\label{mixedinputrhofin}
\rho_{0,1,0,1}=\frac{1-\tau}{\tau}\lambda^2\rho_{0,0,0,0}\ .
\end{equation}

Let us investigate how such a state is transformed by iterations of the Gaussification protocol. Firstly, by applying the procedure describe in section~\ref{sec:geninput} we calculate the covariance matrix $\Gamma$ of the Gaussian state to which the state element-wise converges

\begin{equation}
\Gamma=\frac{1}{\tau-\lambda^2}\left(\begin{array}{cccc}\tau+\lambda^2(2\tau-1)&0&2\lambda\tau&0\\
0&\tau+\lambda^2(2\tau-1)&0&-2\lambda\tau\\
2\lambda\tau&0&\lambda^2+\tau&0\\
0&-2\lambda\tau&0&\lambda^2+\tau
\end{array}
\right)\ .
\end{equation}

\begin{figure}
\psfrag{en}{\large $E_\mathcal{N}$}
\psfrag{lam}{\large $\lambda$}
\begin{center}
\includegraphics{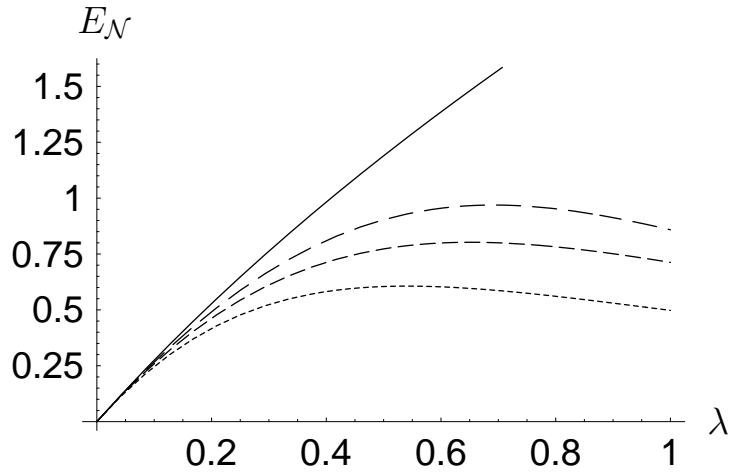}
\end{center}
\caption{\label{fig:t50ent}This figure depicts the logarithmic negativity $E_\mathcal{N}$ as a function of $\lambda$ for, i) the initial state $\rho$ specified in equations~(\ref{mixedinputrhost}) to~(\ref{mixedinputrhofin}) where we set $\tau=0.5$,  ii) the state generated by a single (dashed) and two iterations (wider dashing) of the protocol and iii) the limiting state (solid line). Note that the limiting value only exists for values of $\lambda<\sqrt{\tau}$, since above this value the protocol does not converge.}
\end{figure}

\begin{figure}
\psfrag{S}{\large $S_\textrm{vN}$}
\psfrag{lam}{\large $\lambda$}
\begin{center}
\includegraphics{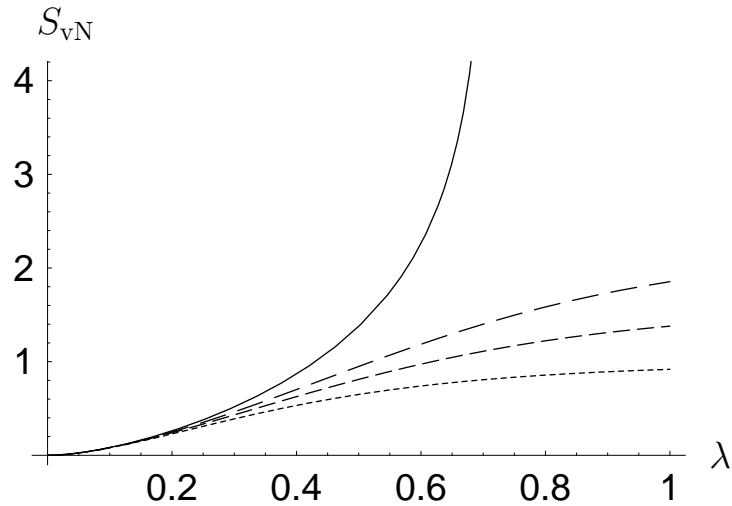}
\end{center}
\caption{\label{fig:t50entropy} Here we plot the von Neumann entropy $S_\textrm{vN}$ for the initial (dotted), first (dashed) and second (wider dashing) iterations and the limiting state (solid line) for the input state $\rho$ specified in equations~(\ref{mixedinputrhost}) to~(\ref{mixedinputrhofin}). We see that the entropy of the limiting state is non-zero and this is reflected in entropy increases at each iteration.}
\end{figure}

\begin{figure}
\psfrag{pr}{\large $P$}
\psfrag{lam}{\large $\lambda$}
\begin{center}
\includegraphics{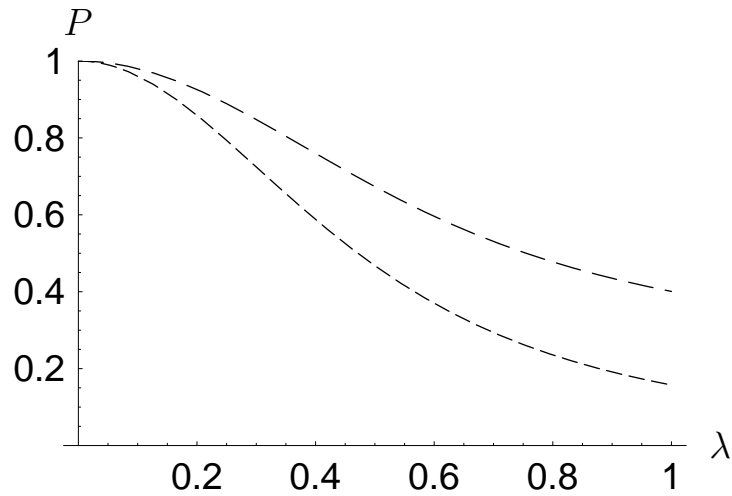}
\end{center}
\caption{\label{fig:t50prob} Here we see the success probability $P$ for the first (dashed) and second (wider dashing) iterations of the protocol with input supply  $\rho$ specified in equations~(\ref{mixedinputrhost}) to~(\ref{mixedinputrhofin}).}
\end{figure}

This covariance matrix only corresponds to a physical state fulfilling $\Gamma+i\Sigma\ge0$ if the condition $\tau>\lambda^2$ is met, thus state-wise convergence can only occur when this is fulfilled. In figure~\ref{fig:t50ent} we plot, as a function of $\lambda$, the logarithmic negativity, $\tau=0.5$ of the input state, the state generated by one and two iterations of the protocol and the limiting state to which the protocol is converging. We see a marked increase in the entanglement after each iteration, in spite of the fact that the input states are not pure. The entropy of these states is depicted in figure~\ref{fig:t50entropy}. We see that the entropy of the final state is, in fact, higher than the starting state, and the entropy increases with each iteration accordingly. In figure~\ref{fig:t50prob} we see the success probability for each iteration. In particular this probability is greater than 50\% for most values of $\lambda$ below the convergence threshold. Comparing this plot to figure~\ref{fig:pureprob} indicates that the success probability is little affected by the mixedness of the input state.

We see a simultaneous increase in  both the entanglement and the von Neumann entropy. This is in  contrast to entanglement distillation protocols for qubits, where an increase in entanglement is usually accompanied by a decrease in entropy, since for any given degree of entropy of the state there is a maximum amount of entanglement which the system may possess \cite{frankandkoenmems,munromems}. This is also true for Gaussian states \cite{italiangaussmems}, although not for  infinite dimensional states in general, but for a given entropy the maximally allowed entanglement is much larger \cite{italiangaussmems}. This explains the seemingly counter-intuitive behaviour seen here. Of course, for application of entangled states  to quantum key distribution, for the highest security and key rate, pure entangled states are desirable. In the next example we investigate the criteria for the convergence of the protocol towards a pure state, and give an example to illustrate this.


\subsection{Convergence to a pure state}

It is a straightforward consequence of the results in section~\ref{sec:geninput} that certain mixed input state supplies will converge to a pure Gaussian state. Any pure two-mode centred Gaussian state $\rho$ may be written $\hat{O}\hat{S}_1(\zeta_1)\hat{S}_2(\zeta_2)\ket{0,0}$ where $\hat{O}$ is a general passive linear two-mode transformation and $\hat{S}_i(\zeta_i)$ is a single mode squeezing operation \cite{wolfentpower}. Since the squeezing operators are quadratic in annihilation and creation operators, matrix elements $\rho_{1,0,1,0}$, $\rho_{0,1,0,1}$ and $\rho_{1,0,0,1}$ are zero. Using the map between these matrix elements and the covariance matrix $\Gamma$ introduced in section~\ref{sec:geninput} we can verify that these conditions alone suffice to ensure that $\Gamma$ has unit determinant. Thus these matrix elements being zero is a  necessary and sufficient condition  ensure that $\rho$ is pure.

Equations~(\ref{firstiter}) and this condition leads directly to the following conditions on the input supply for convergence to a pure Gaussian state,

\begin{equation}\begin{split}
&\rho_{1,0,1,0}^{(0)}=|\rho_{1,0,0,0}^{(0)}|^2/\rho_{0,0,0,0}^{(0)}\\
&\rho_{0,1,0,1}^{(0)}=|\rho_{0,1,0,0}^{(0)}|^2/\rho_{0,0,0,0}^{(0)}\\
&\rho_{1,0,0,1}^{(0)}=\rho_{1,0,0,0}^{(0)}\rho_{0,0,0,1}^{(0)}/\rho_{0,0,0,0}^{(0)}
\end{split}
\end{equation}
provided $\rho_{0,0,0,0}$ and the conditions of convergence described in section~\ref{sec:geninput} are fulfilled.

Let us consider an simple example of a mixed input state which fulfills these criteria.
\begin{eqnarray}
    \rho_{0,0,0,0}&=&1/(1+\varepsilon^2),
    \,\,
    \rho_{1,1,0,0} =\rho_{0,0,1,1} = \varepsilon/(2+2\varepsilon^2)\nonumber\\
     \rho_{1,1,1,1}&=&\varepsilon^2/(1+\varepsilon^2)\label{topure}
\end{eqnarray}
with $\varepsilon\in [0,1)$, and $\rho_{a,b,c,d}=0$ otherwise.

\begin{figure}
\psfrag{en}{\large $E_\mathcal{N}$}
\psfrag{lam}{\large $\varepsilon$}
\begin{center}
\includegraphics[width=10cm]{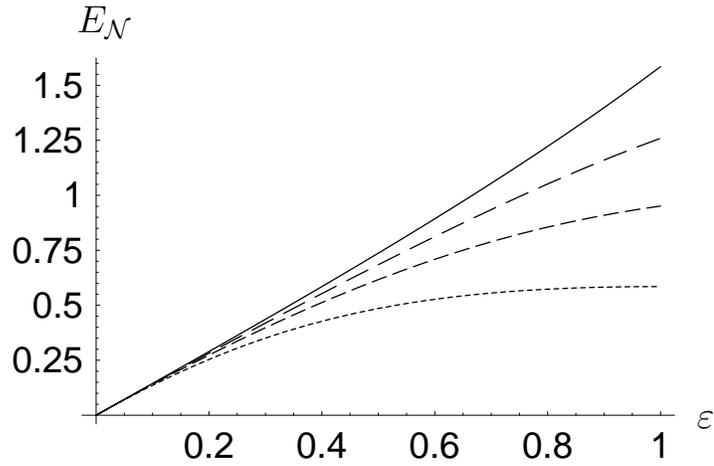}
\end{center}
\caption{\label{fig:topureent}This figure plots the logarithmic negativity $E_\mathcal{N}$ as a function of $\varepsilon$ for, i) the initial state $\rho$ specified in equation~(\ref{topure})  ii) the state generated by a single (dashed) and two (wider dashing) iterations  iii) the limiting pure two-mode squeezed state (solid line).}
\end{figure}

\begin{figure}
\psfrag{S}{\large $S_\textrm{vN}$}
\psfrag{lam}{\large $\varepsilon$}
\begin{center}
\includegraphics[width=10cm]{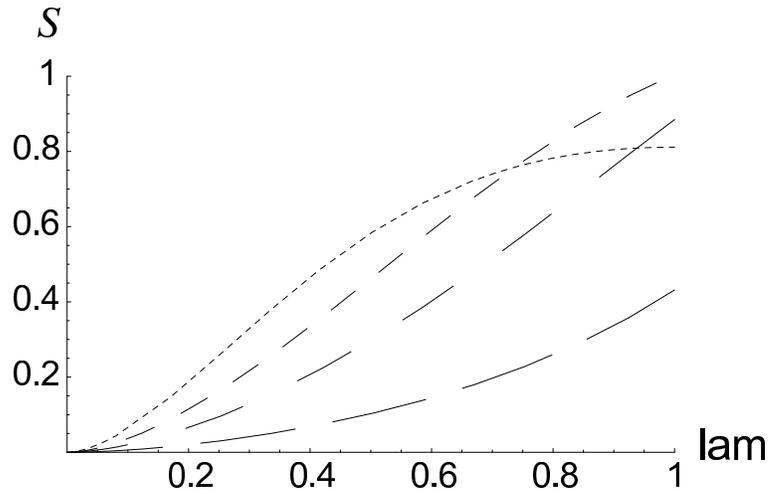}
\end{center}
\caption{\label{fig:topureentropy} Here we plot the von Neumann entropy $S_\textrm{vN}$ for the initial (dotted), first (dashed) and second (wider dashing) and fourth (widest dashing) iterations of the protocol on state $\rho$ described in equation~(\ref{topure}). The limiting state is a pure two-mode squeezed state, the entropy of which vanished.}
\end{figure}

The limiting state, to which repeated iterations of the protocol on this input will tend, is a pure two-mode squeezed state with squeezing $\tanh(r)=-\varepsilon/2$. How the  degrees of entanglement and entropy evolve under repeated iteration of the protocol on this state  are plotted in figures~\ref{fig:topureent} and~\ref{fig:topureentropy}. Although for more highly mixed states the entropy increases after the first iteration, we see that after repeated iterations the entropy decreases for all values of $\varepsilon$, as it tends to zero.


Thus we have demonstrated that under ideal experimental conditions this Gaussification protocol to give a scheme which can distill Gaussian states of higher entanglement, and for certain input states, of lower entropy from a non-Gaussian supply. Combined with a Procrustean entanglement concentration step, as discussed in section~\ref{sec:procmeth},  this gives an  entanglement distillation procedure for Gaussian states based on realistic operations, the only one proposed to date for the continuous variable regime. In the next section we shall continue our analysis of the Gaussification protocol by considering how well it might perform under the imperfect operations which would be implemented in a real laboratory.


%% file: c-9.tex
\section{Imperfect devices}\label{c-9}

In this section we show how the  most likely experimental imperfections affect the performance of the Gaussification protocol, focusing on its ability to increase entanglement. The importance of various imperfections will depend upon the complexity of the implementation. Here we focus on two regimes; firstly the minimal setup implementing a small number of iterations in one shot, and  secondly a full scheme of many iterations which would require the feeding forward of measurement outcomes and the storage of pulses in between iteration rounds. In the minimal setup, there is one obvious imperfection which cannot with current technology be surmounted. That is the fact that photo detectors have finite efficiency, and this is the first imperfection which we will consider. 

For a full implementation, storing the light pulses in between iterations will clearly cause some degradation in their states, and it is these effects which we will focus on in  section~\ref{sec:decoh-betw-iter-1}. In section~\ref{sec:other-potent-probl} we shall discuss other challenges which could affect the scalability of the scheme, such as the difficulty of mode-matching systems with many optical elements.

\subsection{Inefficient detection}

An important experimental imperfection for all implementations is that no photon detector exists which functions with unit efficiency. With current technology the best detection efficiencies at the frequency ranges typically employed in experiment are between 50\% and 60\%. 
This means that for every photon incident upon the detector there is a finite probability $\eta$ that its presence will be registered. Current photo-detectors are also unable to distinguish between different numbers of photons incident in a single pulse, but this is not a problem for the Gaussification protocol which only requires that vacuum is distinguished from the presence of one or more photons. However, with today's inefficient detectors even this ability is clearly diminished and one would expect that a protocol which relies on detecting the vacuum outcome would perform very badly if inefficient detectors are employed.

We model an imperfect detector as follows, the usual projection operator onto the vacuum is replaced by

\begin{equation}
|0\rangle\langle 0|\longmapsto 
\sum_{k=0}^\infty
(1-\eta)^{k}|k\rangle\langle k|,
\end{equation}
where $\eta\in[0,1]$ is the detector efficiency. This formula is equivalent 
to a beam-splitter of 
transmittivity $T=\sqrt{\eta}$ placed in front of a detector 
with unit efficiency.

We investigate how the protocol with imperfect detectors by constructing, as above, a recurrence relation for the matrix elements of the state $\rho^{(i)}$ after the $i$th repetition of the protocol. We find, after some rather tortuous algebraic manipulation, that
\begin{eqnarray}
        &&\rho^{(i+1)}_{A,B,C,D}   =
        \sum_{k,l=0}^\infty
        \sum_{a=0}^{A+k}
        \sum_{b=0}^{B+l}
        \sum_{c=0}^{C+k}
        \sum_{d=0}^{D+l}\\
&\times &
        \sum_{s=0}^{\min\{a,A\}}
        \sum_{u=0}^{\min\{b,B\}}
        \sum_{s'=0}^{\min\{c,C\}}
        \sum_{u'=0}^{\min\{d,D\}}
        \nonumber\\
        &\times &
        N_{A,B,C,D}^{a,b,c,d,s,u,s',u'}
        \rho^{(i)}_{a,b,c,d}\,\, \rho{(i)}_{A+k-a,B+l-b,C+k-c,D+l-d}\ ,\nonumber
\end{eqnarray}
where
\begin{eqnarray}
        N_{A,B,C,D}^{a,b,c,d,s,u,s',u'}    & =&
        2^{-(A+B+C+D+2k+2l)/2}\nonumber\\
        &\times &(-1)^{ A-a+B-b+C-c+D-d+2k +2l} (1-\eta)^{k+l}\nonumber \\
        &\times &\left[
        \binom{a}{s}
        \binom{A}{s}
        \binom{A+k-a}{A-s}
        \binom{k}{a-s}\right]^{1/2} \nonumber\\
&\times &\left[
        \binom{b}{u}
        \binom{B}{u}
        \binom{B+l-b}{B-u}
        \binom{l}{b-u}\right]^{1/2} \nonumber\\
        &\times &\left[
        \binom{c}{s'}
        \binom{C}{s'}
        \binom{C+k-c}{C-s'}
        \binom{k}{c-s'}\right]^{1/2} \nonumber\\
        &\times &\left[
        \binom{d}{u'}
        \binom{D}{u'}
        \binom{D+l-d}{D-u'}
        \binom{l}{d-u'}\right]^{1/2} \ .\nonumber  \\
\end{eqnarray}
The case $\eta=1$ describes a 
detector with unit efficiency, and this expression reduces
to the one described in the previous chapter.

To illustrate the performance of the protocol under various conditions we shall use two particular input states as model inputs throughout this chapter. The first is the pure state  $\ket{\psi_a}=(1/\sqrt{1.25})(\ket{0,0}+0.5\ket{1,1})$ and the second is the mixed state $\rho_b=(1/1.5)(\ket{0,0}\bra{0,0}+0.5(\ket{0,0}\bra{1,1}+\ket{1,1}\bra{0,0})+0.25(\ket{1,1}\bra{1,1}+\ket{0,1}\bra{0,1})$. This is the state in equations~(\ref{mixedinputrhost}) to~(\ref{mixedinputrhofin}) with parameters $\lambda=0.5$ and $\tau=0.5$. Both of these states can be generated by the Procrustean protocol described in section~\ref{sec:procmeth} from a supply of two-mode squeezed states, for $\ket{\psi_a}$ having been distributed through two non-absorbing channel, and for $\rho_b$ through two  channels which absorb 50\% incident intensity.

\begin{figure}
\psfrag{en}{$E_\mathcal{N}$}
\psfrag{eta}{$\eta$}
\psfrag{S}{$S_\textrm{vN}$}
\hspace{-0.5cm}(a)\hspace{8.0cm}(b)
\vspace{-1cm}
\begin{center}
\includegraphics[width=8cm]{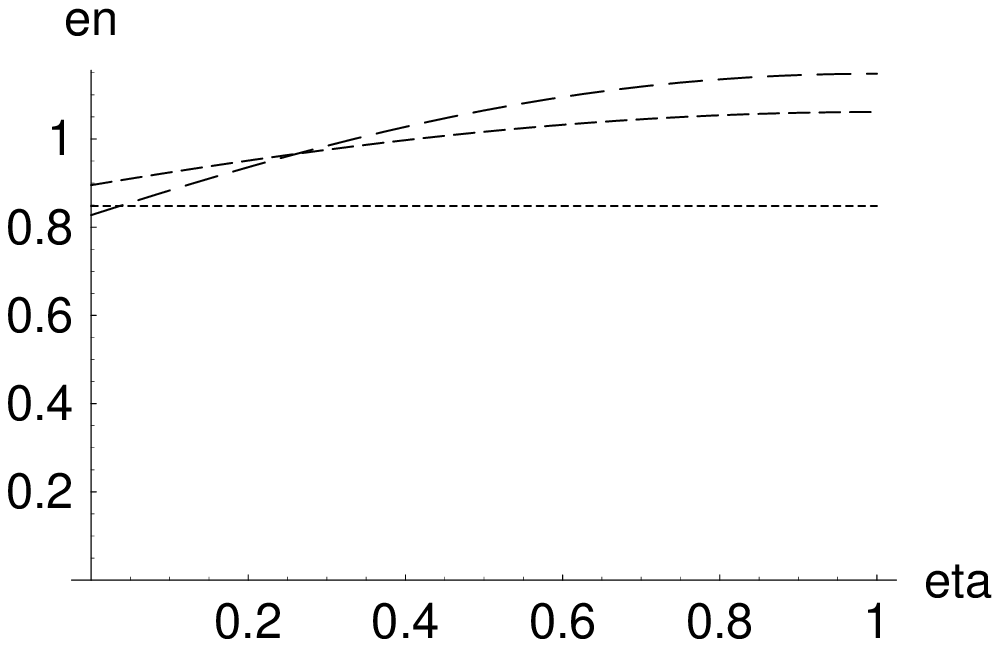}\includegraphics[width=8cm]{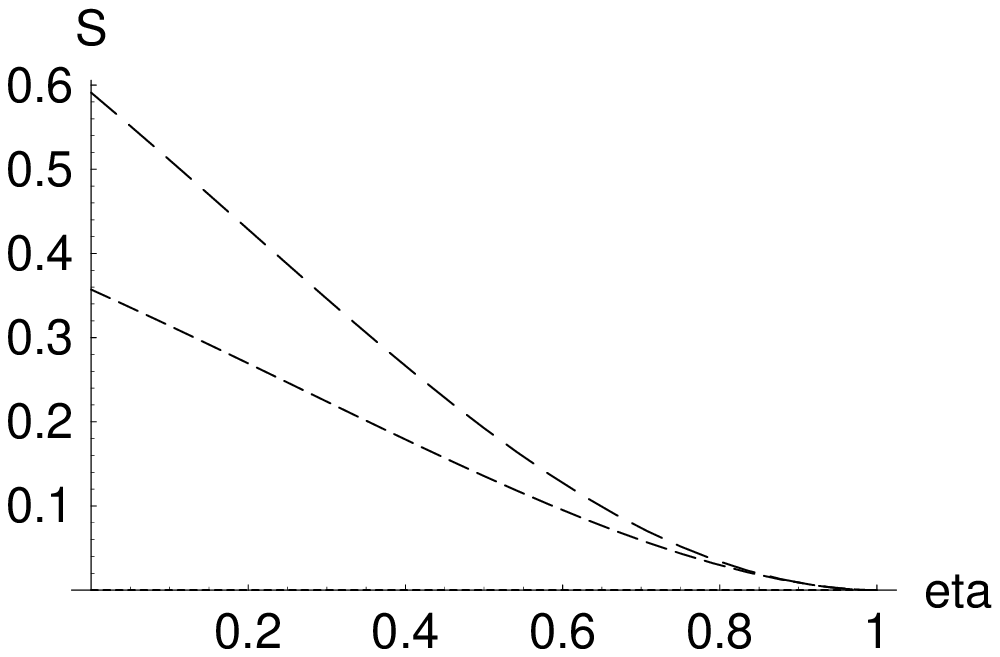}
\end{center}

\caption{\label{fig:impdetpsialogn}\emph{Imperfect detectors.} Plotted here are (a) the logarithmic negativity $E_\mathcal{N}$ and (b) the entropy $S_\textrm{vN}$ of the states generated after zero (dotted), one (dashed) and two (wider dashing) iterations when  detectors with efficiency $\eta$ are used. The input state is $\ket{\psi_a}$ defined in the text.}
\end{figure}

\begin{figure}
\psfrag{en}{$E_\mathcal{N}$}
\psfrag{eta}{$\eta$}\psfrag{S}{$S_\textrm{vN}$}\hspace{-0.5cm}(a)\hspace{8.0cm}(b)
\vspace{-1cm}
\begin{center}
\includegraphics[width=8cm]{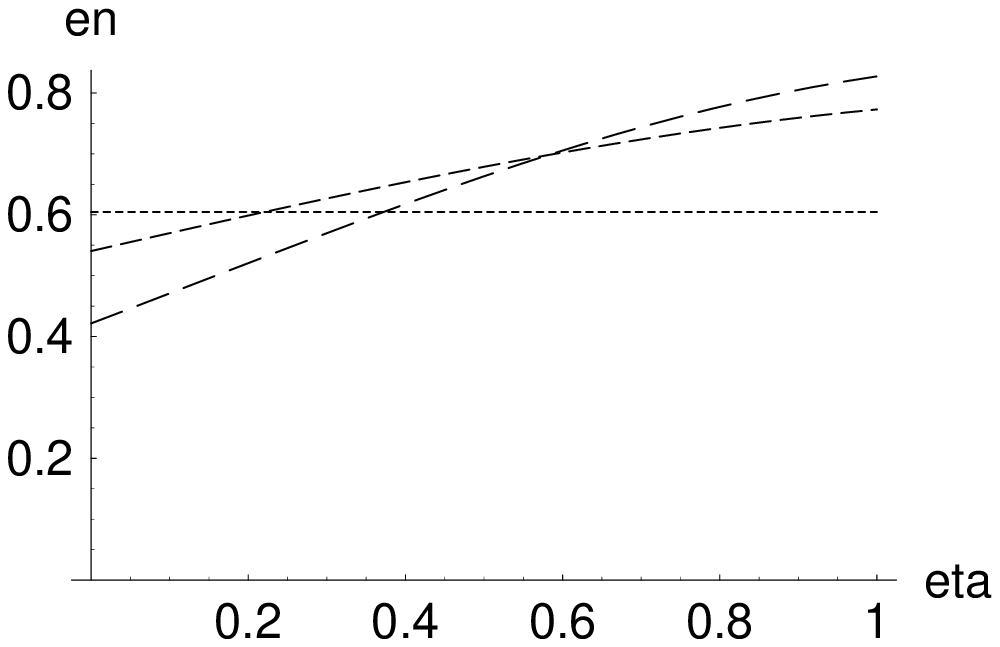}\includegraphics[width=8cm]{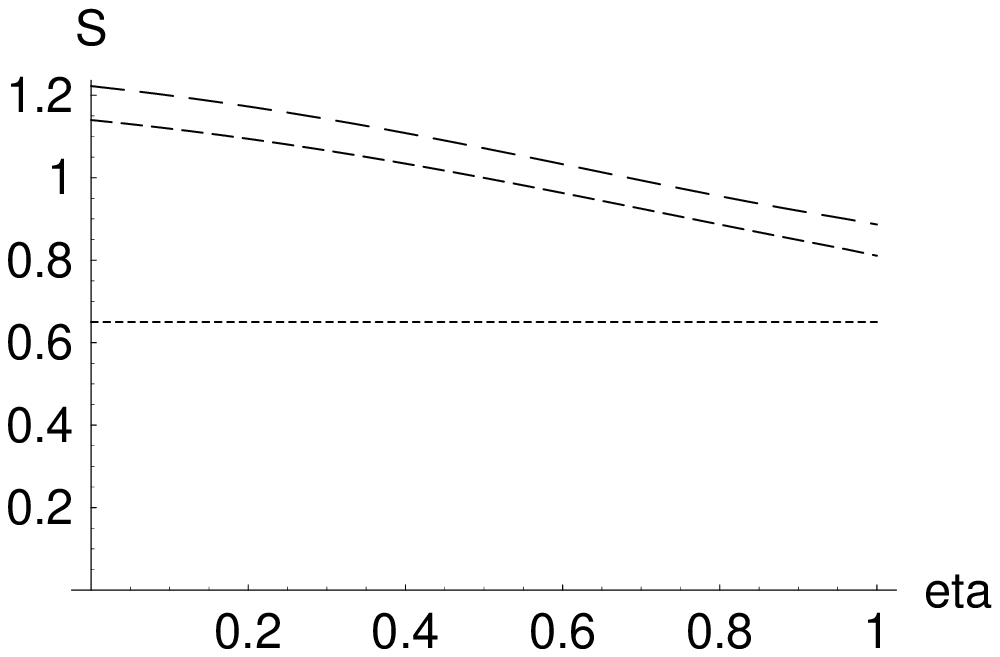}
\end{center}
\caption{\label{fig:impdetrhoblogn}\emph{Imperfect detectors.} Here we see how the scheme copes with inefficient detection when a mixed input state is used. The degree of entanglement (a) and the entropy (b) are plotted against detector efficiency for two iterations as in figure~\ref{fig:impdetpsialogn},  with mixed state $\rho_b$ as input.}

\end{figure}

\begin{figure}
\psfrag{en}{$E_\mathcal{N}$}
\psfrag{eta}{$\eta$}
\begin{center}
\includegraphics[width=10cm]{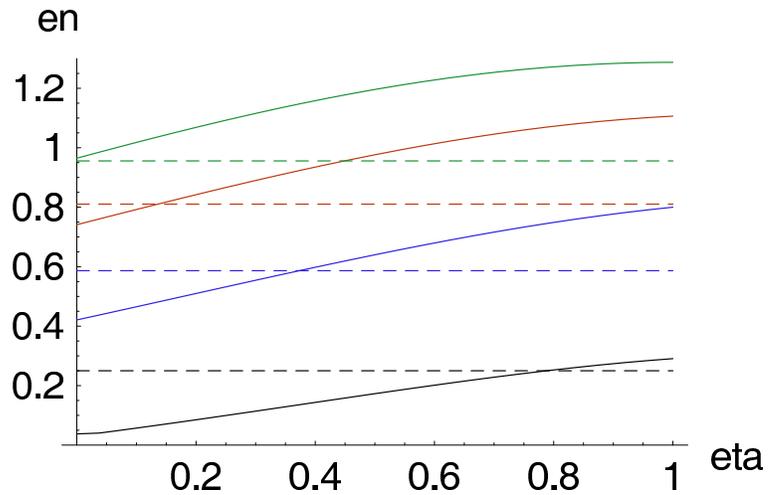}
\end{center}
\caption{\label{fig:noisyimplogn} \emph{Imperfect detectors.} Here we see the trade-off between the mixedness of the input state and the necessary efficiency to see an increase in entanglement in the first iteration. The logarithmic negativity of the initial state (solid) line and after one iteration (dotted line) are plotted as a function of detector efficiency $\eta$ for the state $\rho$ (defined in equations~(\ref{mixedinputrhost}) to~(\ref{mixedinputrhofin})) with $\lambda=1$ and $\tau$ taking the following values: 0.25 (black lines), 0.5 (blue lines), 0.75 (red lines) and 1.0 (green lines).}
\end{figure}

 In figures~\ref{fig:impdetpsialogn} and~\ref{fig:impdetrhoblogn} we see logarithmic negativity of the input state, $\ket{\psi_a}$ and $\rho_b$ respectively, and the states generated by one and two iterations of the protocol. Remarkably, the performance of the protocol is not greatly degraded by the use of imperfect detectors. The only effect we see is a gentle decline in the entanglement with decreasing efficiency. This decrease is more severe for $\rho_b$, and strongest for its second iteration. This may be attributed to the significant increase in entropy of the state at the first step. In fact, there is a trade-off here between the detection efficiency and the mixedness of the input upon which the protocol can still increase entanglement. This is seen clearly in figure~\ref{fig:noisyimplogn} 
where we plot the logarithmic negativity for input states and first iteration,  for the example mixed input state $\rho$ defined in equations~(\ref{mixedinputrhost}) to~(\ref{mixedinputrhofin}) with parameters $\lambda=0.7$ and varying $\tau$. The mixedness of the input state increases with decreasing $\tau$ and we see correspondingly that a higher detector efficiency is required to achieve any gain in entanglement.

Let us discuss why the scheme still functions well with imperfect detectors even though it relies on vacuum detections. One important reason is that in the input states we have considered, the mean photon number of the pairs of pulses is low, so  when one detector does fail, it is not very likely that a photon is present to be missed by the detector. This feature will be generic, as the scheme is designed for, and only converges for, input states which are mostly vacuum. Other symmetries in the input states we have chosen enhance this effect. For example, with the input state $\ket{\psi_a}$, there are correlations in the photon numbers of the detected modes which mean that errors caused by only one detector failing are suppressed. If one of the detectors correctly registers the vacuum, the chance of a single photon being present in the other detected mode is zero and the chance that a two-photon state is present is only 1.4\%. This means that in a demonstration experiment, if input states generated by a Procrustean methods as described in chapter~\ref{ch:7} were employed, a large entanglement increase could be observed in an implementation of the Gaussification protocol with current detector efficiencies.

While inefficient detectors are clearly a very important experimental imperfection for the scheme, there are other obstacles which would make the implementation of the full iterative scheme difficult. In the next section we discuss the effect of decoherence in the pulses when they are stored between iterations.

\subsection{Decoherence between iterations}\label{sec:decoh-betw-iter-1}
For a demonstration experiment of a small number of iterations of the protocol, one could avoid the need for storing pulses between iterations by sending through all states to their next iterations, and then repeating the experiment many times until all measurement outcomes are successful in ``one shot''. The probability of success will decrease approximately exponentially with increasing number of steps so this is not practical for many iterations. 
To avoid the exponential decrease in success probability  one would have to store pulses resulting from a successful iteration until another pair  has been successfully generated. Additionally storage would also be required for every single pulse generated by a successful iteration, as Alice and Bob would need to wait for the signal from each other before they knew if the iteration had been successful on both sides. If Alice and Bob are not too far from each other, however, this storage time need not be long, so we shall neglect it here.

So far, there have been several  proposals towards the storage of the quantum states of light pulses. The first is to simply divert the light pulse into a loop of optical fibre until it is needed. The use of loops of fibre to temporarily store light pulses is commonplace in current experiments. However, the ability to switch a pulse into and out of storage on demand on the time-scales which would be required here is a difficult technological challenge.

A second, and more radical approach is to use electro-magnetically induced transparency effects to transfer the quantum state onto collective excitations in  atomic vapours\cite{eitprop}. The first experiments on this effect have been  promising \cite{eitexp}, but it remains a complicated procedure, and currently realised protocols are suitable only for the storage of qubits, not higher-dimensional states. Nevertheless, the storage of quantum states of light pulses is of vital importance for many proposals of optical quantum information processing, including quantum computation\cite{klm} and is thus a very active research area.

  In this section, therefore, we shall not restrict ourselves to considering a particular method of storing light, but shall consider three kinds of generic decoherence processes; absorption, dephasing and Gaussian phase defusion. The first of these would be particularly important in fibre loop storage and the second  could well be an important decoherence mechanisms in atomic vapour storage\cite{fleisch}. The third process, phase defusion, would be due to noise in the laser source itself, which could mean that pairs of pulses produced at different times are no longer exactly phase coherent.

In order to model these effects,  the recurrence relation in equation~(\ref{recurrence1}) needs only minor modification for the case where two different states $\rho^{(i)}$ and $\sigma^{(i)}$ are used as input to the $(i+1)$ iteration,

\begin{equation}\label{recurrencediff}
        \rho_{a,b,c,d}^{(i+1)} =
        \sum_{s=0}^a
        \sum_{t=0}^b
        \sum_{n=0}^c
        \sum_{m=0}^d
        M_{a,b;c,d}^{s,t;n,m} 
        \rho^{(i)}_{s,t;n,m} \sigma_{a-s,b-t;c-n,d-m}^{(i)}
\end{equation}
where the coefficients  $M_{a,b;c,d}^{s,t;n,m}$ are given in equation~(\ref{nl1}). Let us  assume that one of the pair of input states $\rho^{(i)}$ was created directly before the iteration  and therefore has undergone no decoherence. The other state $\sigma_{(i)}=\mathcal{D}[\rho{(i)}]$ is the result of the same state undergoing decoherence captured by super-operator $\mathcal{D}$. For simplicity we assume symmetry between all attempts and also that the same  decoherence occurs at each iteration.

The first decoherence process we consider is  absorption.

\begin{figure}
\psfrag{en}{$E_\mathcal{N}$}
\psfrag{th}{$\theta$}
\psfrag{S}{$S_\textrm{vN}$}
\hspace{-0.5cm}(a)\hspace{8.0cm}(b)
\vspace{-1cm}
\begin{center}
\includegraphics[width=8cm]{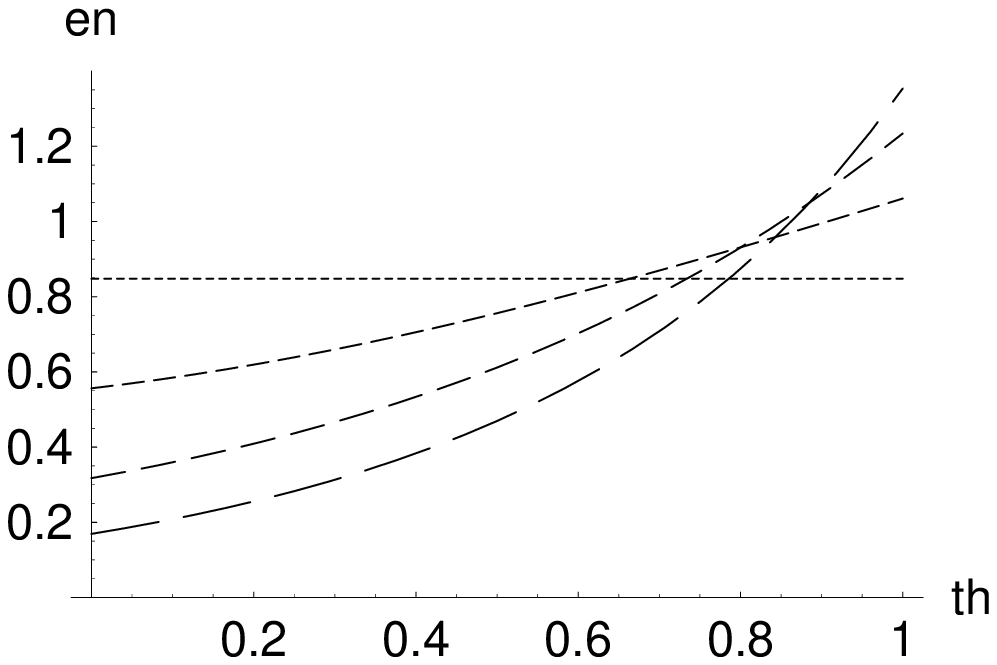}\includegraphics[width=8cm]{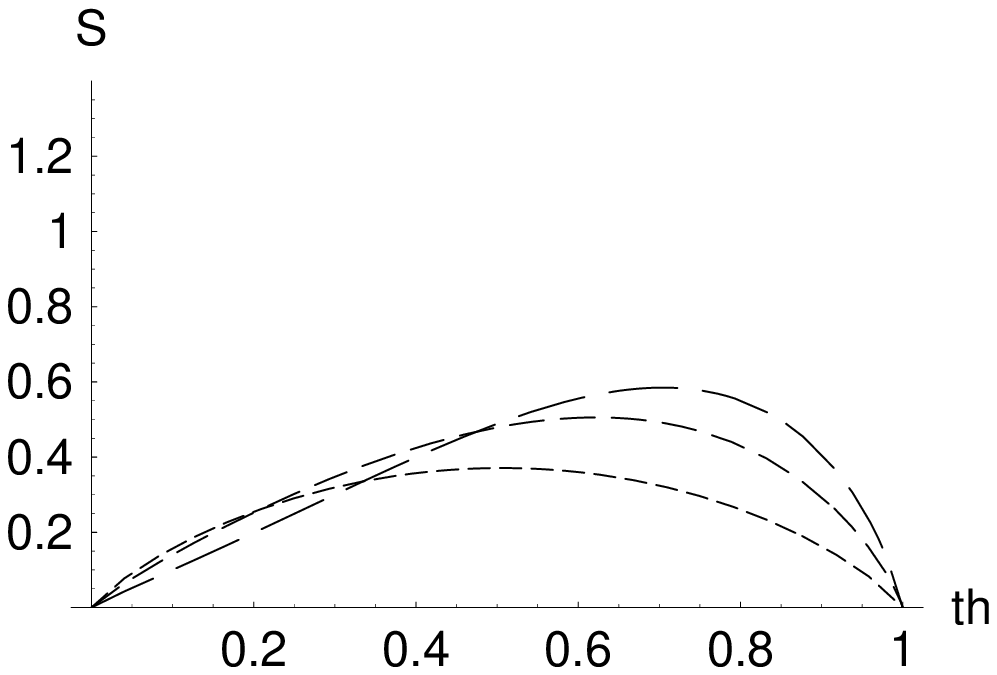}
\end{center}
\caption{\label{fig:psiagauss} \emph{Absorbing channel.} We plot here (a) the logarithmic negativity and (b) the entropy of the state $\rho{(i)}$ generated by $i$ iterations of the protocol for $i=0$ (solid), $i=1$ (dotted), $i=2$ (dashed) and $i=3$ (wider dashing) where one of the inputs at each stage was stored in a Gaussian absorbing channel with transmission coefficient $\theta$. The initial state is $\ket{\psi_a}_b$ defined above.}
\end{figure}

\begin{figure}
\psfrag{en}{$E_\mathcal{N}$}
\psfrag{th}{$\theta$}
\psfrag{S}{$S_\textrm{vN}$}
\hspace{-0.5cm}(a)\hspace{8.0cm}(b)
\vspace{-1cm}
\begin{center}
\includegraphics[width=8cm]{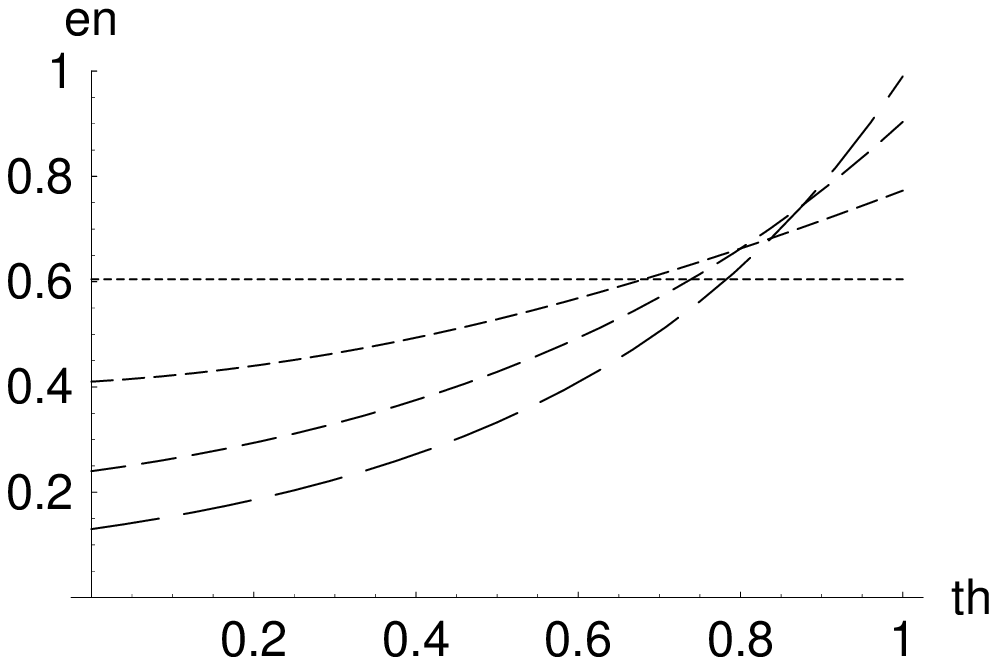}\includegraphics[width=8cm]{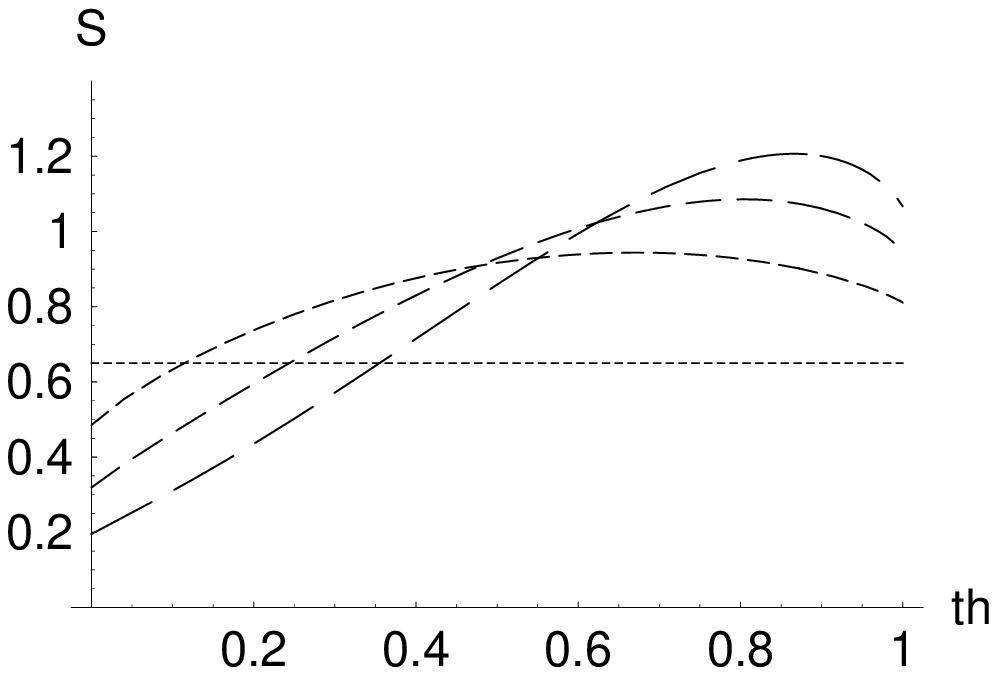}
\end{center}
\caption{\label{fig:rhobgauss} \emph{Absorbing channel.}  Now we see the effect of the absorbing channel for photon storage when the input state in $\rho_b$. For this input state we plot (a) the entanglement and (b) the entropy with teh above conventions.}
\end{figure}

\subsection{Absorbing channel}
 Gaussian absorption has already been  described in section~\ref{sec:entdistrib}. If we set the transmission coefficient of the storage loops  to $\theta$, the channel  is described on the matrix element level via the following relation
\begin{equation}\label{eq:decoherenceloop}
\sigma_{a,b,c,d}^{(i)}=\sum_{j,k}\sqrt{{a+j\choose j}{b+k\choose k}{c+j\choose j}{d+k\choose k}}\sqrt{\theta}^{a+b+c+d}(1-\theta)^{j+k}\rho^{(i)}_{a+j,b+k,c+j,d+k}
\end{equation}

Combining equations~(\ref{recurrencediff}) and~(\ref{eq:decoherenceloop}), we simulate two iterations of the protocol for the model states $\ket{\psi_a}$ and $\ket{\rho_b}$. 
The results of these simulations for the test-states $\ket{\psi}_a$ and $\rho_b$ are plotted in figures~\ref{fig:psiagauss} and~\ref{fig:rhobgauss}. We see that the performance of the protocol declines steeply as the transmission coefficient gets lower. As one would expect, this is initially accompanied by an increase in the entropy of the states, but then the entropy is seen to decline with decreasing $\tau$. This is because as $\tau$ gets closer to zero, one of the input states becomes closer to the pure vacuum state.
We see  similar behaviour for the pure input state $\ket{\psi_a}$ and $\rho_b$. This is in contrast to the case of inefficient detectors where mixed input states were increasingly less robust than pure inputs. 

Optical fibres have a high transmission coefficient. For example, infra-red optical fibre with an attenuation coefficient of 3dB/km is commercially available \cite{fibre}. This means that the transmission coefficient of a fibre of $d$ km is $0.93^d$. For a reasonable increase in entanglement, figures~\ref{fig:psiagauss} and~\ref{fig:rhobgauss} show that the absorbing channel with a transmission coefficient of greater than 95\% would be desirable. This corresponds to a length of fibre of around 750m, or a maximum storage time of the order of microseconds. With a pulse repetition rate on the order of GHz and, remembering that the Gaussification iterations typically have success probabilities on the order of 1/2, one would, if perfect switching were available, be able to store pulses for the necessary waiting time between iterations without enough absorption occurring to disrupt the protocol.

\subsection{Partially dephasing channel}

Partial dephasing may be an important decoherence process for quantum states of light trapped in atomic vapour \cite{fleisch} or other novel storage methods.
It can be considered in one of the following equivalent ways; that with probability $\kappa$ a random phase rotation is applied to the state, or that with probability $\kappa$ the environment ``measures'' the photon number of the state. The effect of this is to reduce all off-diagonal elements by factor $\kappa$. Thus it is particularly simple to simulate. When both modes  of a two-mode state pass through such a channel, the matrix elements transform,

\begin{equation}
\sigma^{(i)}_{a,b,c,d}= \kappa +\kappa(1-\kappa)\delta_{a,c} +\kappa(1-\kappa)\delta_{b,d} + +\kappa^2\delta_{a,c}\delta_{b,d}
\end{equation}
where $\delta_{i,j}=1$ when $i=j$ otherwise  $\delta_{i,j}=0$.

\begin{figure}
\psfrag{en}{$E_\mathcal{N}$}
\psfrag{S}{$S$}
\psfrag{X}{$\kappa$}
\hspace{-0.5cm}(a)\hspace{8.0cm}(b)
\vspace{-1cm}
\begin{center}
\includegraphics[width=8cm]{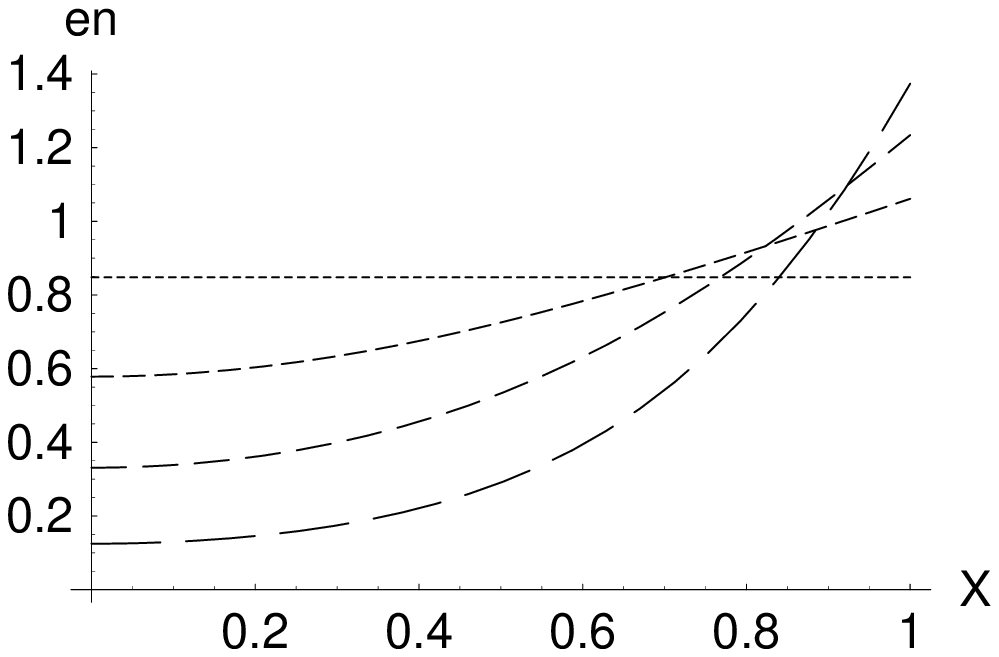}\includegraphics[width=8cm]{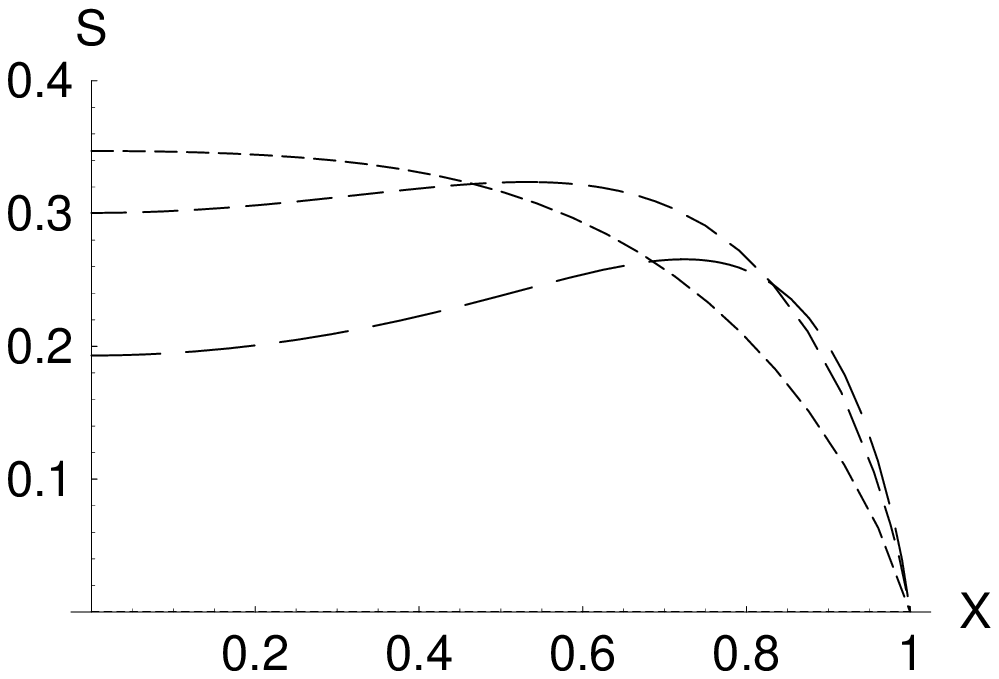}
\end{center}
\caption{\label{fig:dephchpsi} \emph{Partially Dephasing Channel.} Here see the effect of using a partially dephasing channel with parameter $\kappa$ on the entanglement (a) and entropy (b) for pure input $\ket{\psi_a}$ -- Same conventions  as    figure~\ref{fig:psiagauss}.}
\end{figure}

\begin{figure}
\psfrag{en}{$E_\mathcal{N}$}
\psfrag{S}{$S$}
\psfrag{X}{$\kappa$}
\hspace{-0.5cm}(a)\hspace{8.0cm}(b)
\vspace{-1cm}
\begin{center}
\includegraphics[width=8cm]{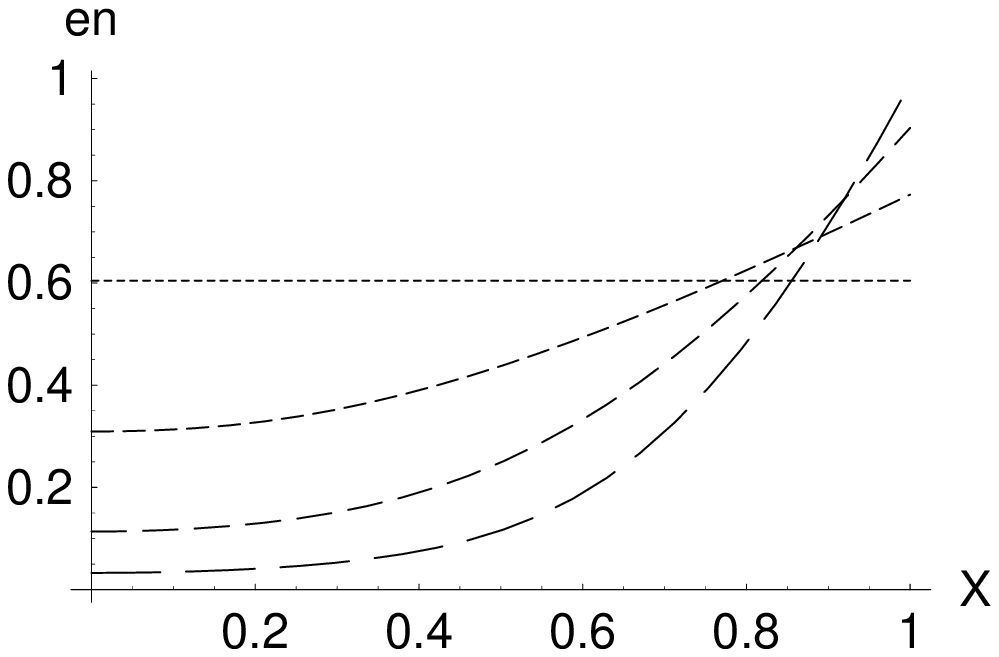}\includegraphics[width=8cm]{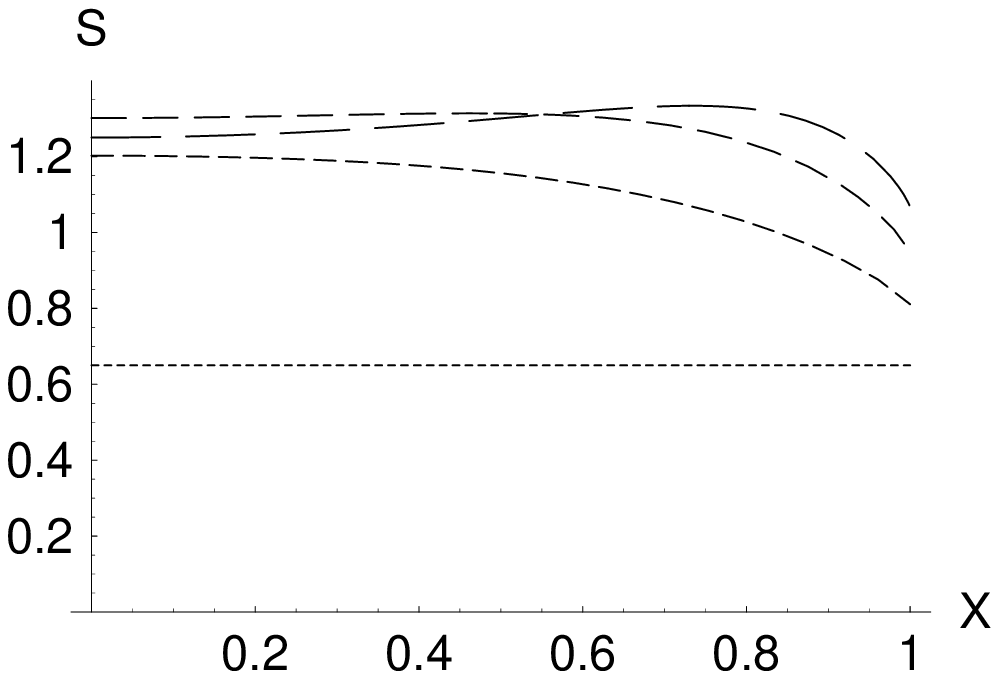}
\end{center}
\caption{\label{fig:dephchrho} \emph{Partially Dephasing Channel.} Here see the effect of using a partially dephasing channel with parameter $\kappa$ on the entanglement (a) and entropy (b) for mixed input $\rho_b$ -- Same conventions  as    figure~\ref{fig:psiagauss}.}
\end{figure}

Results of the simulations on input states $\ket{psi_a}$ and  $\rho_b$ are plotted in figures~\ref{fig:dephchpsi} and~\ref{fig:dephchrho}.

\begin{figure}
\psfrag{en}{$E_\mathcal{N}$}
\psfrag{S}{$S$}
\psfrag{X}{$\kappa$}
\begin{center}
\includegraphics[width=11cm]{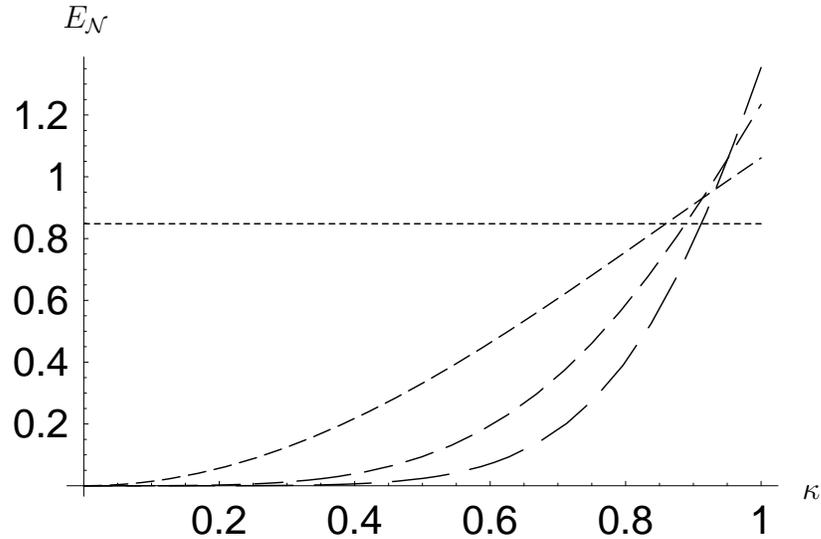}
\end{center}
\caption{\label{fig:dephsymm}Here we plot the logarithmic negativity for three iterations of the protocol on input $\ket{\psi}$, where, before each iteration \emph{both} states pass through a partially dephasing channel with dephasing parameter $\kappa$.}
\end{figure}

\subsection{Phase diffusion channel}

In this section we consider the effect of a random decorrelation in the phase of the laser pulse. As mentioned briefly in section~\ref{sec:laserlight}, noise in the lasing process manifests itself as weak random fluctuations in the phase of the laser source. Thus, the phase of the earlier stored pair of pulses, may not be the same as its partner pair in the next iteration. We model this as phase diffusing channel acting on the stored pulses. We model the phase fluctuations here as a random phase rotation through angle $\phi$ with $\phi$ obeying a Gaussian probability distribution with standard deviation $\upsilon$. Furthermore, since both sides of the state were generated at the same time, we assume that both suffer an equal phase shift. This channel transforms the matrix elements of $\rho^{(i)}$ as follows,

\begin{equation}\begin{split}
\sigma_{a,b,c,d}^{(i)}=&\int d \phi \left[\frac{1}{\upsilon \sqrt{2\pi}}e^{-\frac{\phi^2}{2\upsilon^2}}\right]e^{i\phi(a+b-c-d)}\rho^{(i)}_{a,b,c,d}\\
=&e^{-\frac{(a+b-c-d)^2\upsilon^2}{2}} \rho^{(i)}_{a,b,c,d}\ .
\end{split}
\end{equation}

\begin{figure}
\psfrag{en}{$E_\mathcal{N}$}
\psfrag{S}{$S$}
\psfrag{sig}{$\upsilon$}
\hspace{-0.5cm}(a)\hspace{8.0cm}(b)
\vspace{-1cm}
\begin{center}
\includegraphics[width=8cm]{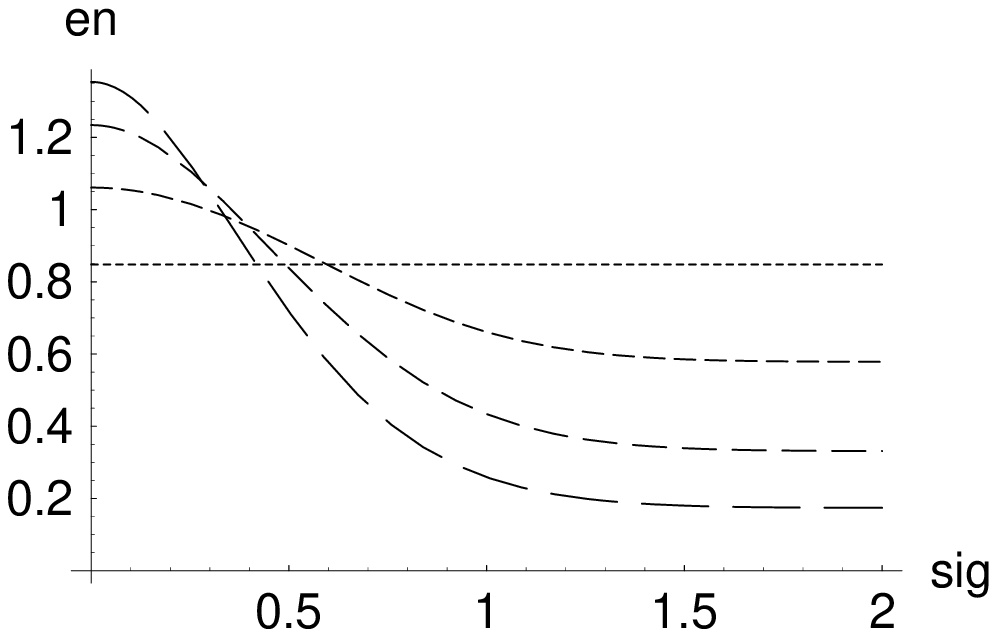}\includegraphics[width=8cm]{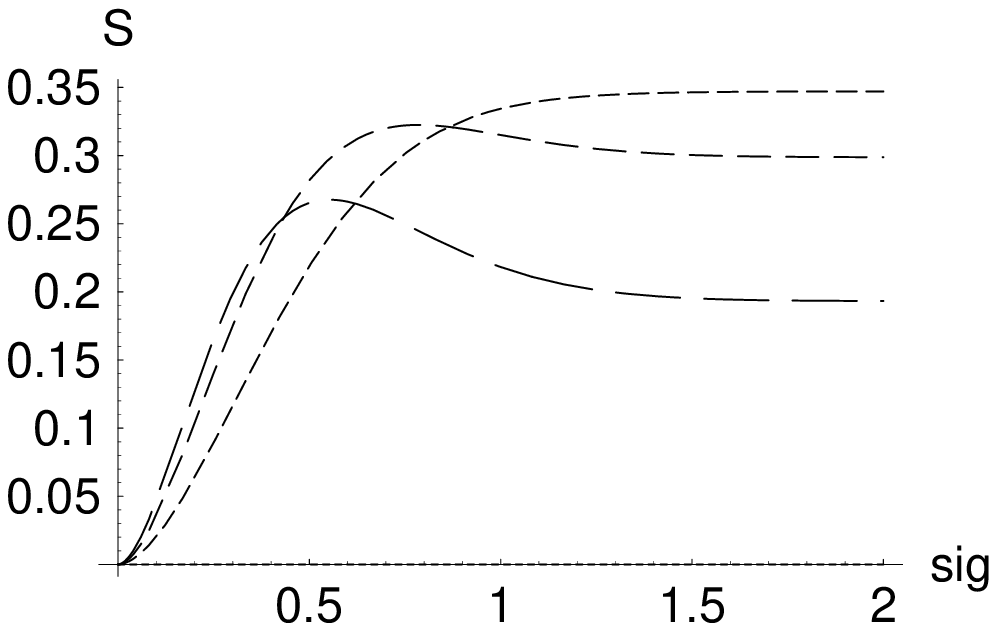}
\end{center}
\caption{\label{fig:phasediffs} \emph{Phase Diffusion Channel.} Here see the effect of storing one pulse in a phase diffusing channel  with standard deviation $\upsilon$ between iterations on the entanglement (a) and entropy (b) for pure input $\ket{\psi_a}$ -- Same conventions  as    figure~\ref{fig:psiagauss}.}
\end{figure}

\begin{figure}
\psfrag{en}{$E_\mathcal{N}$}
\psfrag{S}{$S$}
\psfrag{sig}{$\upsilon$}
\hspace{-0.5cm}(a)\hspace{8.0cm}(b)
\vspace{-1cm}
\begin{center}
\includegraphics[width=8cm]{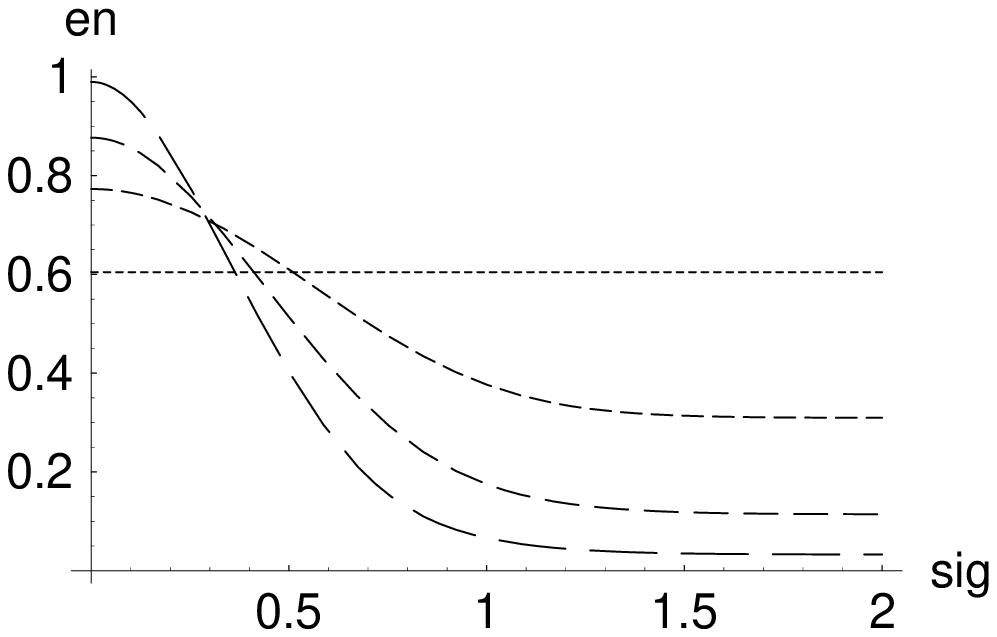}\includegraphics[width=8cm]{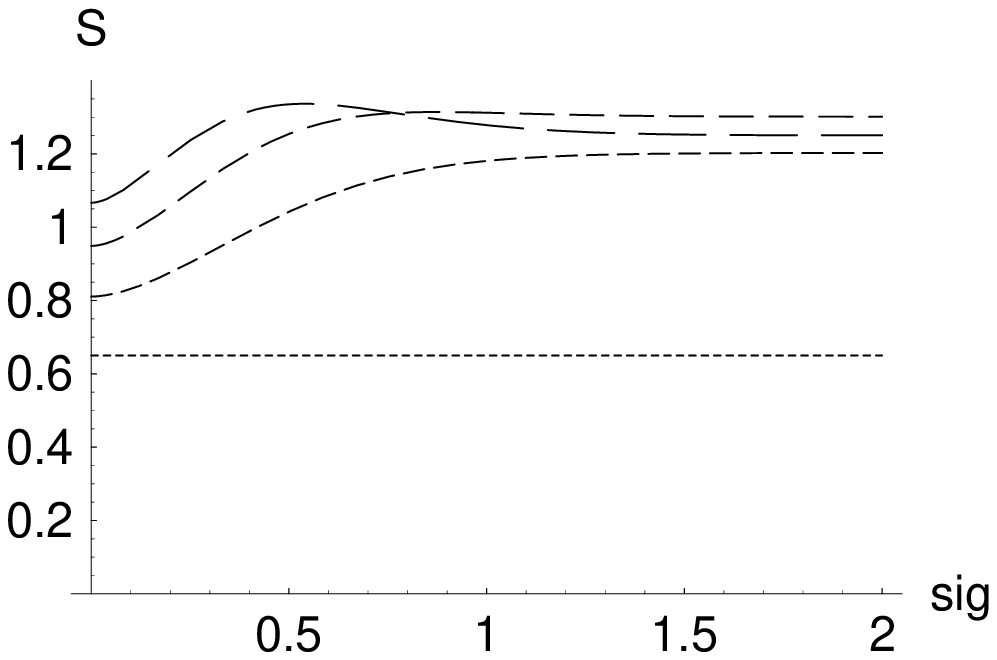}
\end{center}
\caption{\label{fig:phasediffsrho} \emph{Phase Diffusion Channel.} Here see the effect of storing one pulse in a phase diffusing channel  with standard deviation $\upsilon$ between iterations on the entanglement (a) and entropy (b) for mixed input $\rho_b$ -- Same conventions  as    figure~\ref{fig:psiagauss}.}
\end{figure}

The effects of this ``phase diffusion'' on the Gaussification were, again, simulated numerically for input states $\ket{\psi_a}$ and $\rho_b$ and the results are plotted in figures~\ref{fig:phasediffs} and~\ref{fig:phasediffsrho}. Note that, unlike the other two cases, this kind of ``decoherence'' can be prevented by simply making relative phase measurements on the reference pulses for each pulse and correcting accordingly. Of course, this would greatly increase the complexity of the implementation.

\subsection{Comparison of decoherence effects}

As the results of simulations of all three decoherence processes share many similar features we shall discuss them together. Firstly, we notice that in all of these cases, the decoherence has a much more serious effect on the performance of the protocol that we saw for imperfect detectors. 
In each step, we see a sharp decrease in the entanglement gained  for even a small amount of decoherence.  We also notice that, in contrast to the previous case,  the degradation caused to the entanglement gain is similar for both input states $\ket{\psi_a}$ and $\rho_b$ for all three types of decoherence. This is an indication that the special symmetry of $\ket{\psi_a}$ does not shield it from the effect of the decoherence processes. 

In the example we have chosen, the input at each stage is asymmetric. Since the protocol is intended to operate on symmetric pairs of states, we should consider whether degradation in the scheme due to these decoherence processes could be more due to the asymmetry in the input rather than the decoherence itself. To test this we simulated the effect of iterating the procedure on pairs of states which both pass through a partially dephasing channel. The results are plotted in figure~\ref{fig:dephsymm}. We see that the degradation in entanglement for a given dephasing parameter $\kappa$ is much worse than the case plotted in figure~\ref{fig:dephchpsi}. It is therefore reasonable to conclude that it is the decoherence itself, rather than the asymmetry in the input which is degrading the protocol in these cases.

\begin{figure}
\hspace{-0.1cm}(a)\hspace{7.0cm}(b)
\begin{center}
\includegraphics[width=7.0cm]{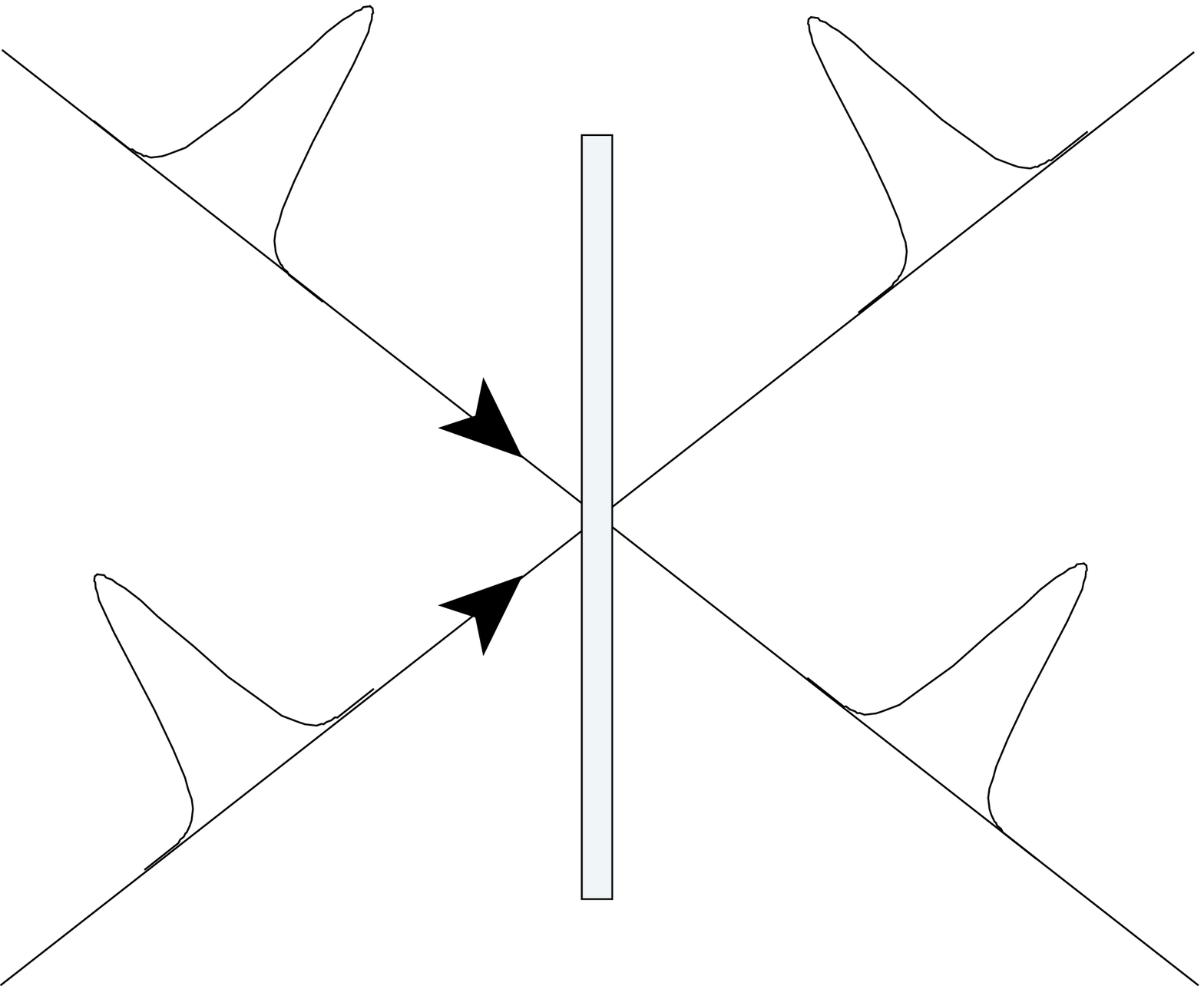}\hspace{1cm}\includegraphics[width=7.0cm]{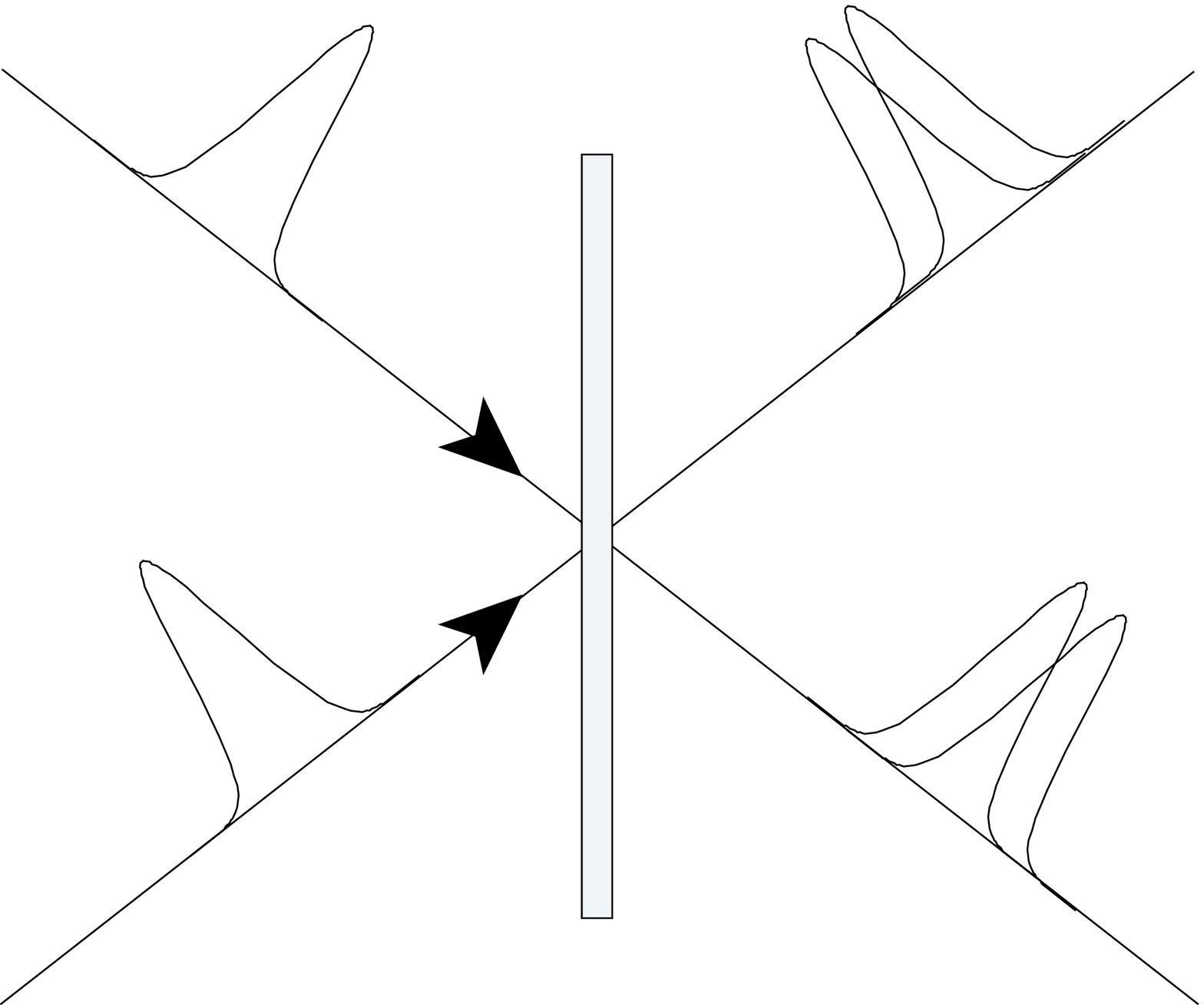}
\end{center}
\caption{\label{fig:mode-match}This figure illustrates the problem of mode-matching optical pulses meeting on a beam-splitter. If  (a) the modes ``match'' spatially and temporally as well as in frequency, then the desired beam splitter operation on the pulses is realised. If (b) the modes do not perfectly match, not only is the beam splitter transformation not perfectly realised, but the shape of the mode envelope for each pulse is disrupted.}
\end{figure}

We have presented a variety of decoherence mechanisms which are simple to calculate. Of course, in a real situation a combination of these effects and others would occur. Nevertheless it is clear that the scheme is not very robust against decoherence between iterations. This emphasises the fact that the development of a method of storing the quantum state of light with a low degree of decoherence is vital for the experimental implementation of complex optical protocols in quantum information science.

\subsection{Other potential problems}\label{sec:other-potent-probl}

The protocol could be implemented in free-space optical system, or in optical fibres. While the free space implementation will lead to less absorption, it suffers from a technical problem which limits the size of any optical network for laser pulses which employs beam splitters. This is the problem of mode matching the pulses in incident on each beam splitter, illustrated in figure~\ref{fig:mode-match}. Essentially, this means that for the desired beam splitter transformation to be obtained between the pulses, they must match perfectly, both in frequency and spatial-temporal envelope. In optical fibres,  \emph{fibre couplers} replace beam splitters, then the mode-matching is much easier, since  spatial mode-matching is guaranteed by the fibre. Thus it is expected that the full implementation of schemes such as the Gaussification protocol will need a fibre-based implementation.

\section{Conclusion}

In this chapter, we have presented an iterative non-deterministic protocol called ``Gaussification''. The protocol allows the generation of approximately Gaussian bi-partite states from a non-Gaussian supply. We have shown that for certain classes of input states, the number of iterations need not be too high, before a near Gaussian form is reached. We have also shown that the entanglement of the states is often increased in the protocol. We have comprehensively analysed the properties of the scheme when all operations are ideal, and also simulated how the scheme may perform when implemented with realistic imperfect operations. We have found that the scheme can still perform well when only detectors of currently available efficiencies are employed, but that the scheme will be badly degraded if the pulses undergo decoherence between iterations. Thus, we believe that while a demonstration of entanglement gain in a few iterations of the protocol can be achieved with current technology, until an effective means of storing the quantum state of light is developed, an implementation with many iterations generating a very highly entangled state will be out of reach. Nevertheless, even the implementation of a simple demonstration experiment would be of enormous importance as it would be the first implementation  of entanglement distillation in the continuous variable regime and a great step forward in our ability to manipulate entanglement in light.


%% file: conc.tex
\chapter{Summary and Outlook}

The generation of highly entangled states in spatially separated  sub-systems promises many useful applications, and will allow tests of some of the fundamental properties of quantum mechanics.
In this thesis, we have presented new proposals for the generation and manipulation of entanglement in quantum optical systems. Here we present  a   summary.

In chapter~\ref{ch:4}, we introduced a proposal for the generation of a maximally entangled Bell state between two spatially separated cavity modes. The basis of the proposal is a very brief interaction between a mediating atom and each cavities. We have seen that this gives the proposal a remarkable robustness against experimental errors such as imperfect control over the atomic speed and path and detection inefficiency.

In chapter~\ref{c-5} we described a novel method of creating Bell states in ions trapped in spatially separated cavities. This proposal uses continual weak driving of the atoms and a detection of a photon after which-path information is erased by a beam-splitter.
This scheme has many advantages over previous proposals. In the strong coupling regime, where decoherence is negligible, and with ideal detectors, it becomes near deterministic in spite of the crucial role played by the detectors. Furthermore, when likely experimental imperfections are taken into account the scheme still allows  high fidelity Bell states to be generated.

In chapter~\ref{ch:7} we showed how simple Procrustean linear optical protocols can generate highly entangled non-Gaussian states from a supply of weakly entangled Gaussian states, and we introduced, in chapter~\ref{c-8}, an iterative procedure which we call ``Gaussification''. This is a simple linear optical protocol, which generates, from a large supply of non-Gaussian states, a small number of Gaussian states which can be significantly more entangled than the input supply. Combining these two procedures gives a  linear optical entanglement distillation procedure. We have assessed the performance of the protocol when important experimental imperfections are taken into account. We showed that the protocol can still produce significant gains in entanglement when detectors with the efficiencies currently available are employed, and thus would be suitable for a demonstration experiment with current detectors. We also showed that  the scheme can be badly  degraded by  decoherence in the pulses when stored between iterations. This highlights the experimental challenge a full many-iteration implementation of the scheme  would pose.

There are many avenues  for future work based upon these results. 
The schemes presented in chapters~\ref{ch:4} and~\ref{c-5} used an approach of weak interaction plus measurement to allow robust entangled state preparation. We have focussed upon cavity QED implementations, but this principle would naturally lend itself to other kinds of system. In particular, in  systems where couplings exist, efficient measurement may be made and robustness is needed,  such as coupled charge qubits \cite{naturecharge}, this approach could be very fruitful.

In chapter~\ref{c-5}, the detection of a photon induces a projection onto a maximally entangled state of the two atoms.  
 It has recently been shown \cite{brownerudolph} that this kind of projection can be used to create special multi-particle entangled states called ``cluster states'' \cite{briegelcluster} which have many interesting properties. In particular, they are a resource for quantum computation  through local measurements alone\cite{raussenbrowne}. It would be a simple matter to apply these ideas to the setup we have outlined, which would then lead to a robust method for generating useful multi-party entangled states in the cavity QED setting.

The entanglement distillation scheme we have presented in chapter \ref{c-9}, while promising in its current form, could still be improved. In particular the Procrustean entanglement concentration step, which incorporates the non-Gaussian operation for the distillation, has a very low success probability. Recent results  have suggested that the cross-Kerr non-linearities one can produce using electro-magnetically induced transparency in atomic vapours \cite{schmidtim}, can be useful for the manipulation of quantum states of light beyond the linear optical regime \cite{billeit,billeit2}.  Such operations are certainly non-Gaussian and it would be worthwhile investigating whether a more effective non-Gaussian preparation step might be developed using these techniques.

We have shown that the Gaussification process generates pure entangled states from certain particular mixed input states. In principle, all input states could be transformed into this special form by an appropriate completely positive map. However, this does not mean that either the map is easy to implement experimentally, or that the state produced after the resultant Gaussification would still have more entanglement than the input. Work on characterising the general operations which can be achieved linear optically is progressing \cite{cambridgedude} and it would be very interesting to apply these ideas to this case.

The most important obstacle to the full implementation of the Gaussification protocol and linear optical quantum computation is that no effective method of storing quantum states of light between non-deterministic steps exists. Whilst important first steps towards this are being made in atomic vapour experiments \cite{eitprop,eitexp}, this work emphasises the importance of  investigating other novel kinds of quantum memory for light. 

The measurement of quantum systems has long been our best connection with the strange phenomena which occur at quantum scales, and it has played an 
 increasingly important and diverse role  as the field of quantum information science has developed. In this thesis we have demonstrated the power that measurement can wield in particular for  the preparation of useful quantum states and the manipulation of quantum information. We hope that these ideas can make a contribution  towards the realisation of quantum information science's extraordinary potential.


%% file: backmatter.tex

\bibliographystyle{thesbib}
\bibliography{thesis}


%% file: acknow.tex
\chapter*{Acknowledgements}

There are many people without whose help, advice and moral support this work  could not have been completed.

I would like to warmly thank my supervisor Martin Plenio, for his constant good natured  support and guidance throughout the past three years. I am very grateful that he has shared his good ideas with me and highlighted such fascinating and enjoyable directions of research.

I also  thank my other direct collaborators in the work in this thesis, Jens Eisert, Susanna Huelga and Stefan Scheel. It has been a pleasure to work with them and I am thankful that they shared  their invaluable experience and mathematical  expertise. I would also like to thank my examiners Peter Knight and Sam Braunstein for their helpful comments and suggestions for this final version of the thesis and for encouraging me to upload this work onto the pre-print arXiv.

The theoretical quantum optics and quantum information group has been a wonderful place to work over the past three years, and I am very grateful to all its members throughout my time there for providing such a friendly and stimulating academic environment; 
Koenraad Audenaert,
Almut Beige,
Adam Brazier,
Hugo Cable,
Angelo Carollo,
Luke Chipperfield,
Ivette Fuentes-Guridi,
Lucien Gaier,
Julian Hartley,
Clare Hewitt-Horsman, 
Elham Kashefi,
Viv Kendon,
Peter Knight,
Manfred Lein,
Yuan Liang Lim,
Christian Lunkes,
Damian Markham, 
Koji Maruyama,
Jiannis Pachos,
Terry Rudolph,
Marcelo Santos,
Ben Tregenna,
P-K Rekdal,
Jesus Rogel-Salazar,
Vlatko Vedral,
Shashank Virmani, 
 the many visitors who enlivened the group during my time there and
all members of our sister groups in the Quantum Optics and Laser Science section.
The group administrators have given invaluable help, in particular Kevin Goddard, Pinku Eeles and Jane Hardy and I would also like to thank Loli Sanchez, the departmental postgraduate secretary, for guiding me through the intricacies of college bureaucracy.

I am grateful to Bill Munro for good advice and assistance throughout my PhD and  to the other members of the  Quantum Information Group at Hewlett Packard Laboratories; Sean Barrett, Pieter Kok, Denzil Rodriguez and Tim Spiller, who, as well as sponsoring this work with a CASE award, provided a very welcoming and engaging place of work during the three months I spent with them in Bristol. I also warmly thank Ros Grimshaw and and Patrick Costeloe and all the residents of 6 Windsor Terrace, Bristol for letting me join their wonderful household while I was working there.

During the latter part of my PhD, I have benefited greatly from interaction with  colleagues in Oxford University, in particular, Christine Silberhorn and Ian Walmsley, to whom I am very thankful  for patiently sharing their experimental expertise and patiently answering my constant naive questioning on very basic matters.

This work would not have been possible without the earlier guidance of Christoph Keitel and the theoretical quantum dynamics group in Freiburg, and Hans Briegel and his group in Munich. In particular, my discussions there with Hans Aschauer, Christoph Gohle, Robert Rau\ss endorf  greatly helped me first get to grips with this diverse and fascinating research field.

I am very grateful to my parents, Chris and Martin, and my sister Tess for all their help and encouragement throughout the long voyage that has been my  education. Finally, I want to express my deep thanks to Lorna Crowhurst who has encouraged and supported me throughout  with patience, care and intelligence.
